\documentclass[IOP,onecolumn]{emulateapj}
\newcommand{\myemail}{Contact OP: o.porth@leeds.ac.uk, RK: }
\usepackage{amsmath}

\graphicspath{ {figures/} }

\slugcomment{Accepted for publication in APJS}

\shorttitle{{\tt MPI-AMRVAC} for Solar and Astrophysics}
\shortauthors{Porth et al.}

\usepackage{tikz}

\newcommand{\delt}{\partial_t}
\newcommand{\boldu}{\mathbf{U}}
\newcommand{\amrvac}{{\tt{MPI-AMRVAC}}\,}

\begin{document}

\title{MPI-AMRVAC for Solar and Astrophysics}

\author{O. Porth, C. Xia, T. Hendrix, S. P. Moschou and R. Keppens}
\affil{Department of Applied Mathematics, The University of Leeds, Leeds, LS2 9JT, UK\\
Centre for mathematical Plasma Astrophysics, Department of Mathematics, KU Leuven, Celestijnenlaan 200B, 3001 Leuven, Belgium}
\email{\myemail}

\begin{abstract}
In this paper we present an update on the open source \amrvac simulation toolkit where we focus on solar- and non-relativistic astrophysical magneto-fluid dynamics.  
We highlight recent developments in terms of physics modules such as hydrodynamics with dust coupling and the conservative implementation of Hall magnetohydrodynamics.  
A simple conservative high-order finite difference scheme that works in combination with all available physics modules is introduced and demonstrated at the example of monotonicity preserving fifth order reconstruction.  
Strong stability preserving high order Runge-Kutta time steppers are used to obtain stable evolutions in multi-dimensional applications realizing up to fourth order accuracy in space and time.  With the new distinction between active and passive grid cells, \amrvac is ideally suited to simulate evolutions where parts of the solution are controlled analytically, or have a tendency to progress into or out of a stationary state.  
Typical test problems and representative applications are discussed, with an outlook to follow-up research.  
Finally, we discuss the parallel scaling of the code and demonstrate excellent weak scaling up to $30\,000$ processors allowing to exploit modern peta-scale infrastructure.  
\end{abstract}


\section{Introduction}

Computational astrophysics witnessed an unprecedented growth in scope and diversity over the last decades. While dedicated, problem-tailored software efforts continuously explore novel solution methods for challenging applications, a fair amount of progress has been realized through open source, community driven software development. The study of astrophysical magneto-fluid dynamics in particular, encompassing pure solar as well as general astrophysical research topics, has seen a flurry of codes generating new insights. Original codes like ZEUS~\citep{zeus}, Stagger~\citep{stagger96} or the Versatile Advection Code (VAC)~\citep{vac,vac96,divbgabor} exploited fairly different discretizations applied to equivalent formulations of the governing PDE system for magnetohydrodynamics (MHD), and already showed the benefit of having the option to select a discretization most suited for the problem at hand. The development of shock-capturing schemes, where in analogy with gas dynamics the hyperbolic structure of the ideal MHD system is exploited, has driven algorithmic improvements to solve especially shock-dominated problems (see e.g. Chapter 19 in~\cite{book2}).  
\amrvac combines these algorithms with automated mesh refinement (AMR) for a fair selection of primarily hyperbolic PDE systems covering (multi-fluid) gas dynamics, non-ideal Newtonian and ideal relativistic MHD.  

A non-exhaustive list of software frameworks that are in active use and development to date includes RIEMANN~\citep{riemann}, BATS-R-US~\citep{batsrus,batsrus12} (a key component of the Space Weather Modeling Framework~\citep{swmf}), Nirvana~\citep{nirvana}, Ramses~\citep{ramses}, AstroBear~\citep{astrobear}, Pluto~\citep{pluto,plutoamr}, HiFi~\citep{hifi11}, Enzo~\citep{enzo14}, Echo~\citep{echo}, FLASH~\citep{flash}, Racoon~\citep{racoon}, CO5BOLD~\citep{co5bold}, Athena~\citep{athena}, Lare3D~\citep{lare3d}, Pencil~\citep{pencil}, A-MAZE~\citep{amaze}, Gumics~\citep{gumics}, SAC~\citep{sac} (based on VAC). In addition, several state-of-the-art codes like MURaM, Stagger or BIFROST~\citep{muram05,stagger96,bifrost} have made significant progress on solar physics applications, where MHD descriptions are combined with radiative transfer treatments. 
Other codes like WhiskyMHD or HARM concentrate on high energy, general relativistic phenomena \citep{whiskymhd,harm}. In this paper, we present an update for the open source software \amrvac, which has evolved alongside codes already mentioned, originally as a tool to study adaptive mesh refinement paradigms for hyperbolic PDE systems such as the ideal MHD set~\citep{amrvac03,amrvac07}, and currently evolved to a mature multi-physics software framework, with possibilities to couple varying PDE models~\citep{jcam14} across the block adaptive AMR strategy. Using the LASY syntax~\citep{lasy}, and with heritage in the solver part to VAC~\citep{vac96}, the software has kept its dimension-independent implementation, while the AMR allows Cartesian, polar, cylindrical or spherical coordinate systems (or generalizations thereoff with simple Jacobians).  Although the emphasis has lately been on handling especially special relativistic hydro to MHD applications~\citep{amrvaczak07,amrvac08,amrvac12},  covering Gamma Ray Burst afterglow physics~\citep{vlasis11,melianigrb}, X-Ray Binary~\citep{remi14} or Active Galactic Nuclei relativistic jet modeling~\citep{jet08,walg13,porth13} and pulsar wind nebulae physics~\citep{porth14}, the present version of the software incorporates many options for Newtonian applications inspired by astrophysical or solar phenomena. Since several of the most recent additions can be of generic interest to the astrophysical community, we here present an overview of both algorithmic and application-driven aspects that have not been documented elsewhere. The details of the parallel, block-adaptive implementation as given in~\cite{amrvac12} remain irrespective of the precise discretization or physics module adopted, and will not be repeated here. Instead, we explain in Section~\ref{fd} how in addition to the many flavors of shock-capturing schemes (e.g. TVDLF, HLL, HLLC, Roe), we now incorporate higher than typically second order accurate spatio-temporal discretizations as well. With respect to new physics applications, we chose to highlight the gas with dust physics module, where an arbitrary number of dust species can be dynamically followed in their size-dependent, drag-modulated evolution through a compressible gas. Recent applications of gas dynamics with relevant dust influences have looked at circumstellar bubble morphologies~\citep{marle11}, Rossby Wave Instability development of vortices in accretion disks~\citep{meheut12}, or Kelvin-Helmholtz developments in molecular cloud context~\citep{tom14}. Since the gas-dust physics module has not been described and tested in detail elsewhere, we here present a stringent suite of tests inspired by recent literature in Section~\ref{dust}. To make contact with modern efforts targeting plasma dynamics in solar conditions, Section~\ref{plasma} provides an update on the treatment of MHD and Hall MHD applications, and how we allow for background potential magnetic fields of significant complexity, based on data-driven extrapolations from actual magnetograms. An update of scaling on massively parallel computers is provided, while the appendices further report on the use of active and passive grid blocks,  the means to generate slices or collapsed views on evolving 3D dynamics during runtime, which are all of general interest to complementary coding efforts.

\section{High order methods}\label{fd}

While a large share of spectacular astrophysical magneto-fluids display shock dominated dynamics, a number of interesting applications can be described as fairly ``smooth''. These may involve the study of wave progression in magnetically structured media, the in-depth study of magnetic reconnection processes when dissipative layers are resolved, and various other typical solar or stellar magneto-convection problems. \cite{radhdwith3} recently intercompared high order finite difference (MURaM, Stagger) with finite volume treatments (CO5BOLD) for hydrodynamic convective layers with radiative transfer, with reasonable agreement but also subtle differences in turbulence properties. Simulations of magnetic reconnection in the regime of chaotic island formation~\citep{pop13} also identified several pros and cons when different discretizations are tried on the same problem. Having the option to choose method depending on the application is obviously beneficial.

Discontinuous flows naturally call for a finite-volume (FV) discretisation that solve the fluid equations in their integral form. Like many open source software in active use today, \amrvac offers a rather large option of FV schemes~\citep{amrvac12}, with Total Variation Diminishing (TVD) type methods like TVDLF or full Riemann solver based solvers as originally described in~\cite{vac96}, extended with variants like HLL or HLLC. To varying degree, these methods require adaptation to the set of governing PDEs at hand (Euler gas dynamics to MHD, Newtonian to relativistic). The pitfall of these methods is however that higher than second order finite-volume schemes (FV) must employ multi-dimensional stencils which considerably increases the computational cost. In \amrvac, all FV schemes can render up to second order accuracy, while some reconstruction procedures with higher than second order capabilities have been incorporated as well, typically to reduce the diffusion in these schemes. Finite difference schemes (FD) on the other hand are well suited for smooth applications and can operate to high order with one-dimensional stencils. In this section we describe the fairly general approach to conservative finite differencing of a hyperbolic conservation law as now available in \amrvac.  The only requirement for implementation to a different physics module is knowledge of the fluxes and of the fastest characteristic velocity.  Thus the finite difference scheme can be applied to all physics modules from hydrodynamics over Hall-MHD to relativistic MHD, or any other PDE set which may be added as a new physics module. 

\subsection{Short primer in conservative finite differences}
Given a set of (near-) conservation laws in cartesian coordinates
\begin{equation}
\delt \mathbf{U} + \mathbf{\nabla}\cdot \mathbf{\bar{F}(\mathbf{U})} = \mathbf{S(\mathbf{U})}
;\hspace{1cm}
\Leftrightarrow
\hspace{1cm}
\delt U_l + \sum_j \partial F_{jl}(\mathbf{U}) /\partial x_j = S_l(\mathbf{U}) \label{eq:conservationlaw}
\end{equation}
we seek the conservative finite difference discretisation of the flux in the $j$ direction $\partial F_{jl} /\partial x_j$.  
Regarding only one component of the solution vector $\mathbf{U}$ and dropping the index $l$, the point wise value at grid index $i$ reads
\begin{equation}
\left.\frac{\partial F_j}{\partial x_j}\right|_i = \frac{1}{\Delta x_j} \left( \hat{F}_{j}|_{i+1/2} - \hat{F}_j|_{i-1/2}\right)
;\hspace{1cm}
\left.F_{j}\right|_{i} = \frac{1}{\Delta x_j} \int_{x_j|_{i-1/2}}^{x_j|_{i+1/2}} \hat{F}_j(\chi)d\chi \, .\label{eq:fdcons}
\end{equation}
We see that the point wise value $\left.F_j\right|_i$ is just the $j$-directional cell-average of the function $\hat{F}_j(x)$.  
For the discretised $\left.\partial F_j/\partial x_j\right|_i$ we require knowledge of the interface values of $\hat{F}_j|_{i\pm1/2}$ which is obtained by reconstruction of the cell-averages of $\hat{F}_j|_{[s]}$ and hence the known point wise values of $F_j|_{[s]}$.\footnote{The index $[s]$ in square brackets should remind us of the index-range given by the stencil of the reconstruction.} 
This is just the hallmark of finite volume reconstruction:  Obtain (point-wise) interface values of the solution vector from its cell-averages.  Instead in finite differences, we can simply apply the finite-volume reconstruction formula to the point-wise \emph{flux}, to obtain
\begin{equation}
\hat{F}_j|_{i+1/2} = \mathcal{R}_{[s]}(F_j)\, .\label{eq:reconstruct}
\end{equation}
All directions are treated in analogue fashion and we form the directionally unsplit time-update operator 
\begin{equation}
\mathcal{L}_{l}(\mathbf{U})=-\sum_j\left.\frac{\partial F_{jl}}{\partial x_j}\right. + S_l(\mathbf{U}) \ ;
\hspace{1cm}
\left.\frac{d U_l}{d t}\right. = \mathcal{L}_l(\mathbf{U})\label{eq:timeoperator}
\end{equation}
which represents an ordinary differential equation (ODE) in time for each spatial grid point. 

\subsection{Applied flux splitting}\label{sec:FVS}

As in finite volume methods, for numerical stability, the flux needs to be upwinded. 
In finite differences, this is achieved with flux vector splitting \citep[FVS; see e.g. the book of][]{Toro1999}.  
Several approaches of flux splitting exist and comprise the Roe flux split, the Marquina's flux split \citep[][and references therein]{Fedkiw98thepenultimate} and the Lax-Friedrichs split \citep{Rusanov1961} to name the most common ones. Here, we settle for the fairly diffusive but easy to implement global Lax-Friedrich splitting.  
Hence for each interface, flux is reconstructed twice, once with left-biased stencil $[L]$ and once with right-biased stencil $[R]$.  In particular, we split the flux as
\begin{align}
F_j^- = \frac{1}{2} \left( F_j - c_{\rm max}\, U\right);
\hspace{1cm}
F_j^+ = \frac{1}{2} \left( F_j + c_{\rm max}\, U\right)
\end{align}
where $c_{\rm max}$ is the grid-global maximal characteristic velocity of the hyperbolic system, and we reconstruct 
\begin{align}
\hat{F}_j^{+L}|_{i+1/2} = \mathcal{R}_{[L]}(F_j^+);
\hspace{1cm}
\hat{F}_j^{-R}|_{i+1/2} = \mathcal{R}_{[R]}(F_j^-)\, .
\end{align}
Finally, the interface flux is obtained as
\begin{align}
\hat{F}_j|_{i+1/2} = \hat{F}_j^{+L}|_{i+1/2} + \hat{F}_j^{-R}|_{i+1/2}.  
\end{align}
This is similar to MHD applications by \cite{JiangWu1999,mignone2010} and to the relativistic HD application by \cite{radice2012}.  
However, note that we omit the projection onto characteristic fields and instead apply two upwinded reconstructions per interface.  
The projection step aims to reduce oscillations in the solution, at the price of greatly increased computational cost.  
It has been observed \citep[e.g.][]{van-Leer1982,Toro1999}, that the Lax-Friedrichs FVS utilised here introduces excessive numerical diffusion at contact and tangential discontinuities.
\footnote{In fact this is true for most schemes that are not based on complete Riemann solutions, with rare exceptions, e.g. the pressure-split schemes (AUSM) in the tradition of \cite{LiouSteffen1993}. 
Thus in problems where contacts and (viscous) boundary layers are of interest, one should resort to Riemann-problem based solvers provided by \amrvac.
This general flaw is less important in flows with few stagnant points and it can be alleviated substantially by adopting high order reconstruction techniques.  }
In the following sections, we demonstrate that the simplified scheme described here turns out to be quite capable in the treatment of ``smooth'' astrophysical flows.  
By design of equation (\ref{eq:fdcons}), the scheme is fully conservative (save for addition of geometric and physical source terms in step (\ref{eq:timeoperator})) and can adopt high spatial order by choice of the reconstruction step \ref{eq:reconstruct}.  
To this end, we provide compact stencil third order reconstruction \cite[][LIM03]{cada2009} and the fifth order monotonicity preserving reconstruction ``MP5'' by \cite{1997JCoPh.136...83S}.  

\subsection{Temporal discretisations}

Apart from the standard one-step, two-step predictor-corrector and third-order Runge-Kutta \citep[][RK3]{gottlieb1998}, we have implemented two multistep high-order strong stability preserving (SSP) schemes introduced by \cite{SpiteriRuuth2002}.  
These yield an explicit numerical solution to the ODE given by eq. (\ref{eq:timeoperator}).  
Adopting a general $s$-step Runge-Kutta scheme in the notation of \cite{SpiteriRuuth2002}, their equation 2.1 (a-c):
\begin{align}
\boldu^{(0)} &= \boldu^n\\
\boldu^{(i)} &= \sum_{k=0}^{i-1} \alpha_{ik}\boldu^{(k)} + \Delta t \beta_{ik} \mathbf{\mathcal{L}}(\boldu^{(k)}),\hspace{0.5cm} i=1,2,\dots,s\\
\boldu^{n+1} &= \boldu^{(s)}
\end{align}
the available optimal $s$-step, $p$-order strong stability preserving (SSP) Runge-Kutta (SSPRK($s$,$p$)) schemes read 
\subsubsection{SSPRK(4,3)}
\begin{align}
(\alpha_{ik}) =
\left(
\begin{array}{cccc}
  1&    -&    -& -\\
  0&    1&    -& -\\
2/3&    0&  1/3& -\\ 
  0&    0&    0&  1
\end{array}
\right);\hspace{1cm}
(\beta_{ik}) =
\left(
\begin{array}{cccc}
1/2&    -&    -& -\\
  0&  1/2&    -& -\\
  0&    0&  1/6& -\\ 
  0&    0&    0&  1/2
\end{array}
\right)\,.
\end{align}
This scheme requires storage of two intermediate steps and is SSP for a Courant number (CFL) of $2$.  
\subsubsection{SSPRK(5,4)}
\begin{align}
(\alpha_{ik}) =
\left(
\begin{array}{ccccc}
  1&    -&    -& - & -\\
0.44437049406734&    0.55562950593266&    -& - & -\\
0.62010185138540&    0&  0.37989814861460& - & -\\ 
0.17807995410773&    0&    0&  0.82192004589227 & -\\
0.00683325884039&    0& 0.51723167208978 & 0.12759831133288 & 0.34833675773694
\end{array}
\right)\\
(\beta_{ik}) =
\left(
\begin{array}{ccccc}
0.39175222700392&    -&    -& - & -\\
  0&  0.36841059262959&    -& - & -\\
  0&    0&  0.25189177424738& - & -\\ 
  0&    0&    0&  0.54497475021237 & - \\
  0&    0&    0&  0.08460416338212 & 0.22600748319395
\end{array}
\right)\,.
\end{align}
This scheme requires storage of four intermediate steps and is SSP for a CFL number of $1.50818004975927$.  

Combined with finite-differences and MP5 reconstruction, the latter scheme theoretically allows fourth-order accuracy in time and space.  We validate the expected convergence behaviour in sections \ref{sec:3Dalfven} and \ref{sec:3DalfvenWhistler} on MHD problems, as realized by the plasma physics module in \amrvac. 

\section{Astrophysical Gas and Dust dynamics}\label{dust}
Before describing the plasma physical module, we first discuss the recently added coupled gas-dust possibilities, which is of interest to a fair variety of astrophysical applications.  We will use both FV implementations and the new FD schemes on a selection of test problems.
In this dusty hydrodynamics module, \amrvac handles the following set of governing equations. 

\subsection{Continuity and momentum equations}
The density $\rho$ and velocity vector field $\mathbf{v}$ combine in a conservation of mass, written as
\begin{eqnarray}
\frac{\partial \rho}{\partial t}+\nabla \cdot \left(\mathbf{v}\rho\right) & = & S_{\rho} \, , \label{masscons}
\end{eqnarray}
where a user may prescribe sink/source terms for mass loss/creation in $S_{\rho}$.
The evolution for the velocity field incorporates inertial effects and pressure gradients, and can include external gravity, viscous forces, friction with multiple dust species, or any user specified force written as

\begin{eqnarray}
\rho \frac{\partial \mathbf{v}}{\partial t}+\rho \mathbf{v} \cdot \nabla \mathbf{v} & = & -\nabla p +\rho \mathbf{g} - \nabla \cdot \left( \mu \hat{\Pi} \right) + \sum_{d=1}^{n_{d}}\mathbf{f}_d + \mathbf{S}_{\mathbf{v}} \, . \label{velocity}
\end{eqnarray}
In this equation, the external gravitational field is quantified by the gravitational acceleration $\mathbf{g}(\mathbf{x})$, viscosity\footnote{The viscous force $\mathbf{F}_{visc}=-\nabla \cdot\left(\mu \hat{\Pi}\right)$ can be rewritten in a variety of ways. Using the split of a tensor in a symmetric and antisymmetric part for $\nabla \mathbf{v}=[\nabla \mathbf{v}]_{symm} + [\nabla \mathbf{v}]_{asymm}$, where $2 [\nabla \mathbf{v}]_{symm}= \nabla \mathbf{v}+ (\nabla \mathbf{v})^T$, the fact that $\nabla\cdot[\nabla \mathbf{v}]_{asymm}=\frac{1}{2}\nabla \times (\nabla \times \mathbf{v})$ and the identity $\nabla \times (\nabla \times \mathbf{v})= \nabla (\nabla \cdot \mathbf{v}) -\nabla^2 \mathbf{v}$, one can rewrite for constant $\mu$ the viscous force as
$$\mathbf{F}_{visc}=\mu \left[ \nabla^2 \mathbf{v}+\frac{1}{3} \nabla \left(\nabla\cdot\mathbf{v}\right)\right]\,.$$ Another point to note is that one sometimes uses the kinematic viscosity coefficient $\nu$, related to the viscosity coefficient $\mu$ through $\mu=\rho\nu$. These observations are relevant when e.g. comparing the detailed handling of the non-ideal terms in other HD or MHD codes.
} 
is quantified by the viscous force written with the aid of the traceless tensor $\hat{\Pi}=-\left(\nabla \mathbf{v} + (\nabla \mathbf{v})^T\right) +\frac{2}{3}\hat{I}\nabla\cdot\mathbf{v}$ (with identity tensor entries $\hat{I}_{ij}=\delta_{ij}$) and the coefficient of the dynamical viscosity $\mu$. A set of $n_{d}$ dust species is coupled to the gas with a drag force that has an essential dependence on $\mathbf{f}_d(\rho, \rho_d, \mathbf{v}, \mathbf{v}_d)$ i.e. on the gas and dust densities and velocity differences. A user defined force would enter through $\mathbf{S}_{\mathbf{v}}$. 
In the {\tt MPI-AMRVAC} code, the mass conservation and velocity evolution equation are actually combined to a momentum (with momentum density $\mathbf{m}=\rho\mathbf{v}$) conservation equation written as
\begin{eqnarray}
\frac{\partial \mathbf{m}}{\partial t}+ \nabla \cdot \left( \mathbf{v}\mathbf{m} +p\hat{I}\right)  & = & \rho \mathbf{g} - \nabla \cdot \left( \mu \hat{\Pi} \right) + \sum_{d=1}^{n_{d}}\mathbf{f}_d + \mathbf{S}_{\mathbf{v}} +\mathbf{v} S_{\rho} \, . \label{momentum}
\end{eqnarray}
The latter two terms then form the user momentum source term $\mathbf{S}_{\mathbf{m}}=\mathbf{S}_{\mathbf{v}} +\mathbf{v} S_{\rho}$.

When dust species are present, each among the $n_d$ dust species obeys a pressureless gas evolution, with dust density $\rho_d$ and velocity $\mathbf{v}_d$, and hence momentum density $\mathbf{m}_d=\rho_d\mathbf{v}_d$, governed by
\begin{eqnarray}
\frac{\partial \rho_d}{\partial t}+\nabla \cdot \left(\mathbf{v}_d\rho_d\right) & = &  0 \, ,\nonumber \\
\frac{\partial \mathbf{m}_d}{\partial t}+\nabla \cdot \left(\mathbf{v}_d\mathbf{m}_d\right) & = &  -\mathbf{f}_d \, . \label{dustset}
\end{eqnarray}

In addition to a (user controlled) addition of $n_d$ dust species (when $n_d=0$ no equations or variables are added), one may opt to add a number of $n_{tr}$ tracer quantities $\theta_{tr}$ (with $tr=1,\ldots n_{tr}$). Each tracer actually adds a simple equation of the form
\begin{eqnarray}
\frac{\partial D_{tr}}{\partial t}+\nabla \cdot \left(\mathbf{v}D_{tr}\right) & = & \theta_{tr}S_{\rho} \, . \label{tracer}
\end{eqnarray}
This form ensures that while we can treat the quantity $D_{tr}=\rho\theta_{tr}$ in a manner similar to the density evolution equation~(\ref{masscons}), we can at each time use them to obtain the actual tracer values from $D_{tr}/\rho$, which in turn obey the simple advection equation
\begin{eqnarray}
\frac{\partial \theta_{tr}}{\partial t}+\mathbf{v} \cdot \nabla \theta_{tr} & = & 0 \, . \label{tracer1}
\end{eqnarray}

\subsection{Closure and energy equation}
The above sets of equations need further closure, which can be one of the following options:
\begin{itemize}
\item prescribe the pressure-density relation, e.g. using a $p=c_{ad}\rho^\gamma$ relation where one can adopt an isothermal ($\gamma=1$) or a polytropic relation. The constant $c_{ad}$ is then either related to squared isothermal sound speed or to constant entropy. Also the case of a zero temperature gas $c_{ad}=0$ is contained in this closure. Under this setting for the equation of state, no further energy equation is needed.
\item adopt an ideal gas with internal energy density $e=p/(\gamma-1)$ itself governed by
\begin{eqnarray}
\frac{\partial e}{\partial t}+ \nabla \cdot \left( \mathbf{v}e\right)  + p \nabla \cdot \mathbf{v} & = & - \left(\mu \hat{\Pi}\cdot \nabla\right) \cdot \mathbf{v}  + \nabla \cdot \left(\kappa \nabla T\right)  - n_i n_e \Lambda(T) + S_e\, , \label{internal}
\end{eqnarray}
where viscous heating\footnote{This viscous heating may also be approximated by $ - \left(\mu \hat{\Pi}\cdot \nabla\right) \cdot \mathbf{v} \approx \mu | \nabla \mathbf{v} |^2$. It is sometimes also handled as $\rho\nu\sum_{ij} \left(\frac{\partial v_i}{\partial x_j}+\frac{\partial v_j}{\partial x_i}\right)^2$.} represents the first term on the right hand side, isotropic thermal conduction introduces the temperature dependent heat conduction coefficient $\kappa(T)$, and an ideal gas law relates pressure, temperature and density through $p=\frac{k_b}{\mu_m m_p}\rho T$ (with mean molecular weight $\mu_m$, Boltzmann constant $k_b$, and proton mass $m_p$). Dimensionless, the latter writes as $p=\rho T$. Optically thin radiative losses are represented by the term $- n_i n_e \Lambda(T)$, which has a tabulated temperature dependent loss function $\Lambda(T)$ (several tables used in the solar to astrophysics literature are pre-implemented, while the ion-electron density product $n_i n_e$ for a fully ionized hydrogen plasma is $\rho^2/m_p^2$). When optically thin radiative losses are incorporated (a module that is also relevant for Newtonian MHD), the need for AMR in combination with various (explicit, semi-implicit, implicit to exact) local source evaluations was demonstrated in~\cite{vanmarleRL}. A user can add an internal energy source/sink through $S_e$. 
\end{itemize}
In the latter case, the code evolves instead of the internal energy equation~(\ref{internal}), an evolution equation for the (conserved) total energy $E$, consisting of internal and kinetic energy with $E=p/(\gamma-1)+\rho v^2/2=e+\rho v^2/2$. By combining equation~(\ref{internal}) and the velocity evolution equation~(\ref{velocity}) (actually $\mathbf{v}\cdot$ operating on this equation), we obtain the total energy density evolution as
\begin{eqnarray}
\frac{\partial E}{\partial t}+ \nabla \cdot \left( \mathbf{v}(E+p)\right) & = & \rho\mathbf{v}\cdot\mathbf{g} + \nabla \cdot \left(\kappa \nabla T\right)  - \nabla \cdot \left(\mathbf{v} \cdot \mu \hat{\Pi}\right) \nonumber \\
& & + \sum_{d=1}^{n_{d}}\mathbf{f}_d\cdot \mathbf{v} - n_i n_e \Lambda(T) + S_e + \mathbf{v}\cdot\mathbf{S}_{\mathbf v} \, . \label{total}
\end{eqnarray}
We have purposely written the governing equations in the form of Equation~(\ref{eq:conservationlaw}), indicating the terms treated as sources on the right hand side, while fluxes as used in the different shock-capturing discretization schemes can be read off in the left hand sides. 
The implemented equations thus consist of Equation~(\ref{masscons}), Equation~(\ref{momentum}), the optional equation sets of the form~(\ref{dustset}) for each dust species, or Equation~(\ref{tracer}) for each optional tracer, and when an energy equation is taken along, Equation~(\ref{total}) is evolved.

\subsection{Dusty gas test suite}
As described earlier, the dust module is similar to the hydrodynamics module, with the addition of an arbitrary number of dust fluids which can be defined flexibly. Each dust fluid has several properties such as the size of the represented particles, $a_d$, and the internal density of the dust grains, $\rho_p$. By having different values $a_d$ and $\rho_p$ one can for example model the dynamical effect of a dust size distribution or the effect of having particles with different chemical compositions. A dust fluid $d$ typically interacts with the gas fluid through the combined Stokes/Epstein drag force $\mathbf{f}_{d}$ defined as
\begin{eqnarray}
	\mathbf{f}_{d} &=& -(1-\alpha) \pi n_{d} \rho a_{d}^2 \Delta  \boldsymbol{v}_{d} \sqrt{\Delta  \boldsymbol{v}_d^2 + v_t^2}, \label{drag}\\
	\alpha &=& 0.35 \exp{\left(-\sqrt{\frac{T}{500}}\right)}+0.1,\label{sticking}
\end{eqnarray}
with $\alpha$ a temperature dependent sticking coefficient \citep{2006A&A...456..549D}, $T$ being the gas temperature, $n_{d}$ the dust particle density of the $d$-th dust species, $\Delta  \boldsymbol{v}_{d} =  \boldsymbol{v} -  \boldsymbol{v}_{d}$ the difference between the gas and dust velocity and $v_t$ the thermal speed of the gas. Other drag laws are available as well. No interaction between dust fluids is included. In the following we present four tests performed with this dust module of the {\tt MPI-AMRVAC} code. Due to the added complexity of having one or multiple dust species it often becomes difficult or impossible to find analytical results for many test problems. To demonstrate the validity of the implementation we have selected test problems with known solutions, or problems for which the purely hydrodynamical variant has been studied in detail in other works. Notably, three out of our four tests (the {\tt dustybox, dustywave} and the Sedov blast wave) have been presented in the test suite of the dust+gas Smoothed Particle Hydrodynamics (SPH) simulations by \citet{2012MNRAS.420.2345L}. Additionally, a gas+dust variant of the Sod shock tube test \citep{1978JCoPh..27....1S} performed with {\tt MPI-AMRVAC} has been presented in \citet{ASTROproc}. Furthermore, these tests highlight some typical dust features. \\

\noindent In all tests we use the dust fluids to represent a dust mixture made of spherical silicates ($\rho_p = 3.3$ g cm$^{-3})$ with a dust grain size distribution $n(a_d)$ with sizes between $10^{-7}$ cm and $2.5\times10^{-5}$ cm. In the ISM, this size range is typically observed to follow a distribution that goes as $n(a_d) \propto a_d^{-3.5}$ \citep{1984ApJ...285...89D}. We model this distribution by dividing the size range in bins with equal total mass, and representing each bin by a dust fluid with a specific dust particle size. This is explained in detail in~\cite{tom14}. The local amount of dust in the system is often quantified by the dust-to-gas ratio $\delta$, which is the total mass density of the dust divided by the mass density of the gas.

\subsubsection{{\tt Dustybox}}
\label{dustybox}

In this test problem one or multiple dust fluids start with an initial velocity difference with respect to a stationary gas fluid. Due to the drag force, the gas will be accelerated by the dust, and all dust species will decelerate at a rate imposed by the properties of the fluid. \citet{2011MNRAS.418.1491L} presented analytical solutions for multiple drag laws in the presence of a single dust fluid and compared this with simulations in \citet{2012MNRAS.420.2345L}. To demonstrate the validity of our implementation, we compare a simulation with four dust species using the physical  drag force described in equation (\ref{drag}) with the semi-analytical solution of the problem. In the case of one dust species, the analytical solution of the change in velocities of the dust and gas fluid are given in \citet{2012MNRAS.420.2345L}. We expand this approach to give a solution for an arbitrary number of dust fluids with the drag force in equation (\ref{drag}), which leads to a set of $N$ coupled nonlinear ordinary differential equations for the time dependent functions $\Delta v_i (t) = v(t) - v_{d}(t)$, namely 
\begin{equation}
	\frac{d \Delta v_{d}}{dt} = \frac{1}{\rho} \sum_{i=1}^N f_{i} + \frac{f_{d}}{\rho_{d}},
	\label{diffeq}
\end{equation} 
with $N$ the number of dust fluids. Note that the terms $f_{i}$ are functions of $v_{i}$, as can be seen for the Epstein-Stokes drag in equation (\ref{drag}). We can solve this set of differential equations using a Python script to find a semi-analytical solution. For the case of one dust fluid, we recover the analytical solution mentioned earlier.

\begin{figure}
  \centering
\includegraphics[scale=0.3]{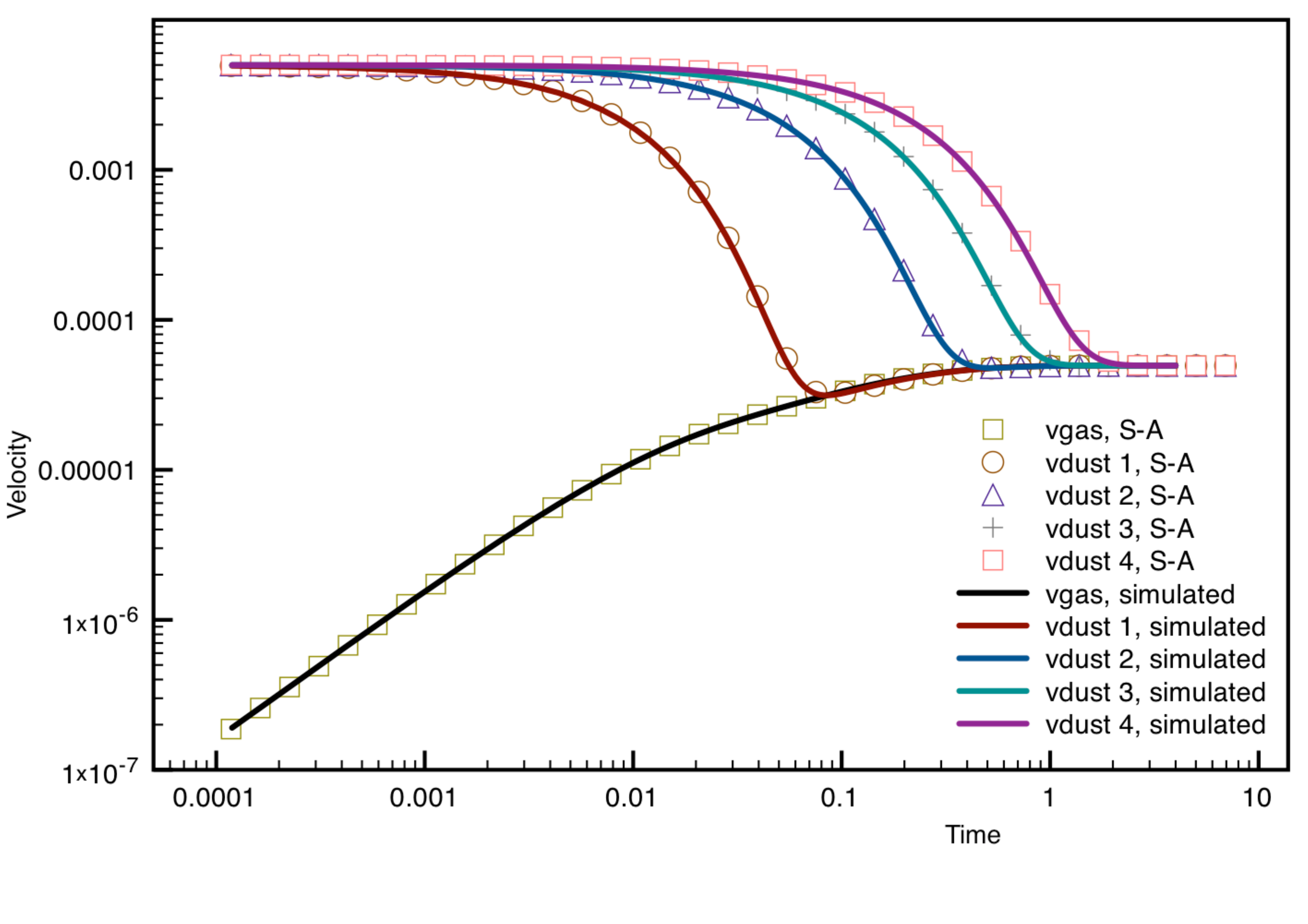}
  \caption{Comparison between the gas and dust velocities in a {\tt MPI-AMRVAC} simulation of the {\tt dustybox} test with four dust species (shown in full lines) and the semi-analytical solution (shown as symbols) found by numerically solving the set of nonlinear differential equations (\ref{diffeq}). Note that the plot is in log-log scale to highlight the features of the simulation. The velocity units are multiples of 10$^7$ cm s$^{-1}$, time units are multiples of 9792 yr.}
\label{fig:dustybox}
\end{figure}

We use a setup in 1D, using a uniform grid with 40 cells and periodic boundary conditions. Note however that the setup itself is resolution independent. We set uniform gas and dust densities, with $\rho = 10^{-20}$ g cm$^{-3}$, and a total dust mass which is a 100 times lower ($\delta=0.01$). Velocities are set to $v=0$ and $v_d = 5\times10^3$ cm s$^{-1}$ for all dust species. The gas temperature is set to $T=100$, giving a sticking coefficient $\alpha = 0.32$ in equation (\ref{sticking}). As numerical scheme we use the TVDLF solver~\citep{vac96} with a two-step time integration and a `Woodward' type slope limiter \citep{1984JCoPh..54..174C}. We use a CFL number of 0.2. 
The result of the simulation with four dust species, compared with the semi-analytical solution, is given in figure \ref{fig:dustybox}. We see that a perfect fit between the simulation and the semi-analytical solution is obtained. The simulation demonstrates how the four dust fluids start with a velocity difference relative to the gas fluid. Due to the interaction with the gas, the dust fluids decelerate. Species 1 represents the smallest particles, and can be seen to decelerate faster than the other dust fluids. Larger dust grains have a higher inertia and take longer before they come to an equilibrium velocity with the gas.

\subsubsection{{\tt Dustywave}}
The {\tt dustywave} problem, described in detail in \citet{2011MNRAS.418.1491L}, describes the propagation of a linear sound wave in a uniform and stationary medium with one or more dust fluids embedded. The coupling of the waves in the gas and dust fluids as well as the dampening of the waves are strongly dependent on the dust-to-gas ratio and the strength of the drag force. An analytic solution for a mixture with one dust fluid is known \citep{2011MNRAS.418.1491L}, and is used here to demonstrate the accuracy of our simulations.

\begin{figure}
  \centering
\includegraphics[scale=0.38]{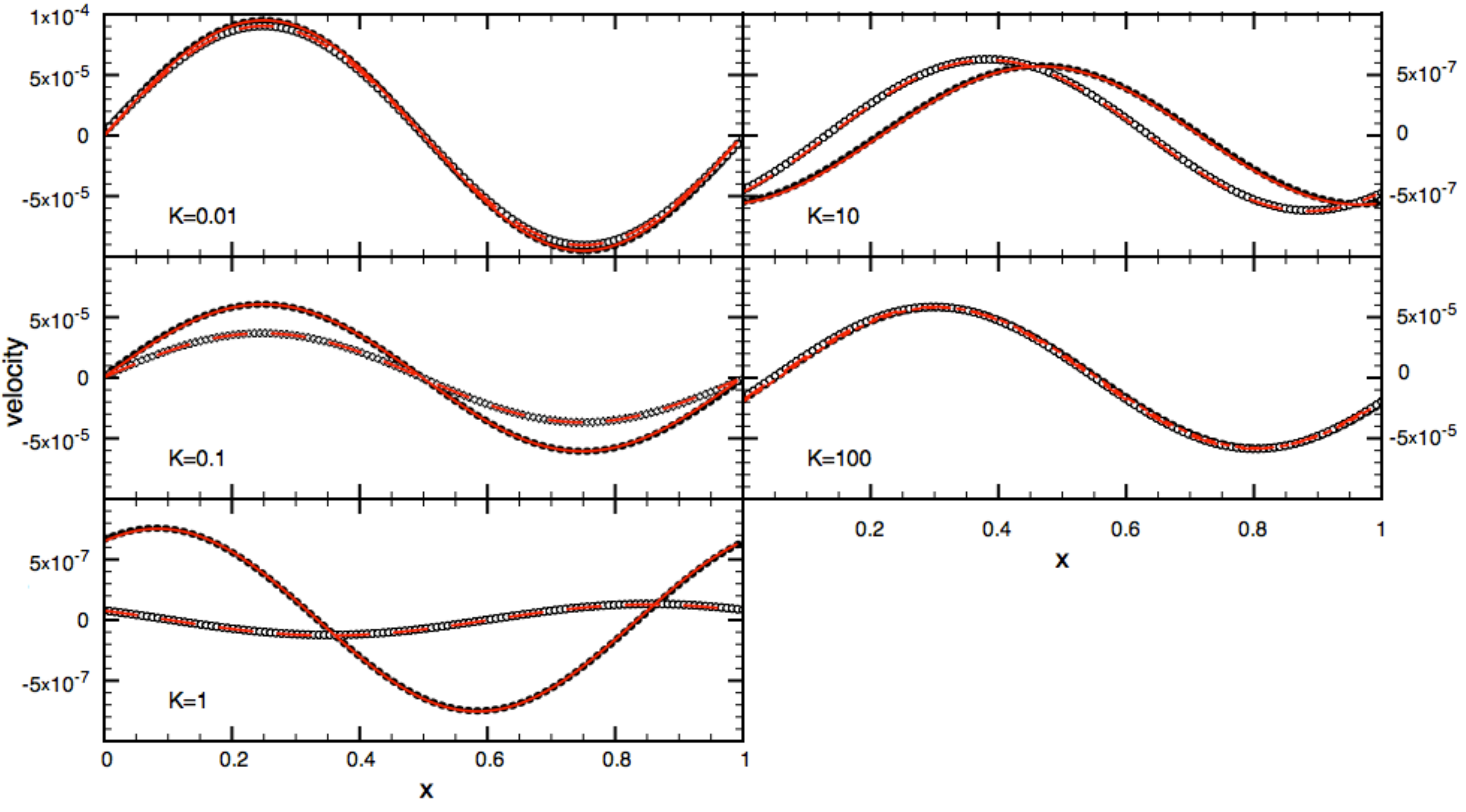}
  \caption{Comparison between the simulation with 120 cells and the analytical solutions for different values of the drag coefficient $K$ at $t=10$. The simulated values of gas and dust velocities are represented by black full and open circles, respectively, while the analytic velocities of gas and dust are given by red full and dashed lines. Note that the vertical scales for $K=1$ and $K=10$ differ from other simulations, as the velocities are damped more effectively in these cases.}
\label{fig:dustywave}
\end{figure}

\begin{figure}
  \centering
\includegraphics[scale=0.42]{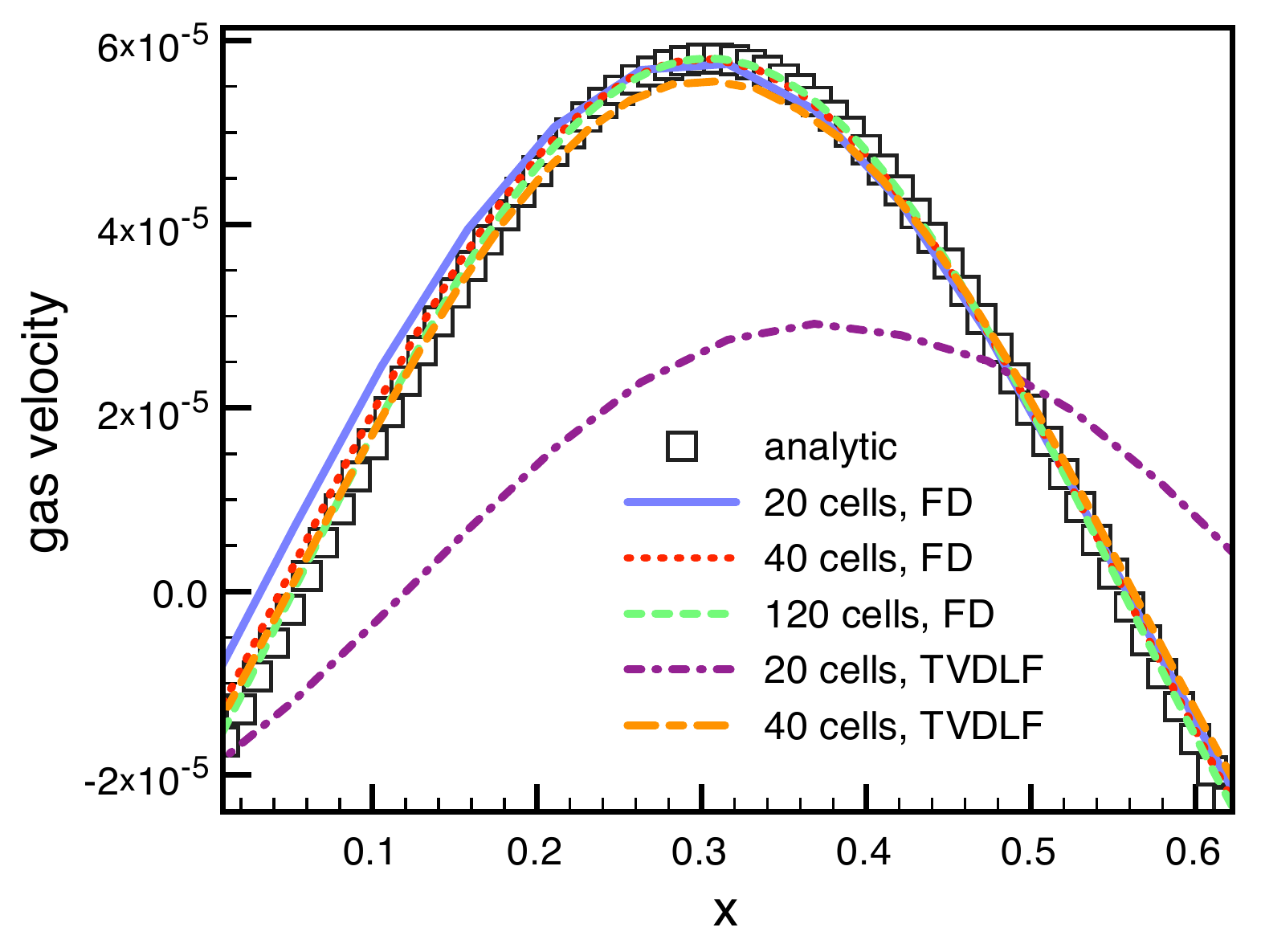}
  \caption{A part of the domain in {\tt dustywave} simulations with $K=100$ at $t=10$ (the whole domain is shown in figure \ref{fig:dustywave}). Different spatial resolutions are compared with the analytic solution. The amplitude of the gas velocity at this time is correctly recovered, even for resolutions as low as 20 cells in the case of the high-order FD method. For the lower order solutions with a TVDLF scheme a lower accuracy in obtained.  }
\label{fig:highDrag}
\end{figure}

Following \citet{2011MNRAS.418.1491L}, we use $\rho = \rho_d = 1$, $v=v_g=0$, and use one dust species. Likewise, we use the isothermal equation of state $p=c_s^2 \rho$ and a speed of sound $c_s=1$. A sine-shaped perturbation in velocity and in both gas and dust densities is added, in all cases the amplitude of the perturbation is $10^{-4}$ and the wavelength is the same as the size of the domain. We simulate this setup in a 1D domain between $x=0$ and $x=1$ with several resolutions. For testing purpose and comparison, we use now a simplified drag force $f_d = K \Delta v$, with $K$ a constant. We use the FD solver together with the fifth order MP5 limiter and the SSPRK(5,4) time integration using a CFL number of 0.5. 
In figure \ref{fig:dustywave} the simulation results at time $t=10$ for simulations with weak ($K=0.01$) up to strong drag ($K=100$) are compared with the analytical solution. All simulations have the same resolution ($\Delta x = 8.33 \times 10^{-3}$, i.e. 120 cells with no AMR). For intermediate coupling ($K$ between 0.1 and 10), the solutions for the dust and gas velocity can be seen to be out of phase. This phase difference causes strong damping in the setups with $K=1$ and $K=10$. All cases are in good agreement with the analytic results and errors are typically below $0.5 \%$. Importantly, in other approaches such as the SPH method in \citet{2012MNRAS.420.2345L} overdamping of the velocity is seen for $K=100$ due to the high drag force when the spatial resolution is low, leading them to propose a resolution criterion $\Delta x \lesssim c_s t_s$, with $t_s$ the stopping time $t_s = \frac{\rho \, \rho_d}{K\left( \rho + \rho_d\right)}$, which would in this test mean about 200 cells. However, figure \ref{fig:highDrag} demonstrates that by using high-order schemes we obtain results without overdamping with as little as 20 cells. In contrast, if we use a TVDLF scheme with `Woodward' type limiter \citep{1984JCoPh..54..174C} and a two-step time advance with a CFL number of 0.2, figure \ref{fig:highDrag} shows that stronger dampening is observed for the case with 40 cells ($\sim4\%$ at peak value, compared to only $0.2\%$ for 40 cells with the high-order schemes). Lowering the resolution to 20 cells, a strongly dampened and shifted solution is found.

\subsubsection{Sedov blast wave with dust species}

The Sedov blast wave problem is a classical problem in which a high energy perturbation is introduced in a static background, causing a shockwave to propagate through the external medium. It is often used to test codes, see for example \citet{2008MNRAS.390.1267T} who compare the ability of several fluid and SPH codes to simulate the Sedov blast wave problem. A version with one dust fluid is discussed in \citet{2012MNRAS.420.2345L}. In the gas-only case, an analytical solution for the location of the blast wave is known.

\begin{figure}
  \centering
\includegraphics[scale=1]{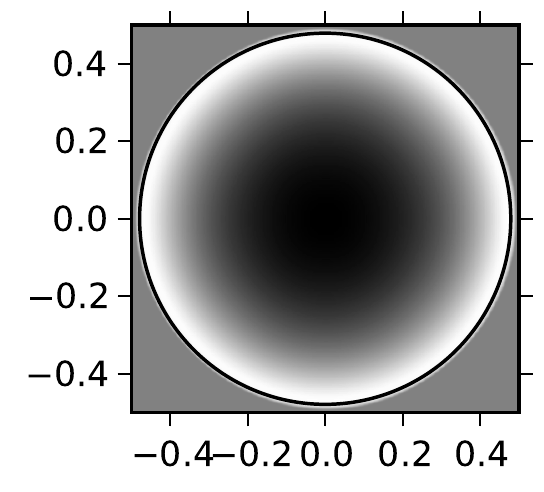}
\includegraphics[scale=1]{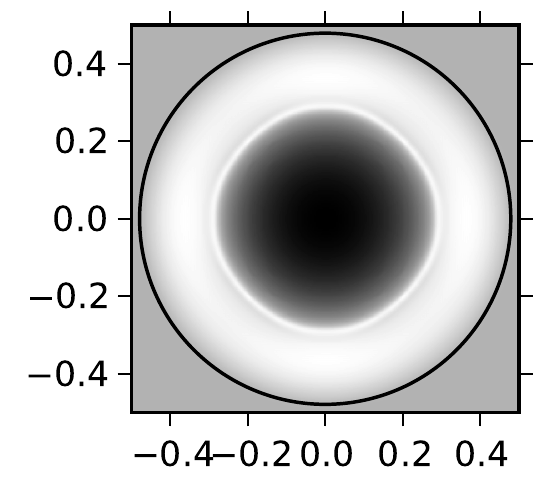}
  \caption{Left: Gas density in a 3D simulation of the Sedov blast wave, integrated along the line of sight. The theoretical distance of the shock location is indicated by a black circle. Right: integrated density for the third dust species. The black circle indicates the theoretical position of the gas shock.}
\label{fig:SedovGas}
\end{figure}

Our ambient medium has uniform gas density $\rho_0 = 6 \times 10 ^{-23}$ g cm$^{-3}$ and a dust-to-gas ratio $\delta= 0.01$ using four dust species. The pressure is set to $p=1.44 \times 10^{-14}$ dyn cm$^{-2}$, except in the middle of the domain where we introduce a high pressure ($p=7.49\times10^6$ dyn cm$^{-2}$) in a spherical region with a radius of 0.01 parsec. The gas fluid has an adiabatic index of $5/3$. The simulations are performed in 3D with Cartesian coordinates, in a cubical domain with sides of one parsec. The boundaries have open outflow conditions. We use three levels of AMR, resulting in an effective resolution of 400$^3$. With this resolution the middle region is covered in 280 cells, resulting in a total central energy $E_0 = 2.098 \times 10^{51}$ erg. We use the TVDLF solver with a three-step time integration and a `Woodward' type slope limiter. We use a CFL number of 0.4. 
Figure \ref{fig:SedovGas} shows a 2D output of the gas density in the 3D domain integrated along the line of sight using the collapse feature of {\tt MPI-AMRVAC}, as described in the appendix~\ref{collapse}. 
The position of the shock front at time $t$ has been calculated analytically in \citet{Sedov} and \citet{Landau}, and is found as
\begin{equation}
	r(t) = \left( \frac{E_0}{\beta \rho_0} \right)^{1/5} t^{2/5},
	\label{sedov}
\end{equation}
with $E_0$ the energy in the central region, $\beta = 0.49$ for an ideal gas with $\gamma = 5/3$ \citep{2008MNRAS.390.1267T} and ambient density $\rho_0$. We simulate up to $t=3.16\times10^{18}$ s (10 years), at which time equation~(\ref{sedov}) predicts a distance of 0.483 pc. Figure \ref{fig:SedovGas} demonstrates that the same radius is obtained in our simulations with an addition of dust with $\delta=0.01$.

 \begin{figure}
  \centering
\includegraphics[scale=0.62]{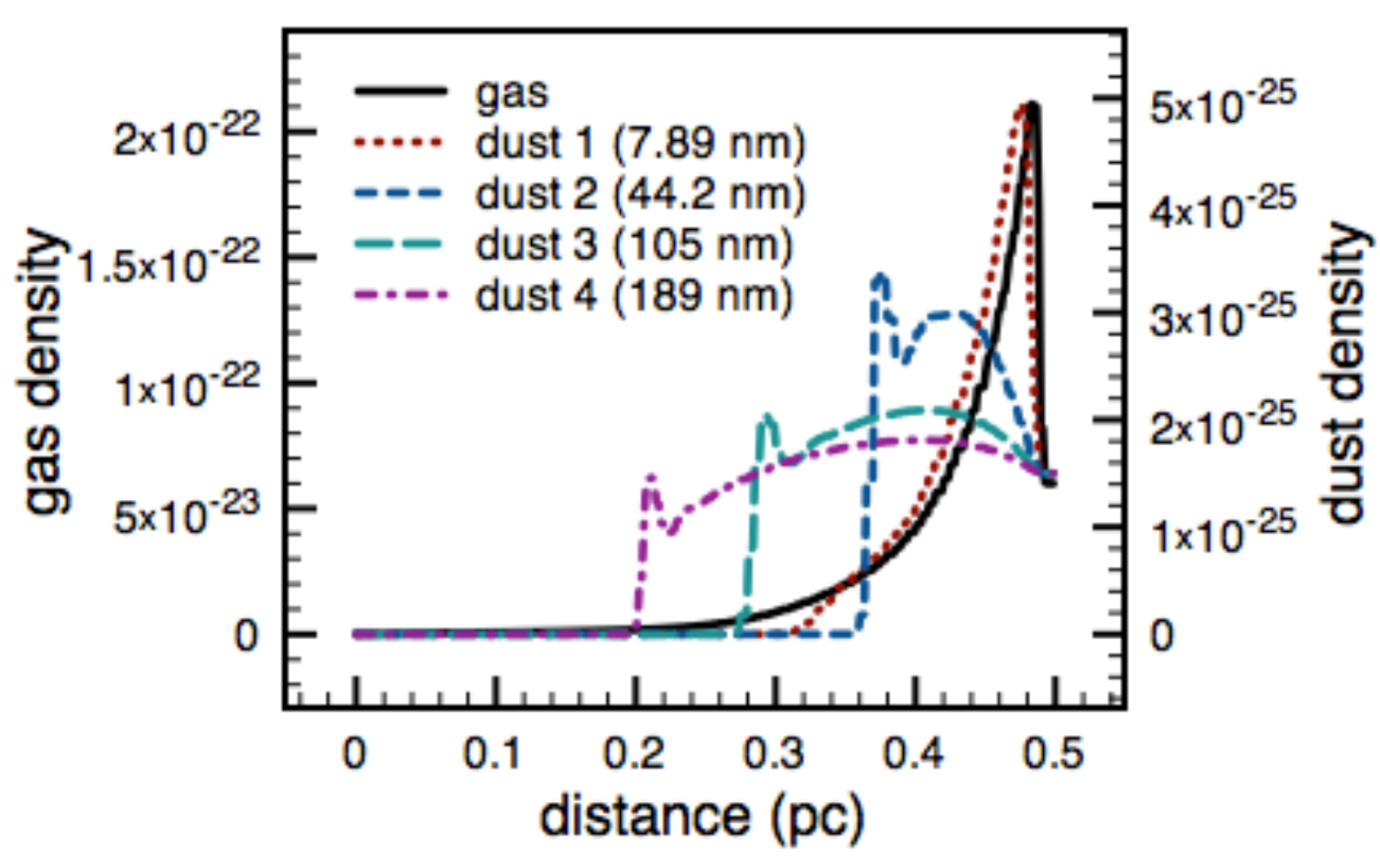}
  \caption{A 1D cut from the 3D Sedov blast wave at $t=3.16\times10^{18}$ s (10 years). All densities are given in g cm$^{-3}$. Dust species one is coupled to the gas density. Two density peaks are seen for dust species two, three, and four. }
\label{fig:sedovcut}
\end{figure}

In figure \ref{fig:sedovcut}, a 1D cut is made, showing clearly the distance the gas and dust fluids have propagated. As the gas shock propagates through the ambient medium, dust is accelerated as well. Small dust particles are more strongly coupled to the gas, resulting in a density peak close to the location of the shock. The density of dust species one is closely coupled to that of the gas. We see in figure \ref{fig:SedovGas} how the dust separates in regions dependent on the size of the particles. Figure \ref{fig:sedovcut} shows how dust species two, three, and four have two peaks. The one closest to the shock is due to the steady acceleration of ambient dust particles by the gas shock. The second peak is the result of the initially high velocity of the gas, which accelerates dust to a velocity which depends on the particle size, as large dust particles take longer to accelerate. As this high velocity dust moves outward, it sweeps up the dust in front of it, causing the second peak. Dust species one only has one peak, as the initially accelerated dust moves along with the gas.

\subsection{Cloud shock in gas-dust settings}
\label{cloudshock}

Our final gas-dust application models the interaction of a high density structure with a shock wave. This test is clearly of relevance in the interstellar medium (ISM), where dusty clouds are often seen to interact with supernova shocks. In the interstellar environment the stability of high density structures is of importance in estimating the rate of stellar formation. Numerically, the cloud is often modeled as a spherical high density structure embedded in a lower density ambient medium, with a planar shock wave propagating through the domain \citep{2005ApJ...633..240P,2006ApJS..164..477N,2007MNRAS.380..963A}. The interaction between the shock wave and the cloud can cause several instabilities, which may lead to the disruption of the cloud.  Here, we demonstrate the ability to add dust to the setup in both the cloud region and the ambient medium.

 \begin{figure}
  \centering
\includegraphics[width=\textwidth]{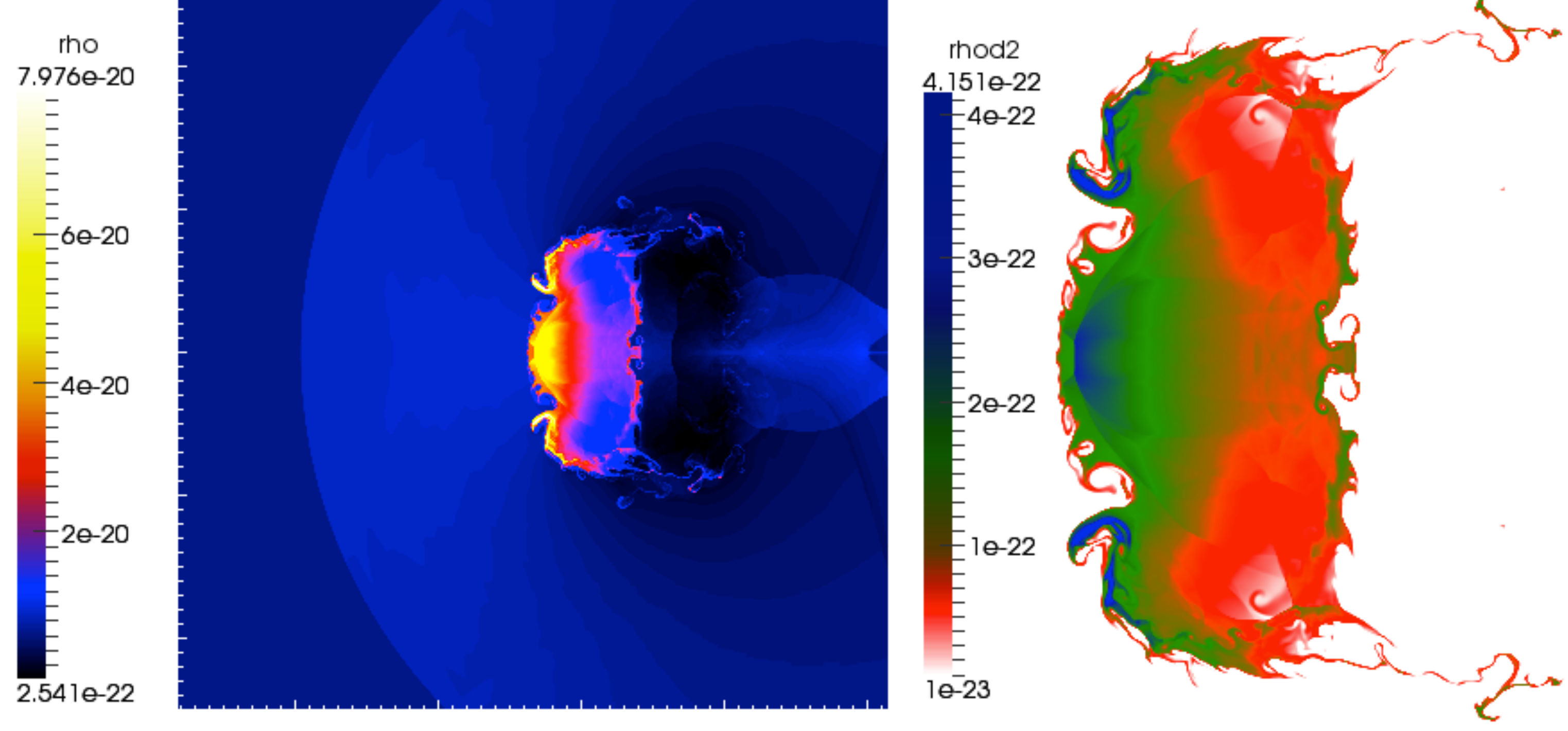}
  \caption{Left: density distribution at $t=3.01\times10^{5}$ yr. The entire domain in shown. Typical features such as the bow shock, the disturbed cloud with Richtmyer-Meshkov features on the front side, and a low density region Rayleigh-Taylor instability behind the cloud. Right: a zoomed in look at the dust density distribution of species two in the cloud region. The dust can be seen to be tightly coupled to the dynamics of the cloud.}
\label{fig:cloudshock}
\end{figure}  
 
The cloud shock test is simulated in a 2D Cartesian domain with size (3.34 pc)$^2$. We use six levels of AMR to obtain an effective resolution of 5120$\times$5120. The ambient medium and the cloud, which has a radius of 0.57 pc, are initially stationary. In the surrounding medium we have $\rho_R = 10^{-21}$ g cm$^{-3}$ and in the cloud $\rho =  10\rho_R$. The pressure is set using 
\begin{equation}
	p_R = \frac{\rho_R k_b T_{R}}{\mu_m \, m_p},
\end{equation}
with $k_b$ the Boltzmann constant, $m_p$ the hydrogen mass and $\mu_m = 2.3$ the molecular weight of the ISM at $T_R=200$ K. On the left side of the simulation we introduce a shocked region with values $\rho_L, p_L$ and $v_L$ calculated from the Rankine-Hugoniot conditions, viz.
\begin{eqnarray}
	\rho_L &=& \rho_R \frac{ \theta + p_{rat} }{1+ \theta p_{rat} }, \\
	v_L &=&  c_R M \left(1-\frac{1+\theta \, p_{rat}}{\theta+p_{rat}} \right), \\
	p_{L} &=& \frac{p_R}{p_{rat}}, \\ 
	p_{rat} &=& \frac{1}{1 + 2\left(M^2-1\right) \frac{\gamma}{\gamma+1}}, \\
	\theta &=& \frac{\gamma+1}{\gamma-1}, \\
	c_R &=& \sqrt{\frac{\gamma p_R}{\rho_R}}, \\
\end{eqnarray}
 with Mach number $M=10$ and $\gamma=5/3$. This results in the initial values $p_L=1.25\times10^{2} p_R$, $\rho_L=3.88 \rho_R$ and $v_L = 7.42 c_R$ with $c_R$ representing the speed of sound in the ambient medium on the right hand side. In this simulation we use two dust species with $\delta=0.01$ everywhere. We use the TVDLF solver (CFL number 0.1) with a `Koren' type limiter \citep{Koren} and a three-step time integrator.
In this simulation the gas can be seen to follow the typical evolution expected from the interaction of a supersonic shock, as shown on the left side of figure \ref{fig:cloudshock}. While the dust also interacts with the shock, in this case the chosen size and density values of the cloud imply that the two dust species used in the simulation (like before, having sizes between 5 nm and 250 nm) are strongly coupled to the dynamics of the initial gas in the cloud. While the dust itself is not sensitive for the development of the Richtmyer-Meshkov or Rayleigh-Taylor instabilities, clear imprints in the dust distribution are visible. A more detailed discussion of the effect of dust on the latter instability can be found in~\citet{ASTROproc}.

\section{Modules for solar applications}\label{plasma}

For solar physics applications, the MHD module of {\tt MPI-AMRVAC} offers a fairly diverse choice of options to model typically magnetically dominated dynamics. By selecting the appropriate combination of settings for pre-compilation of this physics module, this choice encompasses zero-beta simulations, isothermal MHD at finite plasma beta, MHD in ideal to visco-resistive prescriptions, extensions to Hall-MHD, and many sources and sinks that play a role in the radiative plasma conditions of the solar corona.  We first provide an overview of the implemented equations, and then demonstrate their workings on selected applications. 

\subsection{Magnetohydrodynamics: Maxwell's equations and Ohms law}

We will give the complete set of equations tackled by the MHD module.  Due to the possiblitiy of background magnetic field splitting, the standard MHD equations take on a somewhat unusual guise which adds to the usefulness of this collection. 

Starting with the homogeneous Maxwell's equations 
\begin{eqnarray}
\frac{\partial \mathbf{B}}{\partial t} & =  & - \nabla \times \mathbf{E} \,, \nonumber \\
\nabla \cdot \mathbf{B} & = & 0 \,, \label{maxwel}
\end{eqnarray}
the \amrvac MHD module allows the user to split off a time-invariant potential magnetic field, i.e. writing 
\begin{eqnarray}
\mathbf{B} = \mathbf{B}_0 + \mathbf{B}_1 \,; \hspace{1cm}
\frac{\partial \mathbf{B}_0}{\partial t} = \mathbf{0} \,; \hspace{1cm}
\nabla \times \mathbf{B}_0 = \mathbf{0} \,. \label{split}
\end{eqnarray}
Note that then $\mathbf{J}=\nabla \times \mathbf{B}_1$ and $\nabla \cdot \mathbf{B}_0=0=\nabla \cdot \mathbf{B}_1$.
The most general form for the electric field implemented in the code writes the generalized Ohm's law as
\begin{eqnarray}
\mathbf{E}=-\mathbf{v}\times\mathbf{B}+\frac{1}{e n_e} \mathbf{J}\times \mathbf{B} + \eta \mathbf{J} \,. \label{halle}
\end{eqnarray}

The first RHS term is applicable for an ideal MHD scenario, in a perfectly conducting plasma. The last term is related to resistivity, with resistivity parameter $\eta$. The Hall (middle) term introduces a first ion-electron distinction within a single fluid plasma description, where $\rho=n_i m_i$ is related to the ions, quasi-neutrality dictates $n_e=Zn_i$ for ion number density $n_i$ and charge number $Z$, so we can write 
\begin{eqnarray}
\mathbf{E}=-\left(\mathbf{v}- \frac{\eta_{h}}{\rho} \mathbf{J}\right) \times\mathbf{B} + \eta \mathbf{J} \,, \label{hall}
\end{eqnarray}
where the Hall parameter $\eta_h\propto m_i/e Z$ (dimensionalized\footnote{When the dimensions are fixed through a reference density $\rho_0$, a reference length $L_0$ and a reference field strength $B_0$, we measure speeds with respect to the Alfv\'en speed $V_{A0}=B_0/\sqrt{\mu_0 \rho_0}$ and the unit of mass is $\rho_0 L_0^3$, while the time unit is $t_0=L_0/V_{A0}$. Then, the dimensionalized parameters are actually $\bar{\eta}=\eta/(\mu_0 L_0 V_{A0})$, also referred to as the dimensionless Lundquist number, while $\bar{\eta}_h=V_{A0}/(L_0 \Omega_{i0})$. The latter uses the reference ion gyrofrequency $\Omega_{i0}=eZB_0/m_i$.}) appears next to the resistivity parameter $\eta$. Ideal MHD then sets $\eta_h=0=\eta$, resistive MHD has $\eta_h=0$ at finite resistivity, and Hall MHD has finite values for both parameters.

If we insert the electric field expression~(\ref{hall}) into the Maxwell equations~(\ref{maxwel}), and employ the splitting~(\ref{split}), we obtain as evolution equation for the magnetic field $\mathbf{B}_1$ the following
\begin{eqnarray}
\frac{\partial \mathbf{B}_1}{\partial t} + \nabla \cdot \left[ \mathbf{v}(\mathbf{B}_0+\mathbf{B}_1) - (\mathbf{B}_0+\mathbf{B}_1) \mathbf{v} +\frac{\eta_{h}}{\rho} \left( (\mathbf{B}_0+\mathbf{B}_1) \mathbf{J} - \mathbf{J} (\mathbf{B}_0+\mathbf{B}_1)\right)\right] & =  &  - \nabla \times \eta \mathbf{J} \,. \label{b1eq}
\end{eqnarray}
This directly corresponds to the numerical implementation where the terms in square brackets are treated as fluxes while the resistivity is added as a source.  

The resistive source can be added in two ways.  We note the equivalence of
\begin{equation}
- \nabla \times \eta \mathbf{J} = \eta \nabla^2 \mathbf{B}_1 + \mathbf{J} \times \nabla \eta \,, \label{resist}
\end{equation}
and the RHS lends itself to implementing an alternative evaluation using a compact stencil for the discretized Laplacian ($\mathrm{compactres=T}$).

Note that since the Hall-current directly enters in the flux, additional layers (and ghost zones) are required in the Hall MHD case.  
For finite volume, this implies an additional reconstructed layer, while in a finite difference setting, only the overall stencil is increased. 
For the computation of the currents we have implemented second and fourth order central differencing.

\subsection{$\mathbf{\nabla \cdot B}$ treatments}
As thoroughly investigated by~\cite{divbgabor}, controling solenoidality for magnetic fields in shock-capturing schemes can follow many approaches. Especially those handled by adding source terms~\citep{batsrus} or additional equations that advect and diffuse monopoles~\citep{dedner02} are easily carried over to AMR settings, and several source terms strategies were already intercompared in~\cite{amrvac03}.  In order to control the numerical monopole errors introduced when large gradients arise and nonlinearities in the limited reconstructions exist, equation (\ref{b1eq}) is replaced by one of the following options.  

An error-related source term~\citep{batsrus,janhunen} is added 
when writing
\begin{eqnarray}
\frac{\partial \mathbf{B}_1}{\partial t} + \nabla \cdot \left[ \mathbf{v}\mathbf{B} - \mathbf{B} \mathbf{v} +\frac{\eta_{h}}{\rho} \left( \mathbf{B} \mathbf{J} - \mathbf{J} \mathbf{B}\right) \right] & =  &  - \nabla \times \eta \mathbf{J}  -(\nabla\cdot\mathbf{B}_1)\mathbf{v} \,.  \label{b1b}
\end{eqnarray}
The diffusive approach~\citep{amrvac03} writes
\begin{eqnarray}
\frac{\partial \mathbf{B}_1}{\partial t} + \nabla \cdot \left[ \mathbf{v}\mathbf{B} - \mathbf{B} \mathbf{v} +\frac{\eta_{h}}{\rho} \left( \mathbf{B} \mathbf{J} - \mathbf{J} \mathbf{B}\right) \right] & =  &  - \nabla \times \eta \mathbf{J} + \nabla\left( C_d (\Delta x)^2 (\nabla \cdot \mathbf{B}_1)\right) \,.  \label{b1c}
\end{eqnarray}
The generalized lagrangian multiplier (GLM) $\psi$ appears with an added extra equation in the GLM variants~\citep{dedner02}, which we denote as $\mathrm{glm1}$ being
\begin{eqnarray}
\frac{\partial \mathbf{B}_1}{\partial t} + \nabla \cdot \left[ \mathbf{v}\mathbf{B} - \mathbf{B} \mathbf{v} +\frac{\eta_{h}}{\rho} \left( \mathbf{B} \mathbf{J} - \mathbf{J} \mathbf{B}\right) +\psi \hat{I}\right] & =  &  - \nabla \times \eta \mathbf{J} \,, \nonumber \\
\frac{\partial \psi}{\partial t} + \nabla \cdot \left( c^2_h \mathbf{B}_1 \right) & = & -\frac{c^2_h}{c^2_p} \psi  \,. \label{b1d}
\end{eqnarray}
The second variant $\mathrm{glm2}$ writes as
\begin{eqnarray}
\frac{\partial \mathbf{B}_1}{\partial t} + \nabla \cdot \left[ \mathbf{v}\mathbf{B} - \mathbf{B} \mathbf{v} +\frac{\eta_{h}}{\rho} \left( \mathbf{B} \mathbf{J} - \mathbf{J} \mathbf{B}\right) +\psi \hat{I}\right] & =  &  - \nabla \times \eta \mathbf{J}  -(\nabla\cdot\mathbf{B}_1)\mathbf{v} \,, \nonumber \\  
\frac{\partial \psi}{\partial t} + \nabla \cdot \left( c^2_h \mathbf{B}_1 \right) & = & -\frac{c^2_h}{c^2_p} \psi -\mathbf{v}\cdot\nabla \psi \,. \label{b1e}
\end{eqnarray}
A third variant $\mathrm{glm3}$ based on Eq. (\ref{b1d}) omits all $(\nabla\cdot\mathbf{B}_1)$ and $\psi$ related source terms in the induction, energy and momentum equation.  This simpler scheme is often sufficient and naturally adopts the spatial order from the reconstruction procedure.  

In all GLM treatments, the source term for the $\psi$ variable can be handled in two ways, as originally described in~\cite{dedner02}. One can use the exact solution of $\frac{\partial\psi}{\partial t}=-\frac{c_h^2}{c_p^2} \psi$ to write $$ \psi(t+\Delta t)=e^{\left(-\Delta t \frac{c_h^2}{c_p^2}\right)} \psi(t)$$ and prescribe either the constant factor $c_d=e^{\left(-\Delta t \frac{c_h^2}{c_p^2}\right)}$. 
Another choice is to fix the ratio $c_p^2/c_h=c_r$. 
It is also possible to perform the update implicitly by $\psi^{n+1}=\psi^*/(1+\Delta t \frac{c_h^2}{c_p^2})$.
In any case, we handle the source-update of the $\psi$ function in an operator split fashion.

\subsection{Momentum equation, closure and energy equation}

The momentum equation in MHD has the Lorentz force $\mathbf{J}\times\mathbf{B}$ appearing, which could be added as a source term for HD on the RHS of equations~(\ref{velocity}) or~(\ref{momentum}). Employing the identity $(\nabla\times\mathbf{B})\times \mathbf{B}=-\nabla \cdot (\frac{B^2}{2}\hat{I} -\mathbf{B}\mathbf{B} ) -\mathbf{B} (\nabla\cdot\mathbf{B})$ and using the splitting of the field, we actually implement

\begin{eqnarray}
\frac{\partial \mathbf{m}}{\partial t}+ \nabla \cdot \left( \mathbf{v}\mathbf{m} +(p+\frac{B_1^2}{2})\hat{I}-\mathbf{B}_1\mathbf{B}_1 \right) & & \label{momentummhd} \\
+ \nabla \cdot \left(\mathbf{B}_1\cdot\mathbf{B}_0 \hat{I} - \mathbf{B}_0\mathbf{B}_1 -  \mathbf{B}_1\mathbf{B}_0\right) & = & \rho \mathbf{g} - \nabla \cdot \left( \mu \hat{\Pi} \right) + \mathbf{S}_{\mathbf{m}} - \left(\mathbf{B}_0+\mathbf{B}_1\right)\nabla\cdot\mathbf{B}_1  \, . \nonumber
\end{eqnarray}
The final term on the RHS is only present when the source term monopole approach~\citep{batsrus} is taken.
To close the system, two options exist. 
\begin{itemize}
\item 
We can use an isothermal (e.g. used in the finite beta, stratified solar flux rope formation simulations by~\cite{xia14}) or isentropic closure as $p=c_{ad} \rho^\gamma$. Zero beta conditions prevail when $c_{ad}=0$.
\item  
In the second option, we additionally solve an evolution equation for the partial energy (-density)
\begin{equation}
E_1=e+\frac{B_1^2}{2}+\rho v^2/2 \,. \label{frace}
\end{equation}
This energy is the total energy when no splitting of the field is adopted. When splitting is adopted, the total energy is recovered as $E=E_1+B^2/2-B_1^2/2$.  
The governing equation for $E_1$ is obtained by combining the internal energy equation~(\ref{internal}) (which has an extra $\eta J^2$ RHS contribution from Ohmic heating), the velocity evolution equation ($\mathbf{v}\cdot$ equation~(\ref{velocity}) with the Lorentz force added), and the induction equation for the split off field (in fact $\mathbf{B}_1\cdot$ equation~(\ref{b1eq})). 
\end{itemize}

Collecting all together, this yields

\begin{eqnarray}
\frac{\partial E_1}{\partial t}+ \nabla \cdot \left( \mathbf{v}(E_1+\frac{B_1^2}{2}+p) -(\mathbf{B}_1\cdot\mathbf{v})\mathbf{B}_1 \right)  & & \label{fractional} \\
 + \nabla\cdot \left[ (\mathbf{B}_1\cdot\mathbf{B}_0)\mathbf{v}- (\mathbf{B}_1\cdot\mathbf{v})\mathbf{B}_0 \right] & & \nonumber \\
+\nabla \cdot \left[ \frac{\eta_h}{\rho} \left(( \mathbf{J}\cdot\mathbf{B}_1)\mathbf{B}-(\mathbf{B}_1\cdot\mathbf{B})\mathbf{J} \right) \right] 
 & = & \nabla\cdot(\mathbf{B}_1\times\eta\mathbf{J}) -\mathbf{B}_1\cdot\nabla\psi \nonumber \\ & & + \rho\mathbf{v}\cdot\mathbf{g} + \nabla \cdot \left(\hat{\kappa}\cdot \nabla T\right)  - \nabla \cdot \left(\mathbf{v} \cdot \mu \hat{\Pi}\right) \nonumber \\
& & - n_i n_e \Lambda(T) + S_e + \mathbf{v}\cdot\mathbf{S}_{\mathbf v} -(\mathbf{v}\cdot\mathbf{B}_1) (\nabla\cdot\mathbf{B}_1) \, . \nonumber
\end{eqnarray}
The heat conduction now contains only field-aligned heat transport since we adopt $\hat{\kappa}=\kappa_{\parallel}(T) \mathbf{B}\mathbf{B}/B^2$ (note the total field here). The terms related to monopole control may not all be present (GLM introduces $-\mathbf{B}_1\cdot\nabla\psi$, and source-based may use $-(\mathbf{v}\cdot\mathbf{B}_1) (\nabla\cdot\mathbf{B}_1)$, depending on the importance of strict energy conservation). 
A compact stencil evaluation of the resistive source term (activated with {\tt compactres=T}) may employ $\nabla\cdot(\mathbf{B}_1\times\eta\mathbf{J})= \eta J^2 - \mathbf{B}_1\cdot(\nabla\times\eta\mathbf{J})=\eta J^2 + \mathbf{B}_1\cdot(\eta\nabla^2\mathbf{B}_1 +\mathbf{J}\times \nabla\eta)$.

\subsection{Conservative Hall MHD}\label{sec:hallmhd}
As the Hall MHD module is a new addition to the code, we list here the specifics of its implementation.  
Activation of Hall MHD adds terms proportional to $\eta_h$ to the fluxes in the induction equation (\ref{b1eq}) and in the partial energy equation (\ref{fractional}).  
A straight-forward implementation can be provided for conservative finite differences and finite volumes using an HLL-type Riemann solver or a TVDLF-type scheme where no Riemann problem is solved \citep[see e.g. the review of][]{Yee1989}.  
Recent implementations of Hall MHD were also provided by \cite{LesurKunz2014} for the PLUTO code, by \cite{Bai2012} for ATHENA and \cite{TothMa2008} for BATSRUS.  
The crucial ingredient in Hall MHD is that the current $\mathbf{J=\nabla\times B}_1$ enters in the fluxes resulting in a non-hyperbolic set of PDEs.    
\begin{itemize}
\item For finite volume discretisation, the strategy is to obtain $\mathbf{J}$ from the interface values of the reconstructed magnetic field.  In cartesian coordinates, second and fourth order finite differencing yields for the $i$-component of the current vector
  \begin{align}
    J^i|_{j:l+1/2}&\overset{}{=}\epsilon_{ijk}\frac{1}{2\Delta x_j} \left(B_1^k|_{j:l+3/2}-B_1^k|_{j:l-1/2}\right) + O(\Delta x_j^2) \label{eq:jcd2}\\
    J^i|_{j:l+1/2}&\overset{}{=}\epsilon_{ijk}\frac{1}{12\Delta x_j} \left(-B_1^k|_{j:l+5/2} + 8 B_1^k|_{j:l+3/2} - 8 B_1^k|_{j:l-1/2} +B_1^k|_{j:l-3/2}\right) + O(\Delta x_j^4) \label{eq:jcd4}
  \end{align}
where the notation $|_{j:l+1/2}$ denotes the grid interface $l+1/2$ in the $j$-direction while the remaining directions remain with centered indices everywhere.  The interface magnetic field in these equations is obtained either with left biased $[L]$ or with right biased stencil $[R]$ yielding the currents $\mathbf{J}^{L}$ and $\mathbf{J}^{R}$ respectively.  
This interface current is then used along with the corresponding reconstructed variables to either compute fluxes according to \cite{Rusanov1961} and update the state vector directly (yielding the TVDLF scheme) or to use with an HLL-type Riemann solver (see below).  
\item In finite differences, we merely need to obtain the cell centered current prior to reconstruction of the fluxes as described in section \ref{fd}.  We again use central differencing for the components of the current as in equations (\ref{eq:jcd2}) and (\ref{eq:jcd4}) with the transformation $l+1/2\to l$.
\end{itemize}

For upwinding ($S^\pm$) and the explicit time-step criterion ($c_w$), the new local fastest wave speeds given by 
\begin{align}
  S^\pm(\mathbf{U}) = v\pm\max(c_f,\eta_h \frac{|B|}{\rho}k_{\rm max}); \hspace{1cm}
  c_w(\mathbf{U})  = |v|+\max(c_f,\eta_h \frac{|B|}{\rho}k_{\rm max})
\end{align}
are used, where we take the maximum of the ordinary MHD fast velocity $c_f$ and the fast-type whistler wave in field direction with wave number $k_{\rm max}$. The latter signifies the largest wave number allowed in the grid, $k_{\rm max}=\max_d(\pi/\Delta x_d)$ where the maximum is taken over all grid-directions $d$.  We have found that the value of $k_{\rm max}$ can often be reduced by a factor of two without seriously affecting the stability.  

A HLL-type Riemann solver then follows naturally, giving the flux
\begin{align}
  \mathbf{\hat{F}}|_{i+1/2} = \left\{
\begin{array}{ll}
\mathbf{F}^L &; S^{L} >0 \\
\mathbf{F}^R &; S^{R} <0 \\
\frac{S^{R}S^{L}(\mathbf{U}^R-\mathbf{U}^L)+S^{R}\mathbf{F}^L-S^{L}\mathbf{F}^R}{S^{R}-S^{L}} &; \rm otherwise
\end{array}
\right.
\end{align}
where again superscript L,R signifies the quantity on the interface $|_{i+1/2}$ derived from reconstructed variables with left- and right-biased stencil respectively.  The minimal and maximal signal speeds $S^L,S^R$ are then obtained according to \cite{Davis:1988:SSO} from
\begin{align}
  S^L=\min(S^-(\mathbf{U}^L),S^-(\mathbf{U}^R))\ ;\hspace{1cm} S^R=\max(S^+(\mathbf{U}^L),S^+(\mathbf{U}^R)).  
\end{align}
This parallels the implementation given by \cite{LesurKunz2014}.  The TVDLF scheme follows with the setting 
\begin{align}
  S^R = \max(c_w(\mathbf{U}^L),c_w(\mathbf{U}^R))\ ;\hspace{1cm} S^L = -S^R.
\end{align}

Finally, numerical stability requires a time step satisfying 
\begin{align}
  \Delta t < \frac{\Delta x}{c_w}
\end{align}
which becomes $\propto \Delta x^2$ for small $\Delta x$.  As with explicit integration of diffusive terms, the time step will thus eventually become prohibitively small.  
Using high order finite differencing to obtain high accuracy at moderate resolution can yield some mitigation to this problem.

\subsection{Selected tests and applications}
In what follows, we present a fair variety of tests and applications that make use of the novel additions to the MHD physics module specifically, combined with the algorithmic improvements that are generic to all physics modules. We cover 3D ideal MHD wave tests for demonstrating observed accuracies, the possibilities for using high order FD schemes on shock tube problems, and several novel tests for Hall MHD scenarios. Solar physics applications illustrate the possibilities for splitting of potential magnetic fields, and a typical 3D magnetoconvection study.

\subsubsection{3D Circular Alfv\'en wave}\label{sec:3Dalfven}

As the first standard test case to check the convergence of the conservative finite-difference scheme, we consider a circularly polarised Alfv\'en wave.  This incompressible wave is also a solution to the non-linear ideal MHD equations.  The setup is identical to \cite{mignone2010} and the wave in $x$-direction reads: 
\begin{equation}
\left(
\begin{array}{c}
  v_x  \\
  v_y  \\
  v_z   
\end{array}
\right)
=
\left(
\begin{array}{c}
  0  \\
  A \sin(\phi)  \\
  A \cos(\phi)   
\end{array}
\right)
\ 
;
\hspace{1cm}
\left(
\begin{array}{c}
  B_x  \\
  B_y  \\
  B_z   
\end{array}
\right)
=
\left(
\begin{array}{c}
  v_{\rm A}\sqrt{\rho}  \\
  \mp\sqrt{\rho} A \sin(\phi)  \\
  \mp\sqrt{\rho} A \cos(\phi)   
\end{array}
\right)\label{eq:alfven}
\end{equation}
with the phase $\phi=kx-\omega t$ and phase velocity given by the Alfv\'en speed $v_{\rm A}\equiv\omega/k=1$.  The negative (positive) sign in the magnetic field components indicates a wave propagating in positive (negative) $x$-direction.  
We set for the amplitude $A=0.1$ and use uniform background parameters $\rho=1$ and $p=0.1$.  The wave-vector is given by $k_x=2\pi$ and $k_y=k_z=2k_x$ and we rotate the vectors given by Eq. (\ref{eq:alfven}) accordingly.  The 3D domain is given by $x\in[0,1]$, $y\in[0,k_x/k_y]$ and $z\in[0,k_x/k_z]$ and we run the setup (without AMR) over one period $T=k_x/\sqrt{k_x^2+k_y^2+k_z^2}$.  

\begin{figure}[htbp]
\begin{center}
\includegraphics[width=80mm]{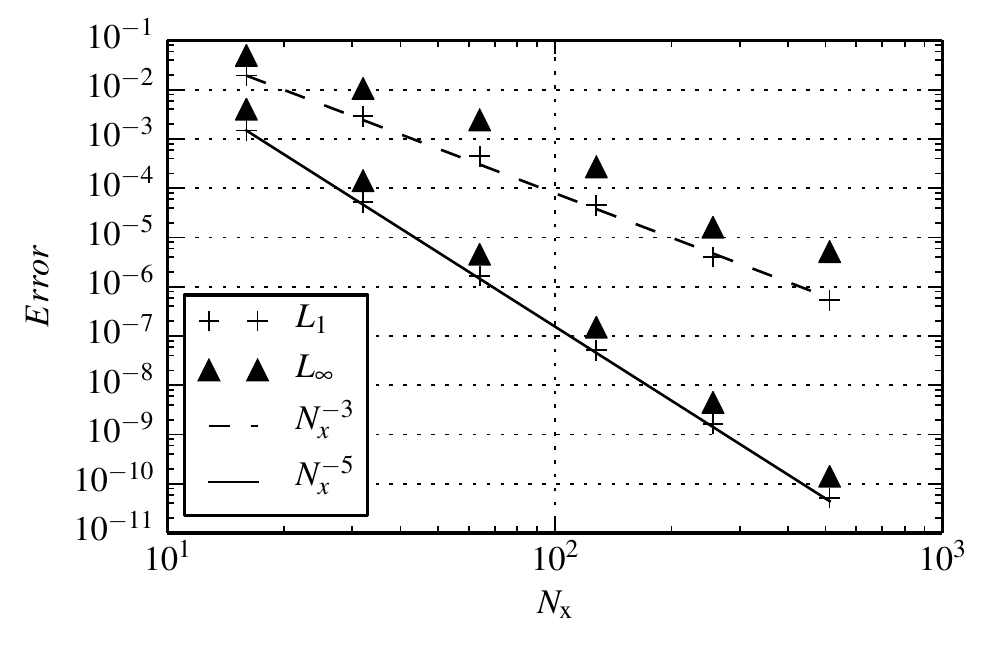}
\includegraphics[width=80mm]{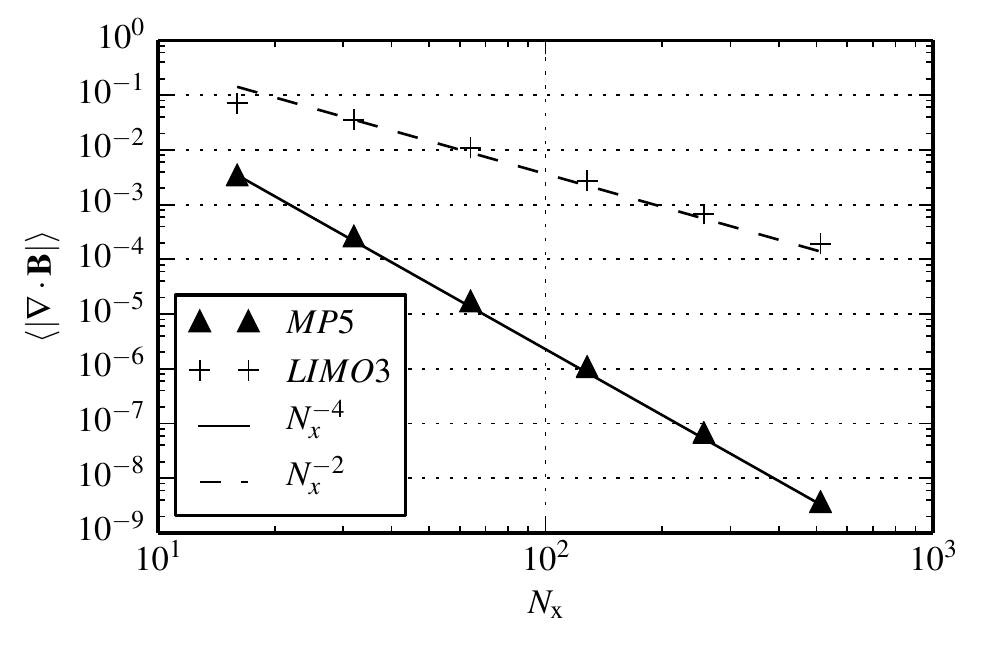}
\caption{\textit{Left:} Convergence of the GLM-MHD state vector using finite difference schemes for the 3D circularly polarised Alfv\'en wave test.  
Upper data points are obtained with three step Runge-Kutta integration and LIMO3, lower data points are obtained with five step Runge-Kutta and MP5 reconstruction. Both $L_1$ and $L_\infty$ norms are shown.  \textit{Right:} Decrease of the domain averaged absolute value of $\mathbf{\nabla\cdot B}$ in the 3D Alfv\'en wave test.  We obtain a second order behaviour in the LIMO3 case and fourth order behaviour with MP5.}
\label{fig:fdConvergence}
\end{center}
\end{figure}

The resulting convergence of the GLM-MHD state vector is shown in $L_\infty$ and $L_1$ norms in the left panel of figure \ref{fig:fdConvergence} for the third and fifth order reconstructions.  
We combine LIM03 reconstruction \citep{cada2009} with the third order TVD time integration (RK3) by \cite{gottlieb1998}.  As expected, third order convergence is achieved in both norms.  
Despite the formal fourth order of the SSPRK(5,4) by \cite{SpiteriRuuth2002}, we obtain fifth order convergence when using MP5 reconstruction.  In the right panel of figure \ref{fig:fdConvergence}, we show the divergence error of the magnetic field for both realisations.  In both cases, we calculate the divergence using fourth order central differences.  
As expected, the divergence of the magnetic field decreases with order given by the order of the spatial reconstruction minus one which demonstrates the effectiveness of the GLM approach.  

\subsubsection{Shocks, discontinuities and high order FD schemes}\label{sec:discont}

\begin{figure}[htbp]
\begin{center}
\includegraphics[width=\textwidth]{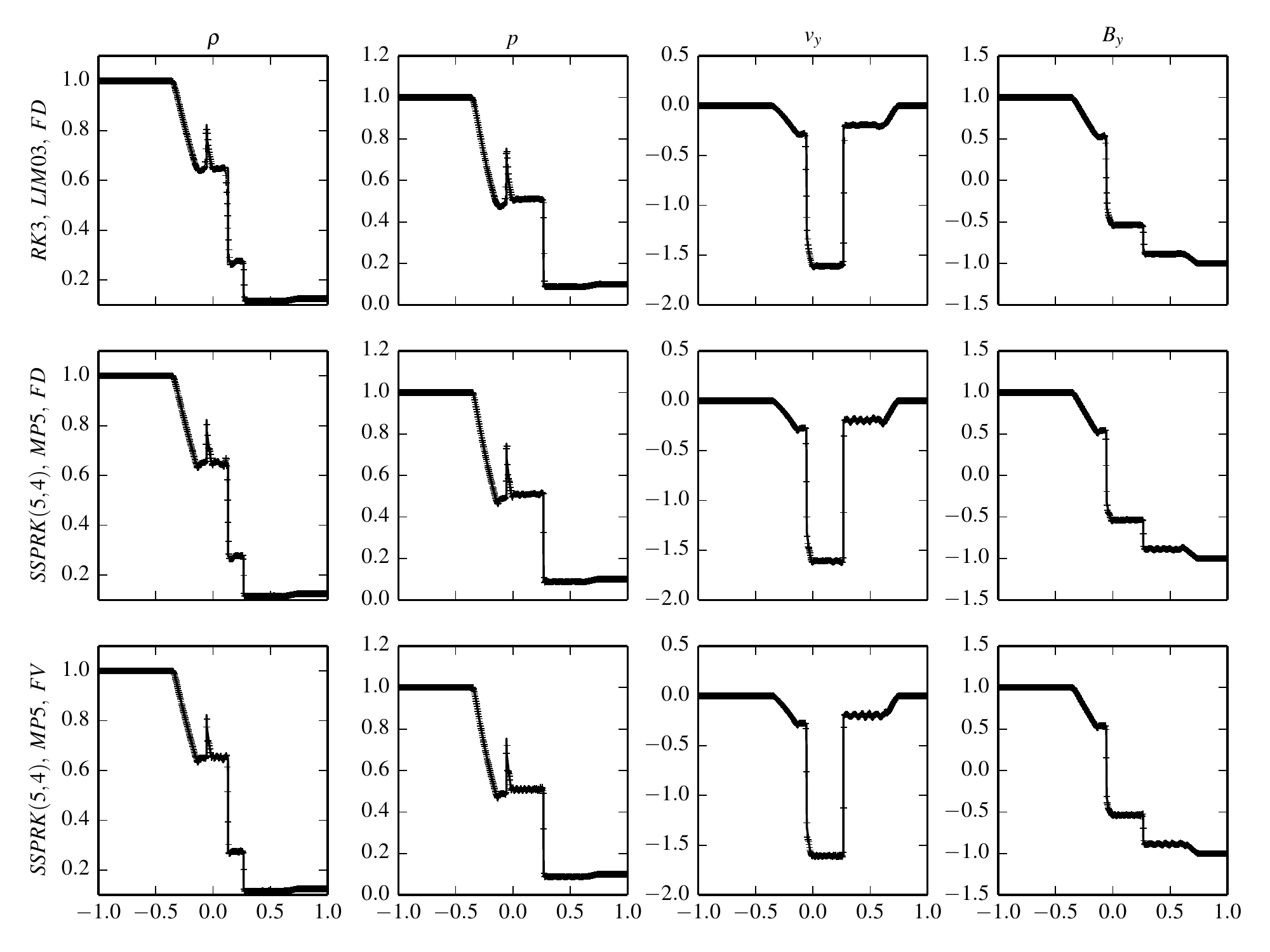}
\caption{1D MHD compound wave shock tube test, case 5a of \cite{ryu1995} at $t=0.2$.  All solutions are obtained with a Courant number of $0.4$.  
All conservative schemes reproduce the overall shock structure well, however the amount of oscillations obtained with MP5 could be considered prohibitive in some applications.  
Note that the amplitude of these oscillations is comparable in FD and FV (using reconstruction of conserved variables and HLLC Riemann solver).  
The solutions obtained with MP5 are indeed more oscillatory than when characteristic fields are used for reconstruction, see \cite{mignone2010}, Figure A.4.  \\
\smallskip}
\label{fig:shocktubes}
\end{center}
\end{figure}

\begin{figure}[htbp]
\begin{center}
\includegraphics[width=50mm]{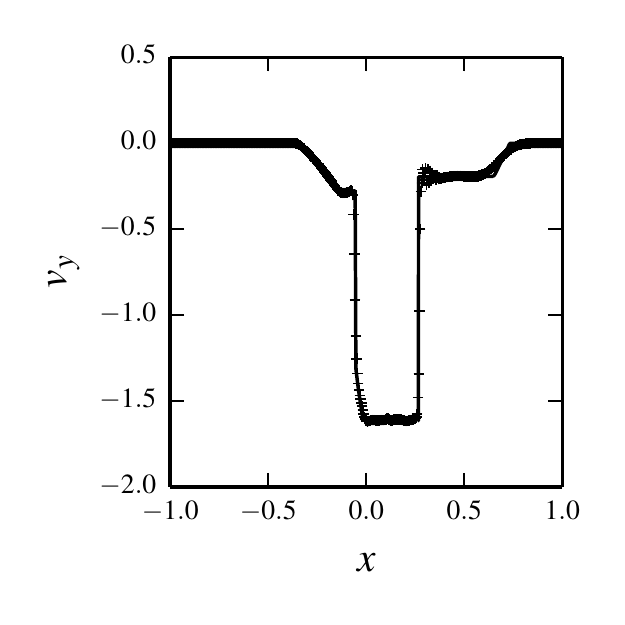}
\includegraphics[width=50mm]{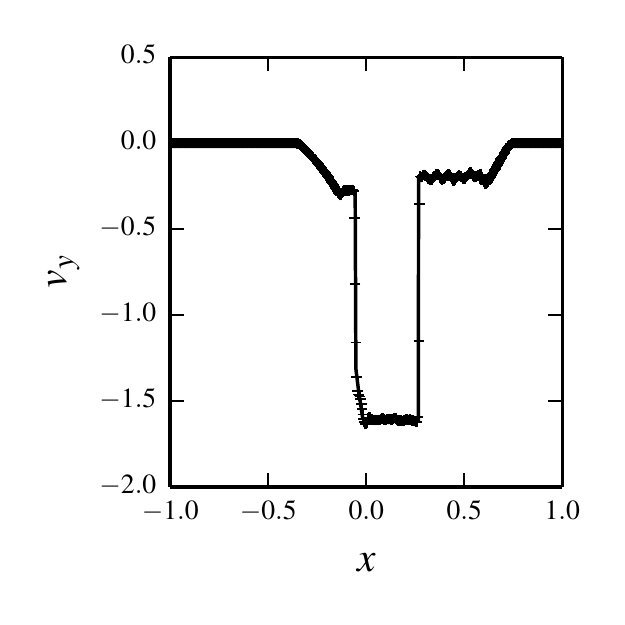}
\includegraphics[width=50mm]{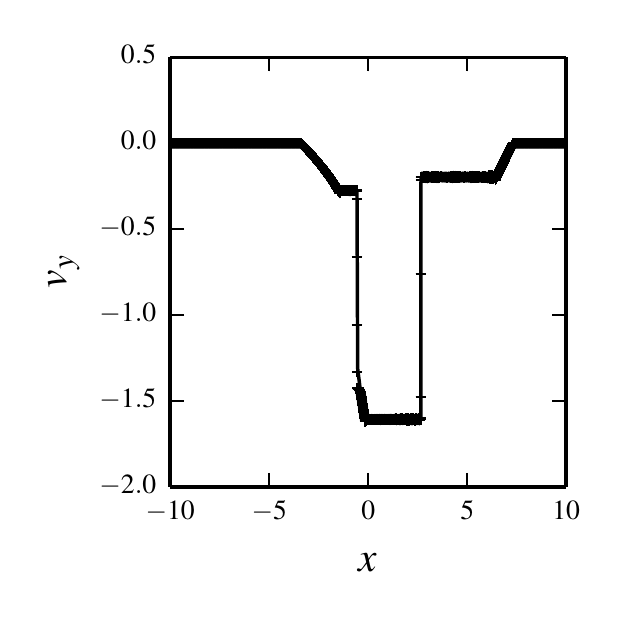}
\caption{Left and center: perpendicular velocity component in the 1D MHD compound wave at maximal Courant number allowed by the time marching scheme.  We show SSPRK(4,3)-LIM03-FD at Courant number of 2.0 (left panel) and SSPRK(5,4)-MP5-FD (center panel) at Courant number of 1.5.  Right:  longterm evolution of SSPRK(5,4)-MP5-FD showing decrease of the oscillation amplitude in time.}
\label{fig:sspshock}
\end{center}
\end{figure}

To investigate how the naive FD scheme handles discontinuous flows, we run a classical 1D MHD shock tube test from \cite{BrioWu1988}.  This standard test was also adopted by \cite{ryu1995,JiangWu1999,mignone2010}.  In terms of the primitive variables, the initial Riemann problem reads
\begin{align}
(\rho,v_x,v_y,v_z,p,B_x,B_y,B_z) = 
\left\{
\begin{array}{ll}
  (1.000,0,0,0,1.0,0.75,+1,0)\ ;& x<0\\
  (0.125,0,0,0,0.1,0.75,-1,0)\ ;& x>0
\end{array}
\right.
\end{align}
and we adopt a ratio of specific heats of $\gamma=5./3$.  
The uniform grid is composed of $512$ cells with $x\in[-1,1]$.  
In addition, a reference solution is obtained with a 2nd order TVD scheme at a ridiculously high resolution of $65\,536$ cells.  
Figure \ref{fig:shocktubes} collects our results with the schemes:  RK3-LIM03-FD, SSPRK(5,4)-MP5-FD, SSPRK(5,4)-MP5-FV.  
Due to their conservative nature, all schemes capture the general shock structure well.  
At the contact discontinuity we obtain over(under)-shooting in the density by $-0.001\%$ for RK3-LIM03-FD, $+2.7\%$ for SSPRK(5,4)-MP5-FD and $+1.6\%$ for SSPRK(5,4)-MP5-FV.  
The level of oscillations in the third order scheme is encouraging despite the omission of characteristic reconstruction.  
With fifth order reconstruction, the oscillations could be considered prohibitive for some applications.  Note that the oscillations visible for example in the profiles of $v_y$ are not a trademark of FD discretisation alone as our FV solution shows a similar behaviour.  
The advantages of the simple FD scheme become apparent when one considers the speedup.  Relative to RK3-LIM03-FD the execution times become $1 : 2.2 : 3.5$ for SSPRK(5,4)-MP5-FD : SSPRK(5,4)-MP5-FV.  Thus the FD scheme is faster than its (HLLC-based) FV counterpart by a factor of $1.6$ as no Riemann problems are solved.  Exploiting the SSP nature of the Runge-Kutta schemes, we ran the shock tube also at the maximal CFL yielding SSP.  The results are shown in figure \ref{fig:sspshock}. Here the third order scheme shows an excessive overshoot while the level of oscillations in the fifth order scheme is comparable to the case with Courant number 0.4.  It is important to note that the amount of oscillations gets dampened in time as illustrated in the right-hand panel of figure \ref{fig:sspshock}. 

\begin{figure}[htbp]
\begin{center}
\includegraphics[width=60mm]{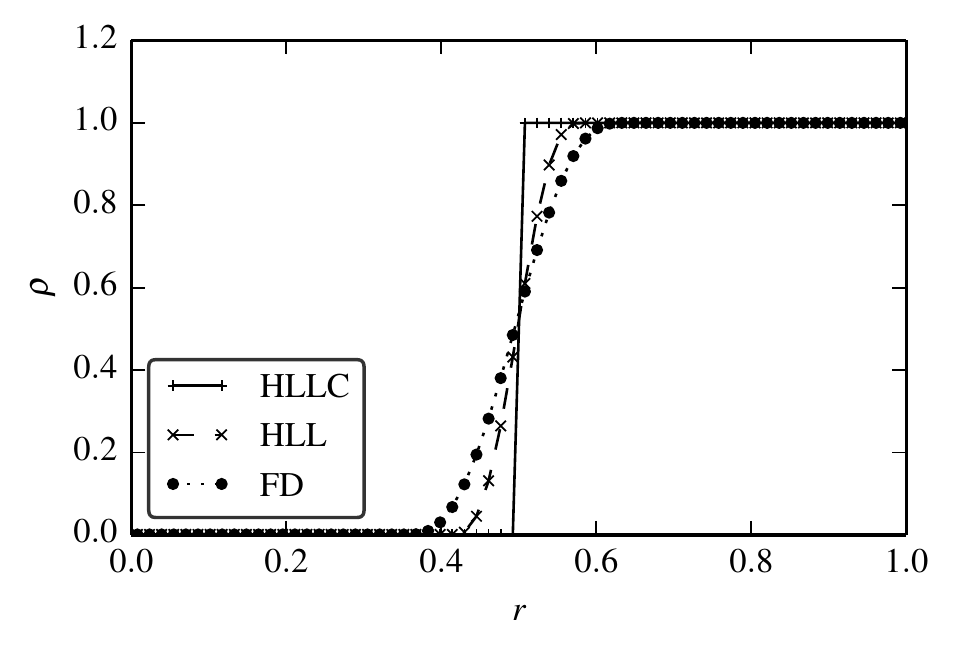}
\includegraphics[width=60mm]{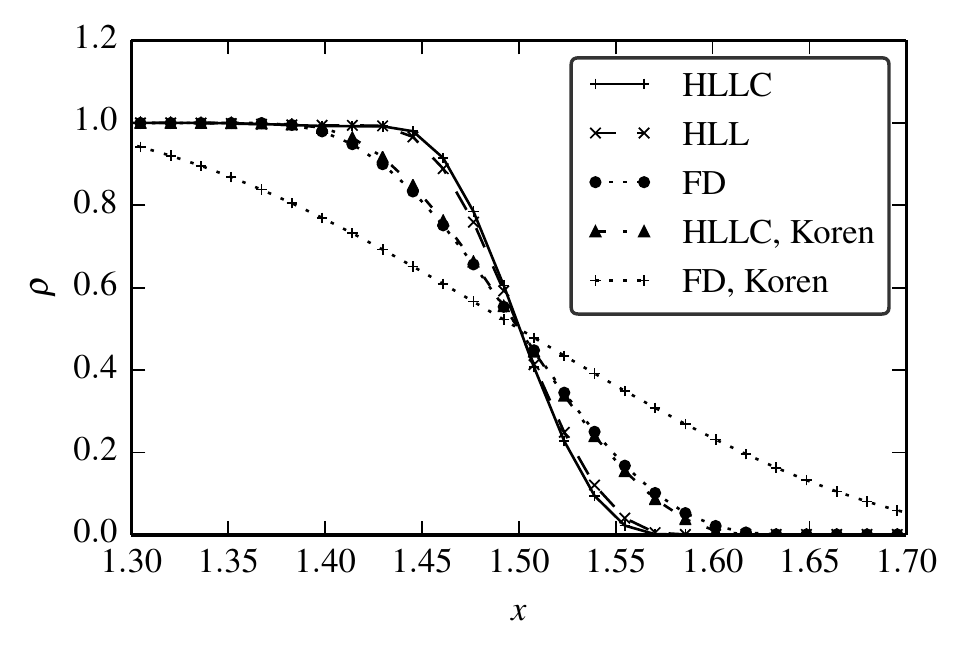}
\caption{Decay of a strong contact discontinuity with density jump from $\rho^L=10^{-3}$ to $\rho^R=1$, $p=1$ at t=10.  Left: Stationary case in spherical symmetry.  
Right: Periodically advected profile in slab symmetry, the advection velocity is $v_x=1$.
}
\label{fig:contact}
\end{center}
\end{figure}

As mentioned in section \ref{sec:FVS}, the FD scheme tends to diffuse tangential discontinuities.  To check the severity of this limitation, we run the following hydrodynamic test: 
We initialise a contact discontinuity with large density jump between $\rho^L=10^{-3}$ and $\rho^R=1$, uniform pressure $p=1$ and ratio of specific heats $\gamma=5/3$.  Thus the sound speed becomes $c^L\simeq40.8$ and $c^R\simeq1.3$.    
The decay of the stationary contact discontinuity is investigated in spherical coordinates where we align the density profile with the $r$-coordinate.  Here, we use a 1D domain with $r\in [0,1]$ with 64 grid points and density jump located at $r=0.5$, the Courant number is set to 0.6.  
For the moving contact, we adopt slab geometry and periodic boundary conditions and choose an advection velocity of $v_x=1$.  In this case, the domain spans $r\in [0,2]$ discretised with 128 grid points.  The density jumps at $x=0.5$ and $x=1.5$.  
In figure \ref{fig:contact}, we compare the states obtained with the HLLC, HLL and FD scheme at $t=10$ (corresponding to several hundred sound crossing times).  In all these cases, MP5 reconstruction and SSPRK(5,4) time stepping is employed. 
As expected, HLLC preserves the stationary contact exactly, while HLL and FD are subject to numerical diffusion.  
The Lax-Friedrich split FD scheme is more diffusive than the HLL solver.  
For the advected discontinuity, HLLC looses its capacity to capture the contact exactly and we find that the results for HLLC and HLL almost coincide.  Again, the FD scheme is the most diffusive of the three.  In this setup, the level of diffusion for the FD scheme is comparable to a second-order HLLC scheme with Koren reconstruction.  The latter is indicated as ``HLLC, Koren'' in the right panel of figure \ref{fig:contact}.  When run without high order reconstruction, dissipation in the FD scheme is clearly excessive (see line labeled ``FD, Koren'').

\subsubsection{Circular Alfv\'en-Whistler wave}\label{sec:3DalfvenWhistler}

In analogy to the MHD case, we can derive the equation for the circularly polarised Alfv\'en-Whistler wave.  One can easily show that the wave is also a solution of the non-linear Hall-MHD system.  
The non-linear circularly polarised Alfven-Whistler wave in $x$-direction reads:
\medskip

\begin{align}
\left(
\begin{array}{c}
v_{x}\\
v_{y}\\
v_{z}
\end{array}
\right)
=
\left(
\begin{array}{c}
0\\
A \sin(\phi) v_A/v_{\rm ph} k\ d\ f_v \\
A \cos(\phi)
\end{array}
\right)
;
\hspace{1cm}
\left(
\begin{array}{c}
B_{x}\\
B_{y}\\
B_{z}
\end{array}
\right)
=
\left(
\begin{array}{c}
B_0\\
-v_y/v_{\rm ph} B_0 + B_0/v_A \ k \ d\ \sin(\phi)\\
-v_z(1/v_{\rm ph} B_0 + B_0/v_{\rm ph} k^2\ d^2 f_v
\end{array}
\right) \,,\label{eq:alfvenWhistlerState}
\end{align}
with 
\begin{align}
f_v = \frac{1}{1-(v_A/v_{\rm ph})^2}
\end{align}
and the phase $\phi = kx-\omega t$ and wave number $k=2\pi m$.  
$A$ is the wave amplitude, not necessarily small.  
The phase velocity follows from the Hall MHD dispersion relation for propagation along the magnetic field 
\begin{align}
(k^2v_A^2-\omega^2)^2 = d^2 k^4 v_A^2 \omega^2 \,,\label{eq:alfvenWhistler}
\end{align}
as 
\begin{align}
v_{\rm ph} \equiv \omega/k = v_A/2\left(d\ k +\sqrt{4 + d^2 k^2}\right) \,,
\end{align}
which reduces to the Alfv\'en speed in the limit $k\ d\to 0$. In this limit, the solution is just the circularly polarised Alfv\'en wave of ideal MHD.  
The parameter $d$ corresponds to the Alfv\'en gyro radius as $d=v_A/\Omega_{\rm i}$ with the ion gyro frequency $\Omega_i$.  Note that the code-parameter $\eta_h$ is connected to $d$ via $\eta_h=\sqrt{\rho}d$.

As the electron velocity depends on the current in the Hall approximation, the Alfv\'en-Whistler wave test is an inherently three-dimensional problem.  
It can be used to test the realisation of the dispersion relation (\ref{eq:alfvenWhistler}) as well as the convergence order of the code in 3D.  

\begin{figure}[htbp]
\begin{center}
\includegraphics[width=80mm]{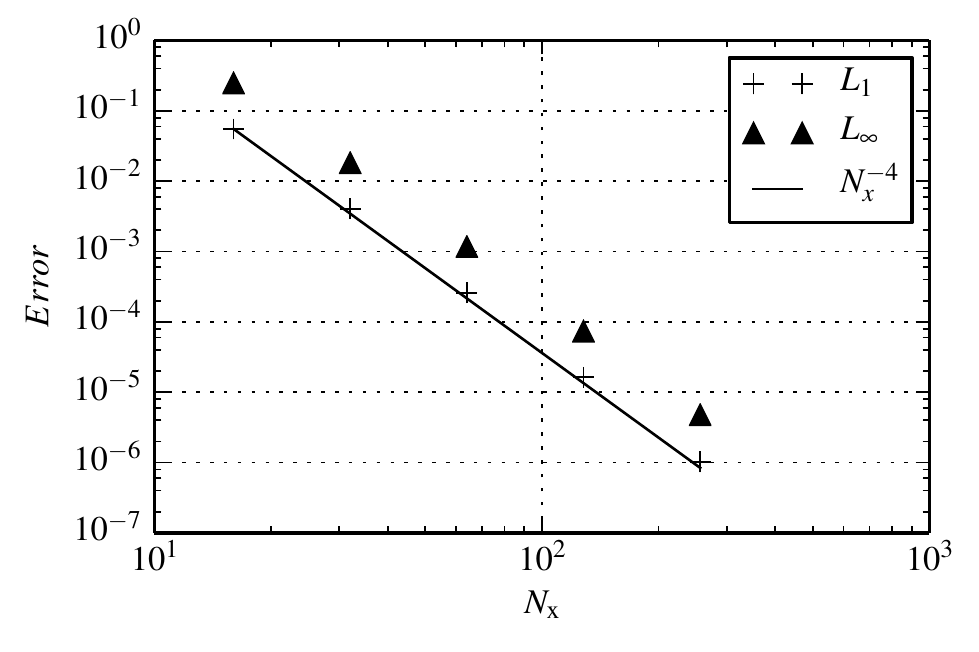}
\includegraphics[width=80mm]{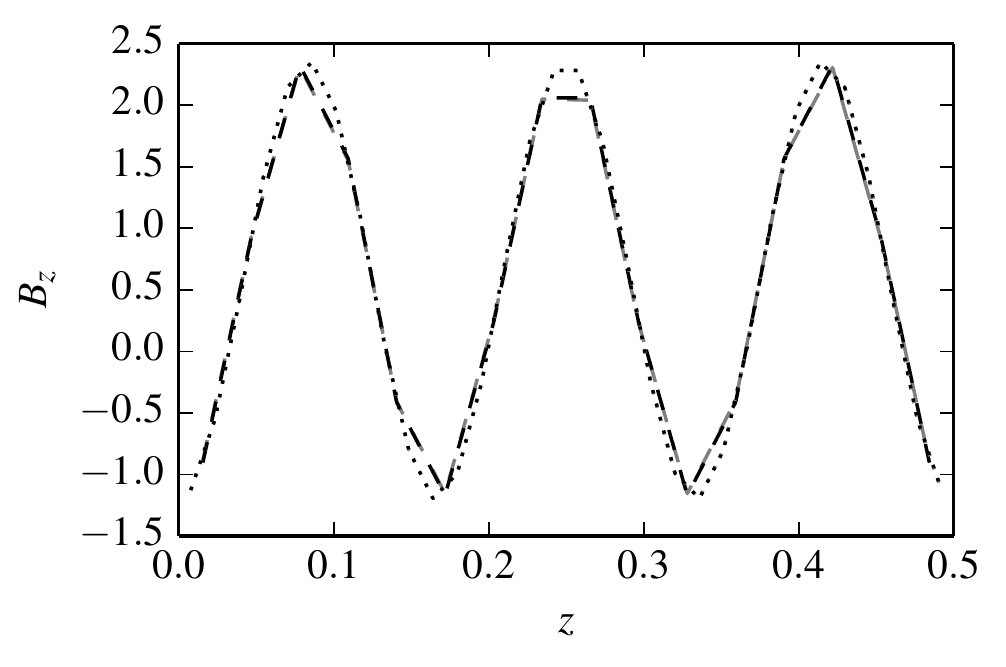}
\caption{Convergence for the 3D circularly polarised Alfv\'en Whistler wave test using finite differencing in Hall-MHD.  Reconstruction is performed with the MP5 algorithm, currents are obtained by fourth-order central differences Eq. (\ref{eq:jcd4}) and time-stepping with SSPRK(5,4) yielding over-all fourth order accuracy (left panel).
The right panel shows diagonal cuts in wave direction $\mathbf{\hat{k}}$ of $B_z$ at $t=0$ (black), and $t=P$ (gray) for the two resolutions $N_x=32$ (dashes) and $N_x=64$ (dots).  }
\label{fig:fdConvergenceHall}
\end{center}
\end{figure}

The setup of the test case is as follows:  We choose $k_x=k_y=k_z=2\pi m$, with $m=2$ and the state Eq. (\ref{eq:alfvenWhistlerState}) is rotated accordingly.  The Alfv\'en ion-gyro-radius is set to $d=1$ and we choose the plasma background parameters to satisfy  $v_{\rm A}=1$.  The wave amplitude is $A=1$.  
A 3D cartesian box with edge length $[1,0.5,0.5]$ is simulated with the finite differencing algorithm for one period 
\begin{align}
P = \frac{1}{\sqrt{3}\, m\, v_{\rm ph}(k)} \,,
\end{align}
using increasing resolution starting at $N_x\times N_y\times N_z = 16\times8\times8$ cells. 
The result of this test is presented in figure \ref{fig:fdConvergenceHall}.  
It shows fourth order convergence as expected based on the use of fourth order central differencing of the current.  In addition, the dispersion relation Eq. (\ref{eq:alfvenWhistler}) is realised by our code with increasing accuracy as the resolution is increased.  
The right-hand panel of figure \ref{fig:fdConvergenceHall} demonstrates the low numerical diffusion obtained with the high-order scheme:  already at a resolution of $N_x=32$, by eye, it is hard to distinguish the profile of the propagated wave (gray) from the analytical expectation (black).  

\subsubsection{Group diagram in Hall-MHD}

\begin{figure}[ht]
\centering
\includegraphics[width=0.45\textwidth]{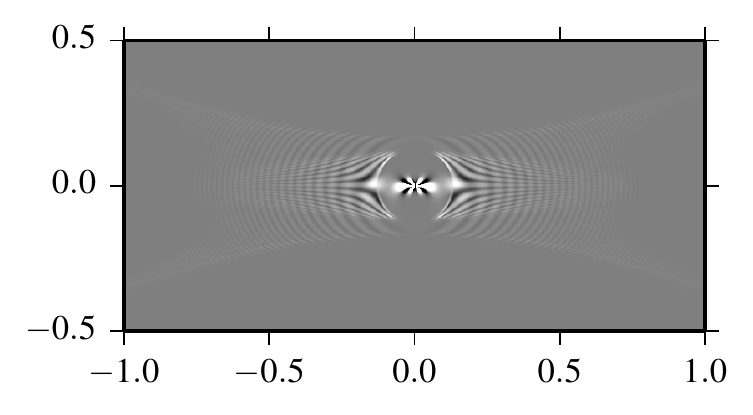}
\includegraphics[width=0.45\textwidth]{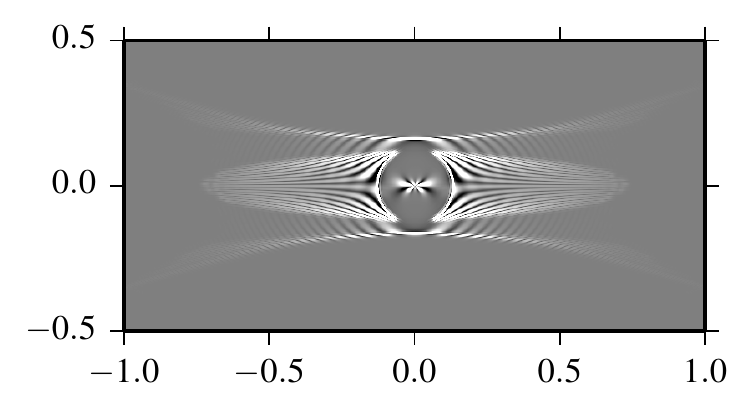}
\includegraphics[width=0.85\textwidth]{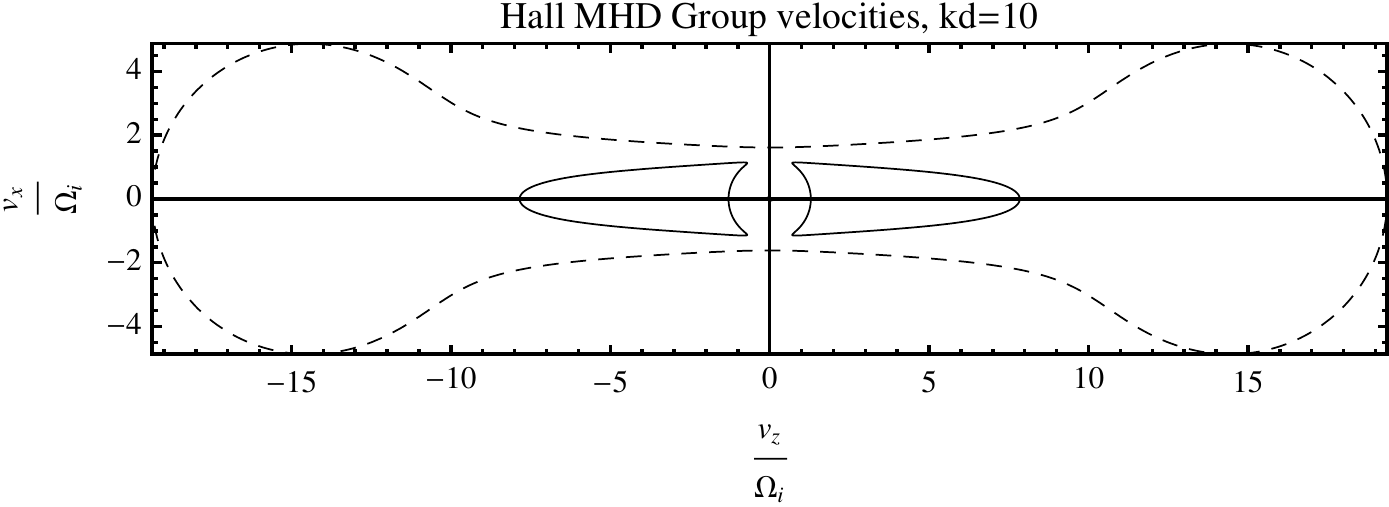}
\caption{Interference pattern resulting from the Friedrich wave diagram test with Hall-MHD.  Top panel: Out-of-plane velocity (left) and pressure (right).
Bottom panel: Analytic Hall-MHD group diagram for $kd=10$ for Alfv\'en-type (solid) and fast-type (dashed) waves.  
One can clearly recongnize the enveloping fast-type and Alfv\'en-type wave families.  }
\label{friedrichhall}
\end{figure}

A further test for the Hall-MHD module is the Friedrich diagram of the group velocity as known from the pure MHD case.  This classical example of MHD wave propagation can be used to study the transition from the ideal to the Hall MHD regime and  provides a qualitative comparison for our Hall-MHD module.  
The Hall-MHD group diagram was first shown in \cite{HameiriIshizawa2005} and differs greatly from the ideal MHD case.  In particular, the Alfv\'en-type ray surfaces -- only points in the ideal MHD case -- change dramatically and develop an extended front as illustrated in the lower panel of figure \ref{friedrichhall}.
Since the Hall-MHD system is not purely hyperbolic, the group diagram yields only an approximate ``envelope'' drawn by the fastest waves present.  

Our numerical realization of the group diagram initialises a (small) point-perturbation of a homogeneous medium threaded by a constant magnetic field in $z$-direction (see also \cite{Keppens2012718}).  The grid is adaptively refined to the sixth level, starting at a base resolution of $120^2$ cells within a domain $z,x\in[-1,1]$.  We again use the solver combination SSPRK(5,4)-MP5-FD yielding formally fourth order accuracy in space and time.  
Information on the perturbed state is transported by slow, Alfv\'en and fast wave packages that form the group diagram.  
We choose the following background state: $v_{\rm A}=0.96824$, $c_{\rm s}=1.29099$ and adopt an Alfv\'en ion-gyro radius of $d=10^{-2}$.  
Due to the dispersiveness of the modified Whistler waves, the group velocities in the Hall-MHD system depends on the wave-number $\mathbf{k}$ and thus our numerical realization consists of the interference of all $\mathbf{k}$ waves triggered by the initial perturbation.  
Analytic envelope functions for fast- and Alfv\'en-type waves can be constructed for the highest $k$-value present in the system, which in our explicit implementation depends on the numerical resolution.  
In figure \ref{friedrichhall}, we show realizations of the Friedrich diagram test in snapshots of out-of-plane velocity and pressure.  
An effective resolution of $3840^{2}$ cells was used for this test, resulting in maximal wavenumbers $kd\sim30$.  
In contrast to the MHD case, the highly anisotropic (fast-type) Whistler waves propagate most rapidly in the direction of the magnetic field.  
Interference between the individual waves scrambles the signal, however we can clearly make out two distinct types of waves.  These are the fast-type and Alfv\'en-type waves as illustrated in the bottom panel of figure \ref{friedrichhall}.  

\subsubsection{Hall MHD reconnection}

Magnetic reconnection plays a key role in plasma physics and many studies ranging from stationary resistive MHD \citep{1957JGR....62..509P} over time-dependent MHD simulations up to full PIC (particle in cell) simulations have been performed to date \citep[see the discussion in][]{2013PhPl...20i2109K}.  
To test our code at a challenging problem, we employ the so-called double-GEM setup adopted from the well-known Geospace Environment Modelling (GEM) challenge.  
Ideal Hall MHD reconnection was first employed to the GEM setup by \cite{ma2001}.  
The main difference in our setup to the classical GEM challenge is that the domain contains two alternating current sheets which allows to employ doubly periodic boundary conditions, facilitating inter comparison between codes and checks on exact conservation properties~\citep{pop13}.  

For completeness, the setup is described below.  
The domain is a 2D cartesian square with dimensions $(x,y)\in [-L/2,L/2]$ with $L=30$ and the current sheets are located at $y_{\rm up}=7.5$ and $y_{\rm low}=-7.5$.  
An ideal gas equation of state is adopted with ratio of specific heats $\gamma = 1.66666667$.  
We employ the magnetic field
\begin{eqnarray}
B_x & = & B_0 \left[-1+\tanh\left(y-y_{\rm low}\right)+\tanh\left(y_{\rm up}-y\right)\right] +\delta B_{x1} \,,\nonumber \\
B_y & = &\delta B_{y1} \,,
\end{eqnarray}
with perturbations 
\begin{eqnarray}
\delta B_{x1}& =& -\psi \frac{2\pi}{L}\cos\left(\frac{2\pi}{L}x\right)
\left[\sin\left(\frac{2\pi}{L}(y-y_{\rm low})\right)+2\left(y-y_{\rm low}\right)\cos\left(\frac{2\pi}{L}(y-y_{\rm low})\right)\right] \nonumber \\
& & \exp\left(-\frac{2\pi}{L}x^2-\frac{2\pi}{L}(y-y_{\rm low})^2\right) 
+\psi \frac{2\pi}{L}\cos\left(\frac{2\pi}{L}x\right) \nonumber \\
& &
\left[\sin\left(\frac{2\pi}{L}(y-y_{\rm up})\right)+2\left(y-y_{\rm up}\right)\cos\left(\frac{2\pi}{L}(y-y_{\rm up})\right)\right] 
\exp\left(-\frac{2\pi}{L}x^2-\frac{2\pi}{L}(y-y_{\rm up})^2\right) \,,\nonumber  \\
\delta B_{y1}& =& +\psi \frac{2\pi}{L}\cos\left(\frac{2\pi}{L}(y-y_{\rm low})\right) 
\left[\sin\left(\frac{2\pi}{L}x\right)+2x\cos\left(\frac{2\pi}{L}x\right)\right] \nonumber \\
& & \exp\left(-\frac{2\pi}{L}x^2-\frac{2\pi}{L}(y-y_{\rm low})^2\right) 
-\psi \frac{2\pi}{L}\cos\left(\frac{2\pi}{L}(y-y_{\rm up})\right) \nonumber \\
& &
\left[\sin\left(\frac{2\pi}{L}x\right)+2x\cos\left(\frac{2\pi}{L}x\right)\right] 
\exp\left(-\frac{2\pi}{L}x^2-\frac{2\pi}{L}(y-y_{\rm up})^2\right) \,.\nonumber
\end{eqnarray}
where the magnitude of the perturbation is set to $\psi=0.1$, a factor of 10 lower than the background field amplitude $B_0=1$.

The density profile is taken as
\begin{equation}
\rho=\left[\rho_{\rm at}+\cosh^{-2}\left(y-y_{\rm low}\right) +\cosh^{-2}\left(y-y_{\rm up}\right)\right] \,,
\end{equation}
and an MHD equilibrium configuration is obtained via the pressure profile
\begin{equation}
p=\frac{B_0^2\rho}{2} \,.\label{eq:gemDens}
\end{equation}
Thus far, the setup differs from \cite{2013PhPl...20i2109K} only in the higher atmospheric  density (outside the current sheets) with a value of $\rho_{\rm at}=0.2$ in equation ($\ref{eq:gemDens}$), compared to $\rho_{\rm at}=0.1$ in the original study.  As the Hall MHD evolution involves low plasma beta regions, the latter proved necessary to assure numerical stability.  This choice of parameters results in plasma $\beta=0.2$, atmospheric Alfv\'en velocity $v_{\rm A}=B_0/\sqrt{\rho_{\rm at}}\approx2.23$ and sound speed $c_{\rm s}=\sqrt{\gamma p_{\rm at}/\rho_{\rm at}}\approx0.91$.  
We choose an Alfv\'en ion-gyro radius of $d=1$ with the setting $\eta_{\rm h} = \sqrt{\rho_{\rm at}}$.  
Resistivity is chosen as $\eta=10^{-3}$ and we adopt a dynamic viscosity of $\mu=10^{-3}$.  In these runs, we employ a base-resolution of $120^2$ cells and add adaptive refinement based on variations in density \citep[following the prescription of][]{lohner1987} to a total of three (low resolution case) and four levels (high resolution case).  

\begin{figure}[htbp]
\begin{center}
\includegraphics[width=80mm]{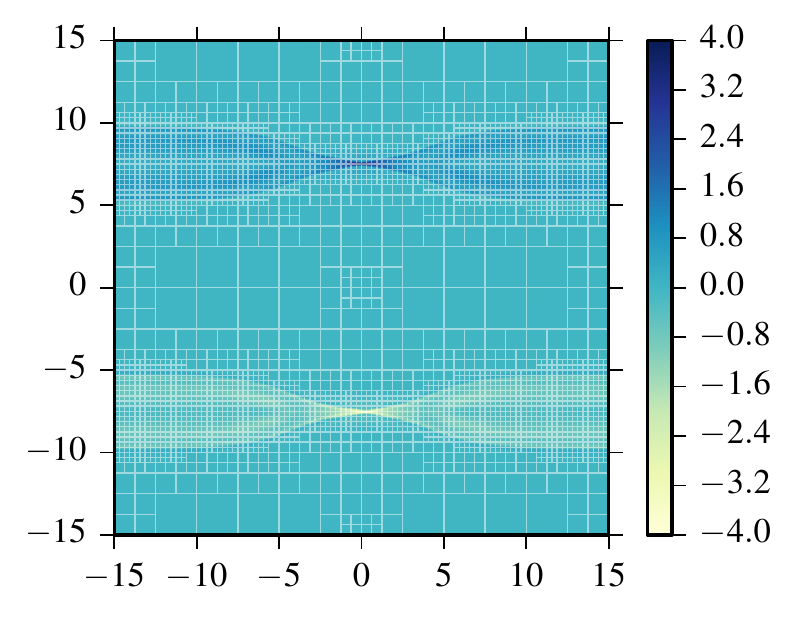}
\includegraphics[width=80mm]{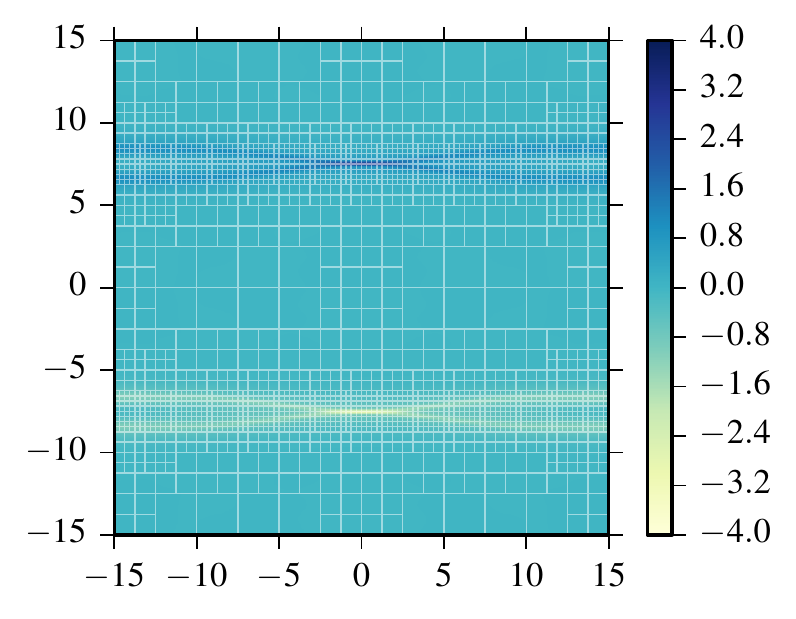}
\caption{Vertical current $j_z$ in the Hall double GEM (HDGEM) setup (left) and in the resistive MHD case with $d=0, \eta=\mu=10^{-3}$ (right).  
Snapshot time is at $t=75$ and we show the grid blocks as white lines each containing $10^2$ cells.  
We adopt an effective resolution of $960^2$ cells and the fourth order finite differencing scheme.}
\label{fig:dgemA}
\end{center}
\end{figure}

The evolution of the high resolution run is portrayed in figure \ref{fig:dgemA}. 
We observe the rapid development into an X-point, through which reconnection of magnetic field proceeds.   
By contrast, the visco-resistive MHD case shown in the right panel of figure \ref{fig:dgemA} develops a near stationary current sheet with well defined aspect ratio.  

\begin{figure}[htbp]
\begin{center}
\includegraphics[width=80mm]{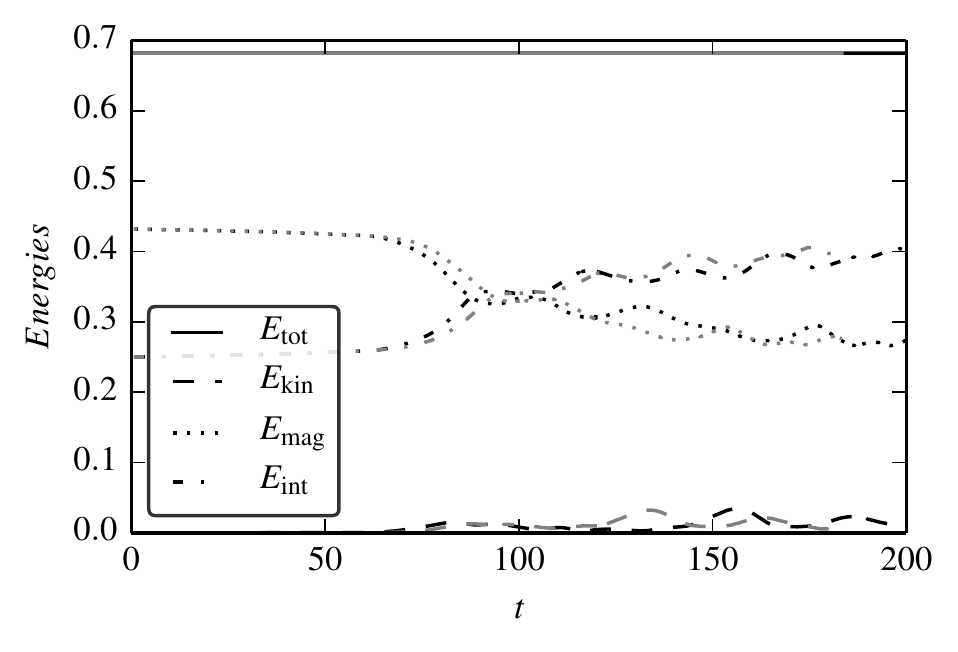}
\includegraphics[width=80mm]{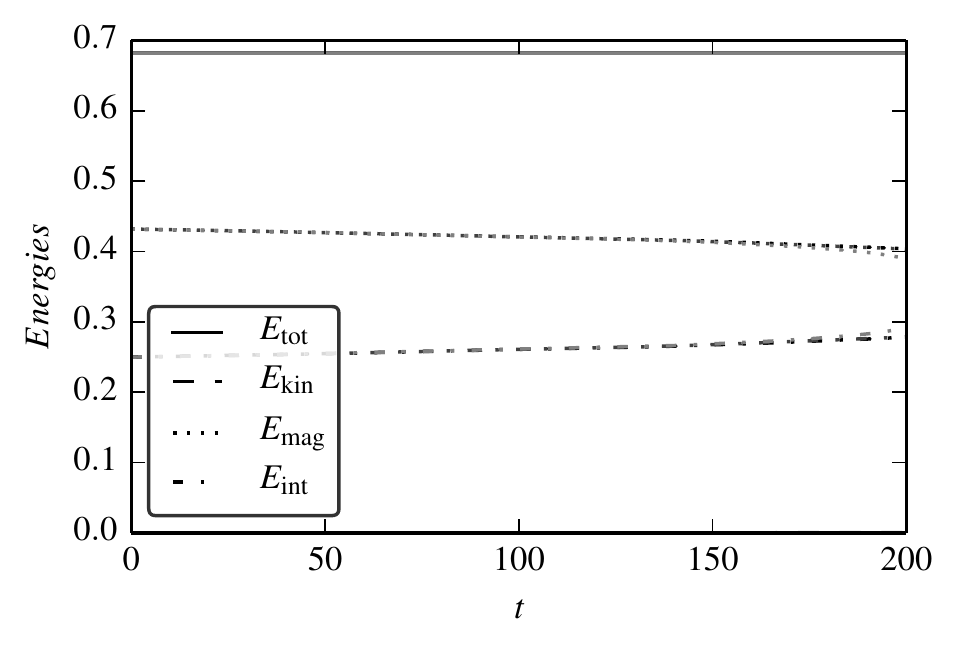}
\caption{Left: Energetics in the HDGEM setup for two different resolutions: $480^2$ (gray) and $960^2$ (black). 
Right: The equivalent evolution in the visco-resistive MHD case with $d=0$ for resolutions $480^2$ (light gray), $960^2$ (gray) and $1920^2$ (black).   
It is hard to distinguish the medium and high resolution cases by eye.  
In the second case, up to the simulated time, the dynamics is entirely governed by Ohmic dissipation.  By comparison, dissipation in Hall MHD is far more effective leading to an increased reconnection rate.  
}
\label{fig:energiesDGEM}
\end{center}
\end{figure}

\begin{figure}[htbp]
\begin{center}
\includegraphics[width=80mm]{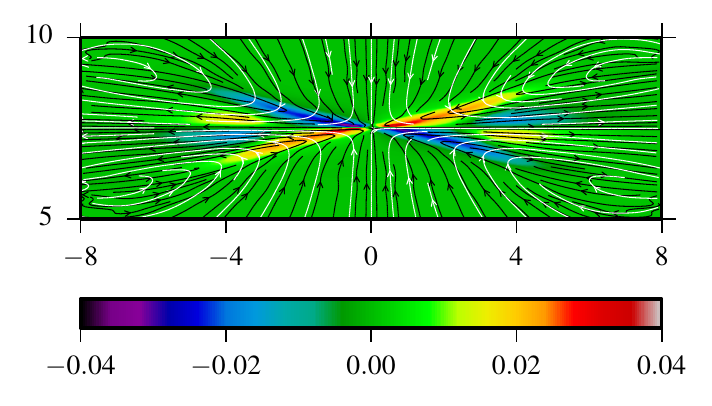}
\includegraphics[width=80mm]{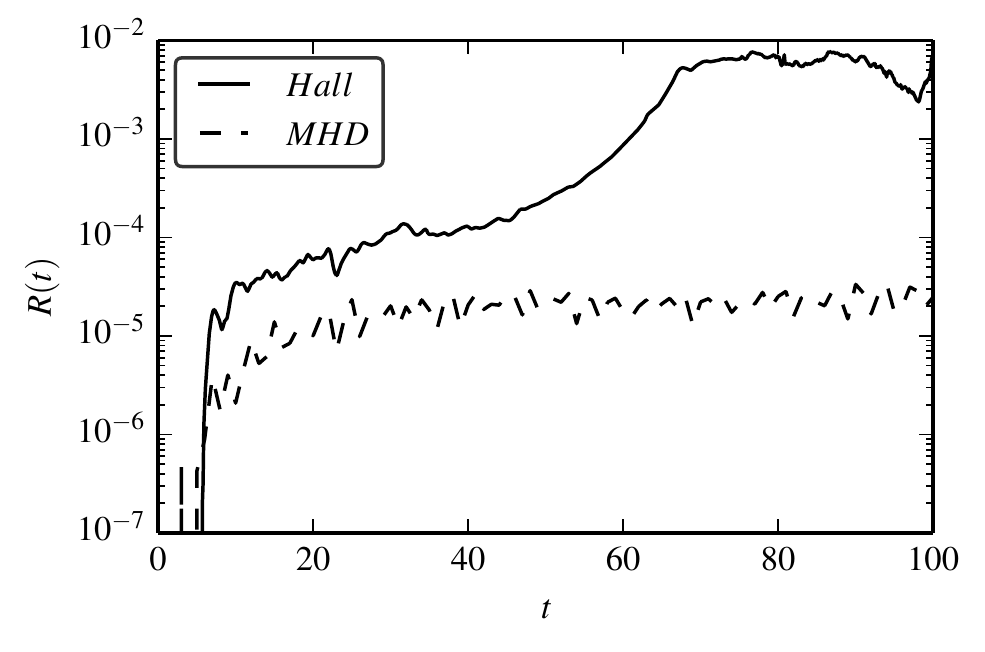}
\caption{Left: parallel electric field component $E_{||}$ and streamlines of electrons (black) and ions (white) in the HDGEM case, $t=70$.  In-plane field direction is traced by black lines.  Right: Rate of reconnected flux for the HDGEM case (solid) and the resistive MHD case (dashed).  The peak reconnection rate in the Hall case is $\sim 400$ times higher than in the resistive run.  
}
\label{fig:DGEMBz}
\end{center}
\end{figure}

The energetics of the reconnection process is shown in figure \ref{fig:energiesDGEM}.  
In the Hall MHD case, the initial magnetically dominated equilibrium reaches equipartition between internal and magnetic energy at $t\approx 80$.  This is also where the dissipation rate peaks.  Afterwards, the thermal energy increases more gradually and we observe fluctuations in the energetics that stem from compressive waves permeating the system.  
On the other hand, in the resistive MHD case the dynamics is dominated by Ohmic heating and the dissipation rate is nearly constant up to $t=200$.  
Conservation of total energy is granted with a relative error of $\Delta E/E=-1.46\times10^{-4}$ in the Hall case and with $\Delta E/E=4.1\times10^{-5}$ in the purely resistive case.  
This small energy error stems from the fact that resistive terms are currently not added in a conservative fashion.  
As noted previously \citep[e.g.][]{ShayDrake2001,birn2001}, inclusion of the Hall term is vital to obtain reconnection rates comparable to full kinetic descriptions.  Indeed, the reconnection rate of the in-plane flux $R(t)$ shown (see e.g. \cite{2004PhPl...11.3961F} for a definition of this rate of reconnected flux) in figure \ref{fig:DGEMBz} (right panel) increases over the resistive MHD case by over a factor of 100.  
In Hall-MHD, electrons and ions decouple on the scale of the ion-gyroradius.  Ideal Hall-MHD retains the frozen-in condition of ordinary MHD, however field-lines are advected only with the electron flow.  The stream-lines in our reconnection setup are drawn in the vicinity of the X-point in the left panel of figure \ref{fig:DGEMBz}.  It shows the decoupling on the scale of the Alfv\'en ion-gyroradius $d=1$ with momentary electron streamlines (black) and ion streamlines (white) on a background of the parallel electric field component $E_{||}=\mathbf{E\cdot \hat{B}}$.  
Upon entering the reconnection region, electron- and ion-flows are well aligned.  In the regions of strong $E_{||}$ at the ``wings'' of the X-point, the incoming electron-flow is deflected sharply towards the O-point.  Eventually, the ion-flow is deflected as well, however owing to its higher inertia with a larger radius of curvature.  

\subsubsection{Options for splitting magnetic fields: Field line extrapolations}

As indicated when describing the MHD equations as implemented, it is possible to split of a potential field solution $\mathbf{B}_0$, and reformulate the evolution equations in terms of the deviation $\mathbf{B}_1$ from this steady background field. This is particularly useful when one wishes to follow both gradual and more violent plasma dynamics in a realistically structured, solar coronal field topology.  To that end, we here demonstrate the available options for generating exact potential field solutions, from actual magnetogram data. In the context of this paper, we demonstrate the availability (as additional open source modules) of frequently used models for global spherical (PFSS) models, as well as for local Cartesian box models (Green function based), and make some observations on their accuracy. 

\subsubsection{Global spherical PFSS model}
The fundamental assumption made in the potential field source surface (PFSS) model~\citep{1969altschuler,1969schatten,1984hoeksema,1992wang,2003derosa} is that the magnetic field $\mathbf{B}_0$ is potential within the coronal volume, allowing a magnetic potential $\Phi$ to be defined such that $\mathbf{B}_0=-\nabla\Phi$. Since $\nabla\cdot\mathbf{B}_0=0$ everywhere, the potential $\Phi$ satisfies a Laplace equation, $\nabla^2\Phi=0$. The solution in spherical coordinates is 
\begin{equation}
\Phi(r,\theta,\phi)=\sum\limits_{l=0}^\infty\sum\limits_{m=-l}^l\left[A_l^m r^l+B_l^m r^{-(l+1)} \right]Y_l^m(\theta,\phi) \,,
\end{equation}
where the $Y_l^m$ indicate spherical harmonic functions of degree $l$ and order $m$ and the coefficients $A_l^m$ and $B_l^m$ are determined by the imposed radial boundary conditions. By definition, $Y_l^m(\theta,\phi)=C_l^m P_l^m(\cos\theta)e^{im\phi}$, where $P_l^m$ are the associated Legendre functions and the constants $C_l^m$ are determined as
\begin{equation}
C_l^m=(-1)^m\left[\frac{2l+1}{4\pi}\frac{(l-m)!}{(l+m)!}\right]^{1/2} \,.
\end{equation}
The photospheric boundary condition for $\Phi$ is 
\begin{equation}
\frac{\partial\Phi}{\partial r}=-B^p_r(1,\theta,\phi) \,,
\end{equation}
where $B^p_r(1,\theta,\phi)$ denotes the radial magnetic field as measured at the photosphere (quantified from a line-of-sight magnetogram). If we denote the spherical harmonic coefficients of $B^p_r(1,\theta,\phi)$ as $F_l^m$, such that $B^p_r(1,\theta,\phi)=\sum\limits_{l=0}^\infty\sum\limits_{m=-l}^lY_l^m F_l^m$, applying the boundary condition on $r=1$ for $\Phi$ leaves us with
\begin{equation}
\sum\limits_{l=0}^\infty\sum\limits_{m=-l}^lY_l^m\left[A_l^m l-B_l^m (l+1) \right]=-\sum\limits_{l=0}^\infty\sum\limits_{m=-l}^lY_l^m F_l^m \,. \label{bc1}
\end{equation}
In principle, the coefficients $F_l^m$ are determined by the equation
\begin{equation}
F_l^m=\int^{2\pi}_{0}d\phi\int^{\pi}_{0}d\theta \,\sin\theta \,Y_l^{m*}(\theta,\phi)\,B^p_r(1,\theta,\phi)=\int^{\pi}_{0}d\theta \,\sin\theta \,C_l^m \, P_l^m(cos\theta)\, F_m(\theta) \,,
\end{equation}
with $Y_l^{m*}(\theta,\phi)$  the complex conjugate of the spherical harmonic functions and $F_m(\theta)=\int^{2\pi}_{0}d\phi \, e^{-im\phi}\,B^p_r(1,\theta,\phi)$ the continuous spherical harmonic transform on the photospheric magnetic field. 
Though, instead of knowing the continuous function $B^p_r(1,\theta,\phi)$, we only know the values of the photospheric magnetic field at $N_{\theta}\times N_{\phi}$ points $(\theta_i,\phi_j)$, with $i=1,2,...,N_{\theta}$ and $j=1,2,...,N_{\phi}$ as obtained by observations. The discrete spherical harmonic transform is then given by $B^p_r(1,\theta_i,\phi_j)=\sum\limits_{l=0}^\infty\sum\limits_{m=-l}^l F_l^m Y_l^m(\theta_i,\phi_j)$, with
\begin{equation}
F_l^m=\sum\limits_{i=1}^{N_{\theta}} \left[ w_iC_l^mP_l^m(\cos\theta_i)\sum\limits_{j=1}^{N_{\phi}}\frac{1}{N_{\phi}}e^{-im\phi_j}B^p_r(1,\theta_i,\phi_j)\right]=\sum\limits_{i=1}^{N_{\theta}} \left[ w_iC_l^mP_l^m(\cos\theta_i)\,F_m(\theta_i) \right]  \,, \label{eq:flm}
\end{equation}
where the weights $w_i$ are obtained with the help of Legendre functions of the first kind.

The outer radial boundary condition is obtained by the ``source surface'' assumption. As source surface we define 
the sphere $r_{\mathrm{ss}}=2.5 r_{\odot}$ which is threaded by a purely radial field giving the von Neumann boundary condition 
$\partial_\phi\Phi(r_{\mathrm{ss}},\theta,\phi)=\partial_\theta\Phi(r_{\mathrm{ss}},\theta,\phi)=0$. 
This is satisfied if $\Phi$ is constant on this sphere and we can choose $\Phi(r_{\mathrm{ss}})=0$.  
Thus we obtain a relation between the expansion coefficients 
\begin{equation}
A_l^m r_{\mathrm{ss}}^l+B_l^m r_{\mathrm{ss}}^{-(l+1)}=0 \,. \label{bc2}
\end{equation}

Once the coefficients $F_l^m$ are determined from the magnetogram using relation (\ref{eq:flm}), the coefficients for $A_l^m$ and $B_l^m$ follow from (\ref{bc1}) using the orthogonality of the spherical harmonics.  We can then determine the magnetic field from the equation $\mathbf{B}_0=-\nabla\Phi$, or written out per component we get:
\begin{eqnarray}
B_r&=&Re\left(\sum_{l,m} Y^m_l \left[ A^m_llr^{l-1}-B^m_l(l+1)r^{-(l+2)} \right]\right) \,,\\
B_{\theta}&=&Re\left(\frac{1}{r\sin\theta} \sum_{l,m} Y^m_l \left\{ R^m_l(l-1) \left[ A^m_{l-1}r^{l-1}+B^m_{l-1}r^{-l} \right] -R^m_{l+1}(l+2) \left[ A^m_{l+1}r^{l+1}+B^m_{l+1}r^{-(l+2)} \right] \right\}\right) \,, \\
B_{\phi}&=&Re\left(\frac{1}{r\sin\theta}\sum_{l,m} i\, m Y^m_l\left[ A^m_lr^{l}+B^m_lr^{-(l+1)}\right]\right) \,,
\end{eqnarray}
where the factor $R^m_l$ is defined as $R^m_l=\sqrt{\frac{l^2-m^2}{4l^2-1}}$. Note that the field components are real, while all $Y^m_l, A^m_l, B^m_l, F_l^m$ are actually complex numbers.

\subsubsection{PFSS extrapolation for Carrington Rotation $CR2029$}

Synoptic magnetograms with resolution $N_{\theta}\times N_{\phi}$ can be used as input to estimate the solar coronal magnetic field. We can routinely use inputs from GONG observations\footnote{The Global Oscillation Network Group (GONG) is a community-based program to conduct a detailed study of solar internal structure and dynamics using helioseismology, see {\tt http://gong.nso.edu/}.} at resolution $180\times 360$ and from {\tt MDI} at resolution $1080\times 3600$. {\tt MDI} is an instrument onboard SOHO, the Solar and Heliospheric Observatory{\footnote{See {\tt http://sohowww.nascom.nasa.gov/}, a project of international collaboration between ESA and NASA to study the Sun from its deep core to the outer corona and the solar wind}. 
We make use of these magnetograms after performing a magnetogram remeshing technique using the Chebyshev collocation method \citep[e.g.][]{Carpenter95spectralmethods}. The latter interpolates the original grid where grid points are spaced equally in $\cos(\theta)$ onto a uniform $\theta$-grid.  Here we present a study for the solar Carrington rotation number $CR2029$ in 2005, using observations from the space telescope instrument {\tt MDI}. We pay particular attention to two active regions within $CR2029$, one located at the North hemisphere $AR10759$ and one at the South hemisphere $AR10756$. These two dominant active regions on each hemisphere will be used to (1) compare the global potential field source surface spherical extrapolation approach and a local potential field Cartesian one, and (2) to understand the influence of raising the number of spherical harmonics. For the latter, we will take active region $AR10756$  as the photospheric region on which we examine the radial magnetic field variation over a line crossing the active region's opposite polarities, as influenced by the number of spherical harmonics taken.
In Figure~\ref{globalviews}, we first present an impression of the global magnetic field topology obtained from the full MDI magnetogram, used to generate a PFSS model up to $r_{\mathrm{ss}}$ using $l_{\mathrm{max}}=720$ and exploiting a 3-level block-AMR grid with effective resolution $240\times360\times720$.

\begin{figure}[htp]
\centering
   \begin{tabular}{cc}

    \includegraphics[width=80mm]{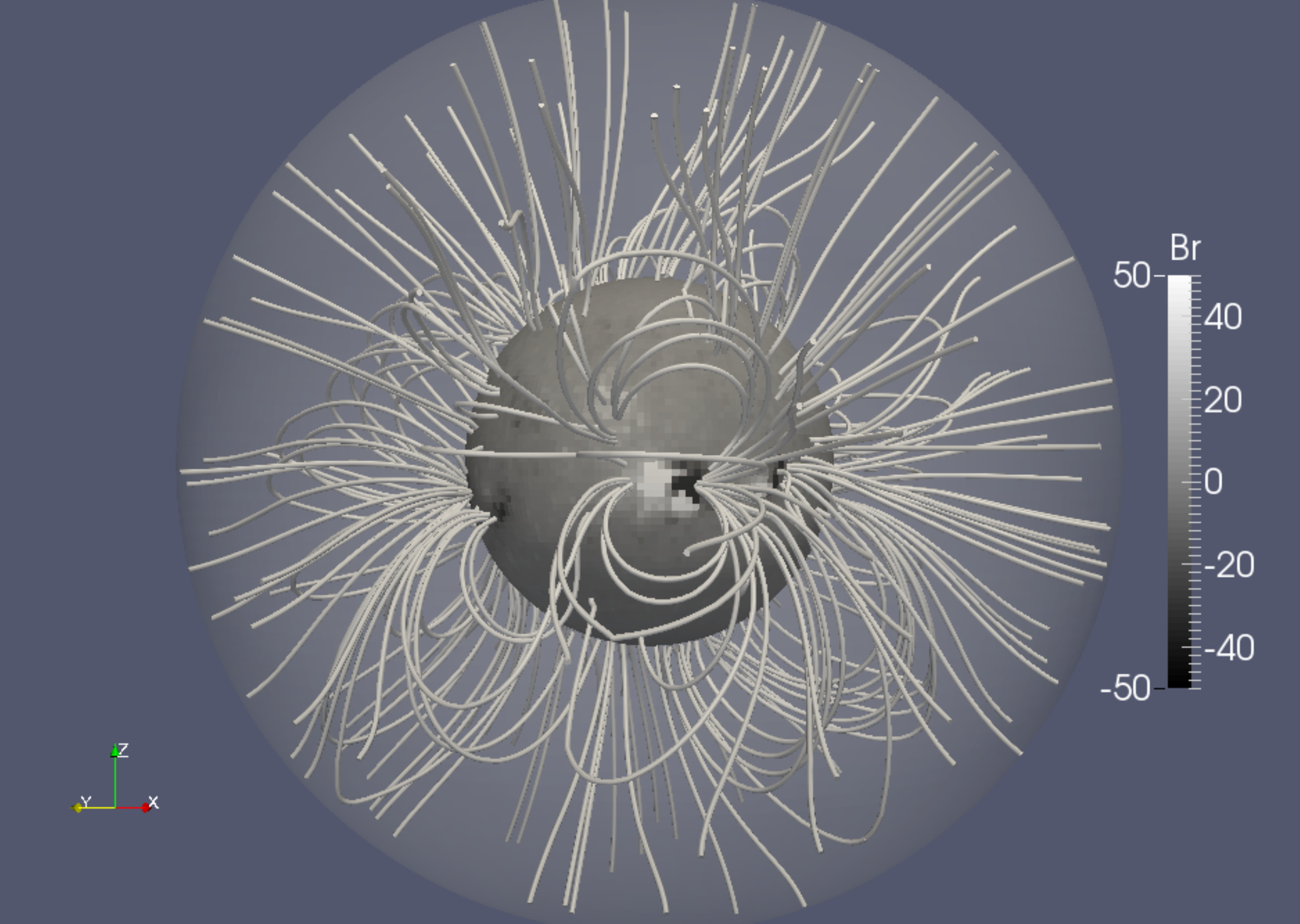}
    \includegraphics[width=80mm]{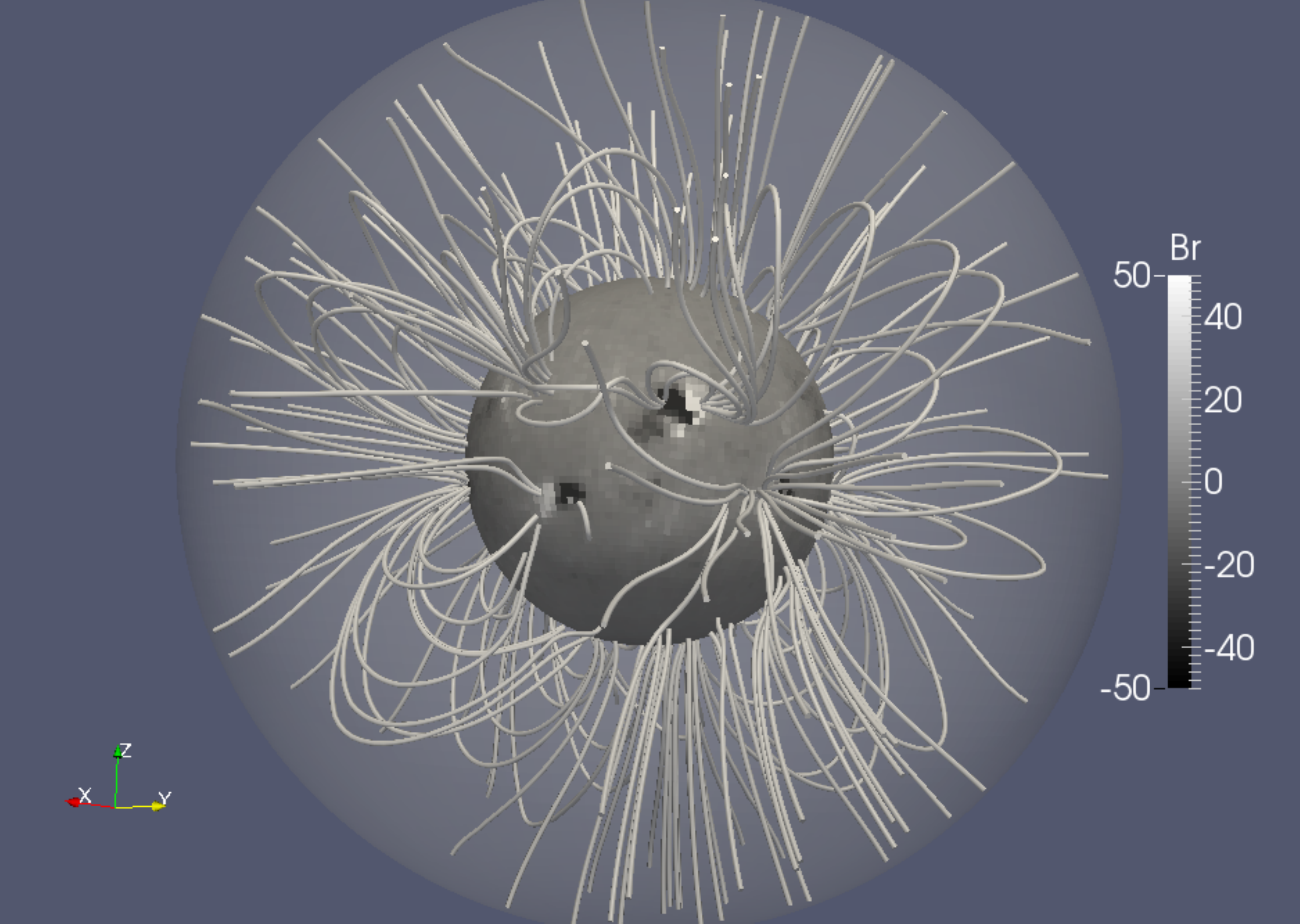}
   \end{tabular}
   \caption{Global view of the magnetic field topology from the photosphere to the source surface at $2.5r_{\odot}$ together with the active region $AR10756$ observed for the synoptic map of $CR2029$ on the 30th of April 2005 as shown on the left panel and $AR10759$ observed on the 14th of May 2005 as shown on the right panel. The two active regions differ in location by about half a solar rotation. These global PFSS models use spherical coordinates, have $l_{\mathrm{max}}=720$ and use a 3-level {\tt AMR} grid with effective resolution $240\times360\times720$.}\label{globalviews}
\end{figure}

In order to investigate the effect of the number of spherical harmonics that we include in our computations on the accuracy of our results,  we construct several PFSS models and compare them to a reference case. For this comparison purpose, all these models exploit a fixed uniform resolution $150\times180\times360$, but we vary $l_{\mathrm{max}}=90, 135, 270, 540, 720$, where $l_{\mathrm{max}}$ is the maximum degree of the spherical harmonic functions used in each case for the magnetic field calculation. The case with $l_{\mathrm{max}}=720$ determines our reference case, since it is the maximum degree we can use for {\tt MDI} magnetograms according to alias-free conditions as mentioned by~\cite{sudafast2002}: $l_{max}\leq min \left( \frac{2 N_{\theta}}{3},\frac{N_{\phi}}{3} \right)\Rightarrow l_{max}\leq 720$. In order to compare the different runs, we examine the radial component of the magnetic field on the photosphere as it varies along a line that crosses active region $AR10756$, as demonstrated in figure~\ref{fluctuations}. As the number of spherical harmonics increases, the magnetic field maximal amplitude grows and gradually approaches the reference case variation. At the same time, the intensity of the ringing effect~\citep{2011toth} affecting the magnetic field value around the active region diminishes and the curve gets smoother.

\begin{figure}[htp]
\centering
    \begin{tabular}{ccc}

    \includegraphics[width=50mm]{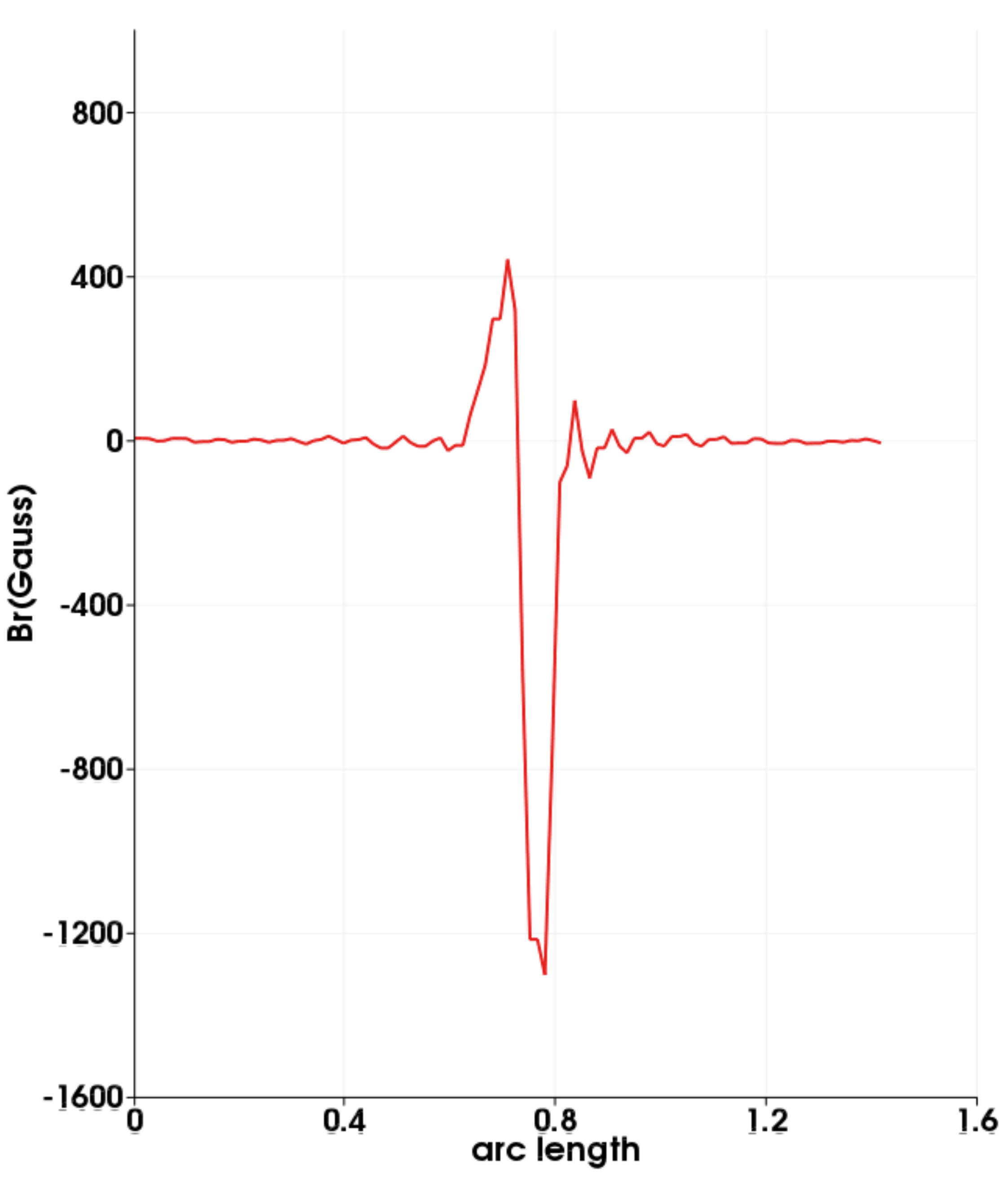}
    
    \includegraphics[width=50mm]{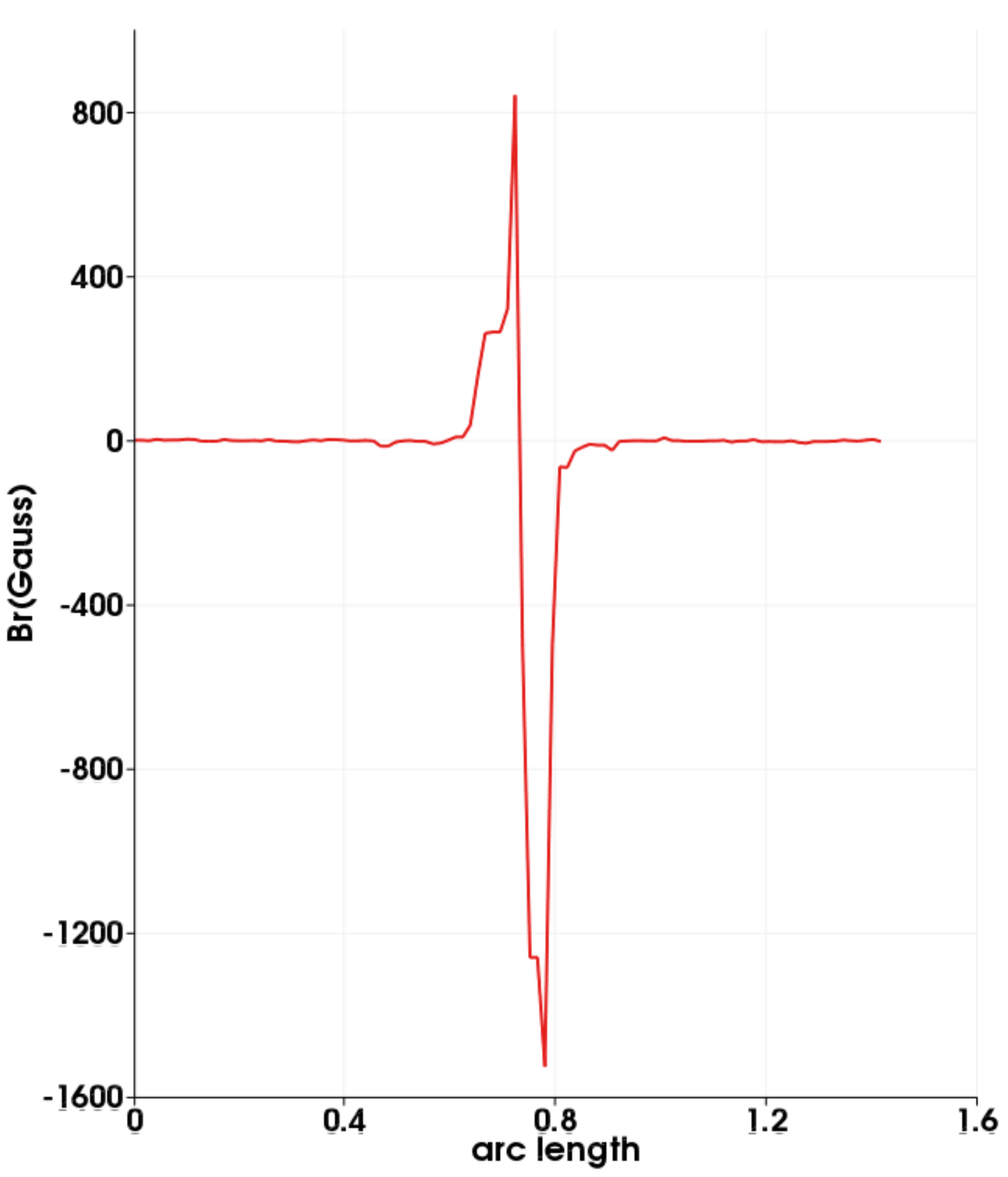}
    
    \includegraphics[width=50mm]{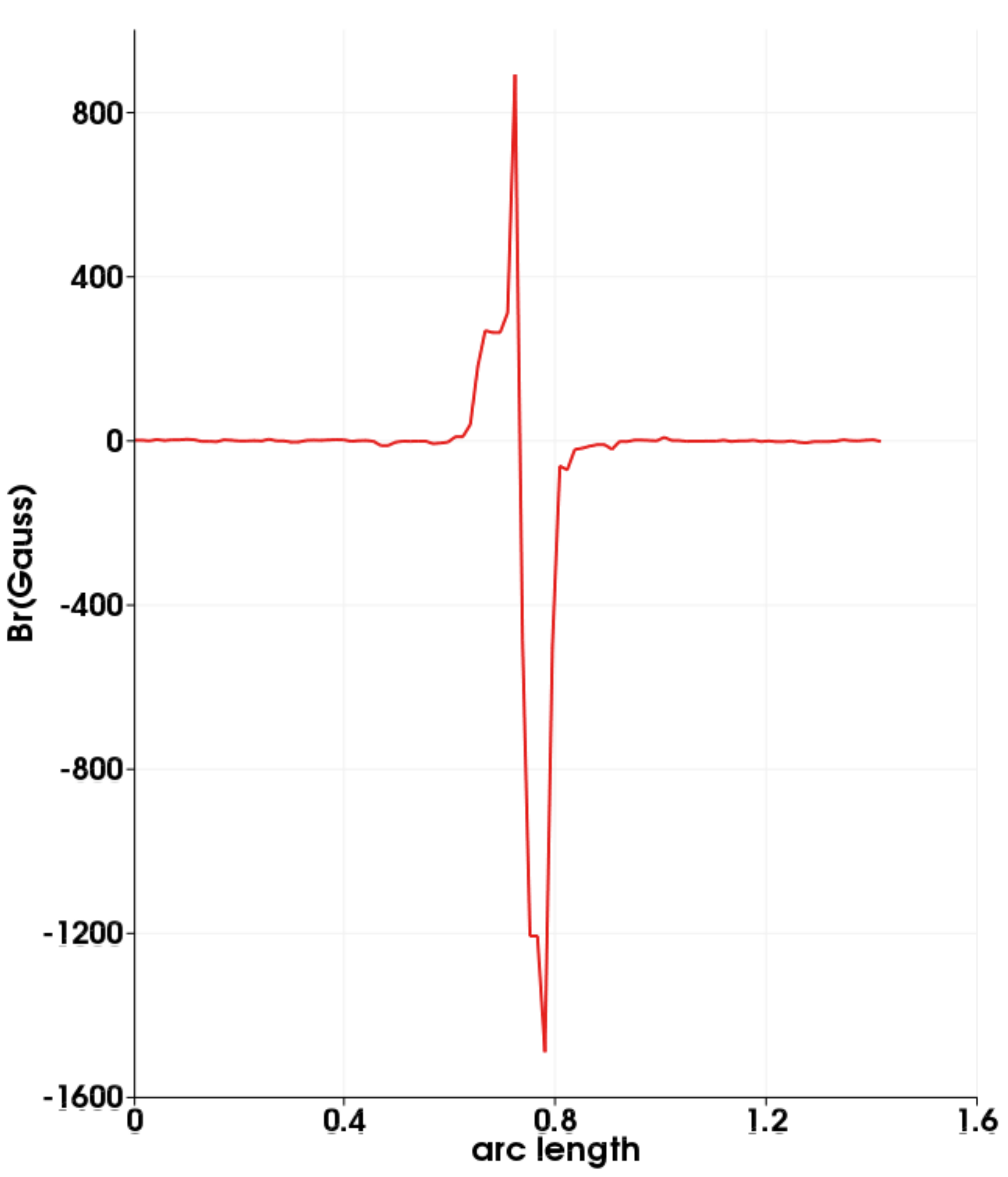}

  \end{tabular}
   \caption{Plot of the radial component of the magnetic field along a line passing through the active region $AR10756$ in $CR2029$. From left to right $l_{\mathrm{max}}=90,270,720$. We notice that as $l_{\mathrm{max}}$ increases, the ringing oscillations visible when $l_{\mathrm{max}}=90$ decrease and both magnitude and variation of the radial component of the magnetic field approach the reference case value at right.}\label{fluctuations}
\end{figure}

In order to quantify this better, we compute the errors $\mathcal{E}$ of the magnetic field magnitude for each of the above models with respect to the reference case with maximal $l_{\mathrm{max}}=720$.  
In figure~\ref{norms}, the error calculation is demonstrated with two norms, the $L_1$ (lowest curves) and the $L_{\infty}$ norm (upper data).
The solid line indicates a power law fitting curve given by $\mathcal{E}=836.3152 \,l_{\mathrm{max}}^{-2.22912}$ and the dashed line has $\mathcal{E}=0.05634 e^{-0.0087419\, l_{\mathrm{max}}}$ as an exponential fit for the $L_1$ norm. For the $L_{\infty}$ norm, the dotted line is the power law fitting curve $\mathcal{E}=25.7 \times 10^6 \, l_{\mathrm{max}}^{-2.30698}$ and the dot-dashed line $\mathcal{E}=1.3\times 10^3 e^{-0.00918513 \,l_{\mathrm{max}}}$ is the exponential fit.  The two norms differ about 4 orders of magnitude for all PFSS models reported in the plot, a fact which indicates that our main errors originate from specific localized regions. This conclusion is realistic as we would expect that the main errors are introduced by the existence of regions where the magnetic field is noticed to show a sudden increase of orders of magnitude, i.e. the active regions.

\begin{figure}[htp]
\centering
\includegraphics[width=0.8\textwidth]{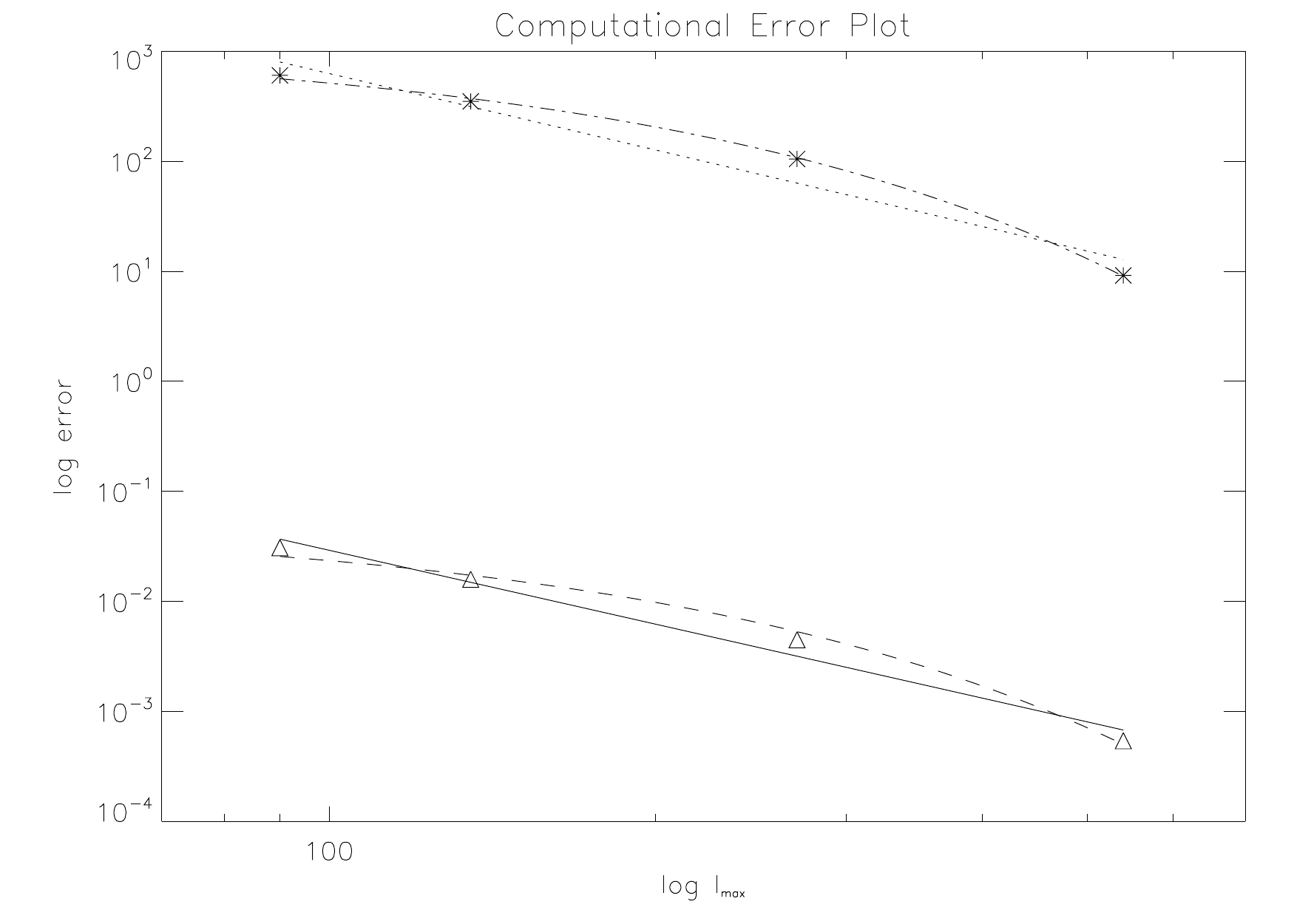}
\caption{The error quantification on the magnitude of the total magnetic field for each PFSS model where $l_{\mathrm{max}}=90, 135, 270, 540$, with respect to the reference run at $l_{\mathrm{max}}=720$, with grid resolution $150\times180\times360$ and uniform grid for every case using $L_1$ and $L_{\infty}$ norms. We also show second order power law and exponential extrapolation curves that capture the observed trends, as quantified in the text.}\label{norms}
\end{figure}

\subsubsection{Global versus local Cartesian extrapolation}

Besides the global PFSS model using full sun synoptic maps, we also implemented a local potential field extrapolation method, useful when interested in simulating specific local active region behavior. This local Cartesian approach is using exact closed-form solutions of the force-free magnetic field boundary-value problem with the help of Green's function method~\citep{1977chiu}. When the photospheric surface corresponds to the $z=0$ plane, the problem still reduces to solving the Laplace equation for the magnetic potential with the photospheric boundary condition $B^p_z(x,y)$ given by magnetograms. When we take as second boundary condition $\mathbf{B}_0\rightarrow \mathbf{0}$ as $z\rightarrow \infty$, the Cartesian components of the magnetic field in Green's function forms are given by
\begin{eqnarray}
\overline{B}_i=\frac{1}{2\pi}\int^{y_b}_{y_a}\,\int^{x_b}_{x_a}\,dx'\,dy'\,\overline{G}_i(x,y,z;x',y')B^p_z(x',y')   \,\, \,\,\, \mathrm{for}\, i=x,y,z \,,
\end{eqnarray}
where $[x_a,x_b]$ and $[y_a,y_b]$ are the boundaries of the extracted magnetogram, and the integrals contain
\begin{eqnarray}
\overline{G}_x&=&\frac{x-x'}{R}\frac{\partial\overline{\Gamma}}{\partial z}+\alpha\overline{\Gamma}\frac{y-y'}{R} \,,\\
\overline{G}_y&=&\frac{y-y'}{R}\frac{\partial\overline{\Gamma}}{\partial z}+\alpha\overline{\Gamma}\frac{x-x'}{R}  \,,\\
\overline{G}_z&=&\frac{z}{r^3}\,\cos(\alpha r)+\frac{\alpha z}{r^2}\,\sin(\alpha r) \,,\\
\overline{\Gamma}&=&\frac{z}{R r}\,\cos(\alpha r)-\frac{1}{R}\,\cos(\alpha z) \,,\\
\frac{\partial\overline{\Gamma}}{\partial z}&=&\left(\frac{1}{Rr}-\frac{z^2}{Rr^3} \right)\cos(\alpha r)-\frac{\alpha z^2}{R r^2}\sin(\alpha r)+\frac{\alpha}{R}\sin(\alpha z) \,,
\end{eqnarray}
with $R^2=(x-x')^2+(y-y')^2$ and $r^2=R^2+z^2$ the position vector squared. The above formulae allow for a constant nonzero value of $\alpha$, then generating a linear force-free field, while for $\alpha=0$ we get the potential magnetic field solution which can be split of. This exact solution does not suffer from the need to truncate at a specific angular degree $l_{\mathrm{max}}$ encountered when using the spherical harmonics. For the integral evaluations, a simple midpoint rule is adopted. 

To qualitatively compare this local extrapolation method with the PFSS model in a global spherical geometry, we can do the following, all shown in figure~\ref{localglobal}. We can start from a synoptic magnetogram of a full Carrington rotation, so that the observational input is in the form of a 2D matrix of size $N_{\theta}\times N_{\phi}$, for MDI this is $1080\times 3600$. The abovementioned Chebyshev  remeshing technique is first used to transform the whole magnetogram into a uniform $(\theta,\phi)$ grid, similar to the global case. This uniform $(\theta,\phi)$ grid can be transformed into a Cartesian grid with each angular degree corresponding to a length equal to $\pi r_{\odot}/180$. Finally, an area of interest is extracted in Cartesian coordinates $\Delta y\times\Delta x$ in the form of a 2D submatrix corresponding to a user selected $\Delta\theta\times\Delta\phi$  angular part of the magnetogram. Here, we take a $30^o\times 30^o$ part containing the $AR10759$ which counts $181\times 301$ grid points, covering a region of $364.8 \times 364.8 \mathrm{Mm}^2$. We use this as a bottom magnetogram for a local extrapolation using the above method, where we use a 4-level AMR grid with effective resolution $384\times 384\times 384$ with in each direction a maximal resolving power  where $\Delta x=\Delta y=\Delta z= 0.95$ Mm. We also take the full MDI magnetogram as bottom boundary for a global PFSS extrapolation, this time using a uniform $180\times270\times 540$ grid in spherical coordinates $(r,\theta,\varphi)$, where the radial range goes up to the source surface. This latter spherical grid ensures an effective resolving power of about $8\times 8 \mathrm{Mm}^2$ on the solar suface. 

The field lines for both kinds of extrapolations are drawn in figure~\ref{localglobal}, where we show a zoomed view on the active region from the global PFSS model, and the local Cartesian result. 
We selected 20 specific points to start drawing the field lines. In the same figure there is also an observational EIT\footnote{Extreme ultraviolet Imaging Telescope (EIT) is an instrument on the SOHO spacecraft, sensitive to four different wavelengths 171, 195, 284, and $304\AA$ with a $17$ minute cadence and spacial resolution of $(1800 \mathrm{km})^2$.} image of the active region $AR10759$ at 195\AA, which corresponds to Fe XII and a temperature of $1.6\times 10^6 K$. The two approaches show similar structure, as expected. We underline that for the local simulation, there is no source surface (the top boundary differs for the exploited Green function), a fact that explains why we have more dominant open field line topology in the right panel of figure~\ref{localglobal}. The remaining differences are due to finite curvature effects, not taken along in the Cartesian approach. The extrapolations agree only qualitatively with the observational data in the extreme ultraviolet. The (magnetically structured) plasma morphology visible at this wavelength shows similarities with the open and closed field lines of both global and local simulations, but may well deviate significantly from potential field conditions. E.g. at the center of the active region in the EIT view, a region with negative polarity differs most from the bottom magnetogram structure, as this filter shows the plasma higher inside the low corona than the photosphere itself. By inspection, the potential field extrapolation misses the implied magnetic connectivity in those regions.

\begin{figure}[htp]
\centering
  \begin{tabular}{ccc}

    \includegraphics[width=0.3\textwidth]{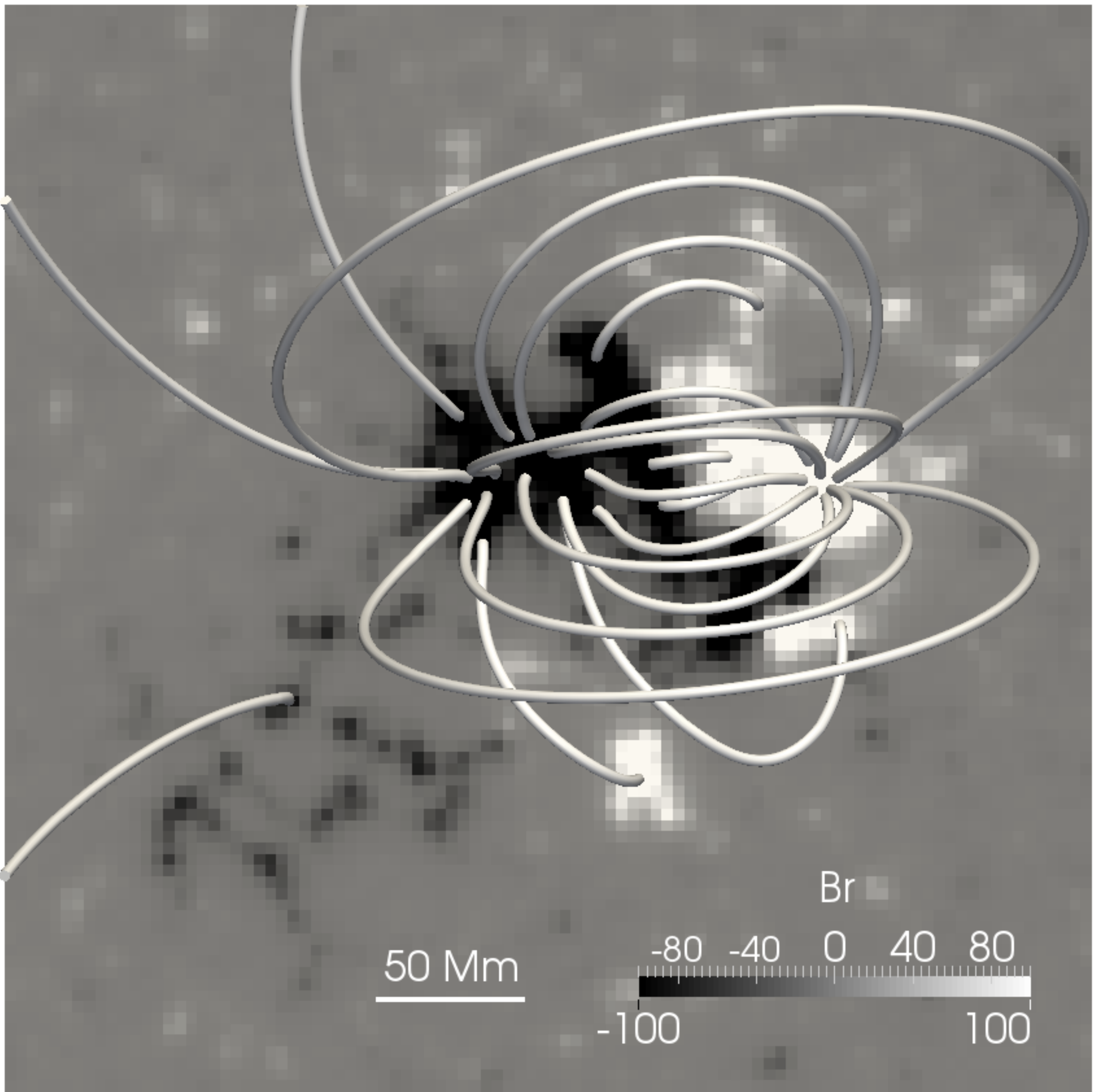}
    \includegraphics[width=0.3\textwidth]{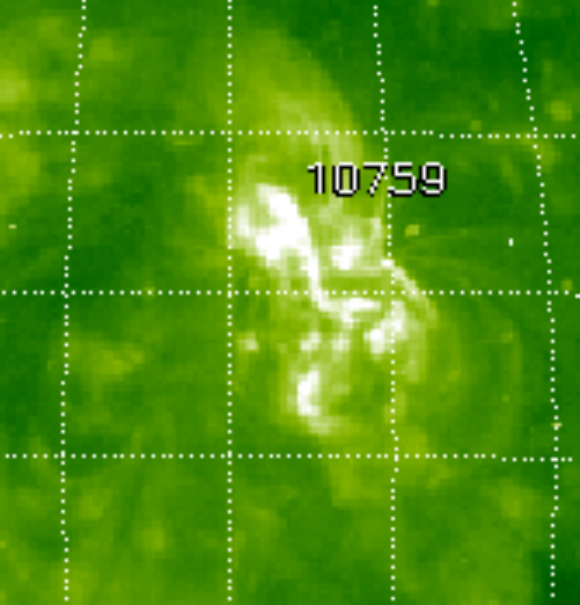}
    \includegraphics[width=0.3\textwidth]{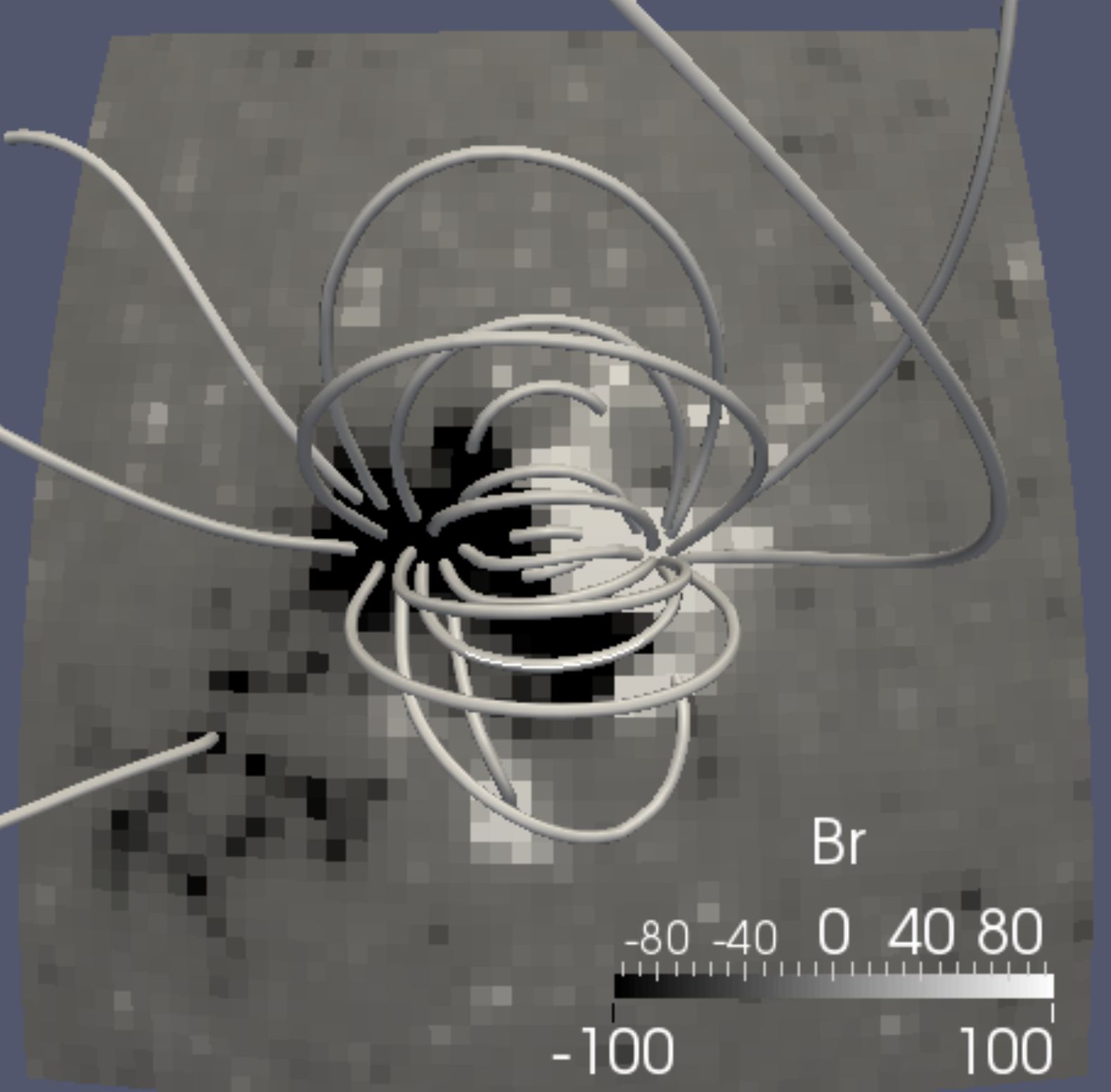}

  \end{tabular}
  \caption{Local potential field extrapolation in Cartesian coordinates (left) versus a zoomed view taken from a global PFSS model in spherical coordinates (right), for $CR2029$ zoomed in to $AR10759$. The effective resolution for the local simulation is $384\times384\times384$ with cell size of 0.95 Mm, while for the global simulation we exploit $180\times270\times540$ with on disk 8 Mm per cell. We visualize and compare the magnetic field topology for the two approaches of the same model. The middle panel is an extreme ultraviolet observation from EIT $195\AA$ for $AR10759$.} \label{localglobal}
\end{figure}

\subsubsection{Magneto-convection}

As a representative, time dependent, solar application where non-ideal MHD processes are incorporated, we simulate compressible magnetoconvection in a strongly stratified layer, following \cite{rucklidgeWeiss2000}. Their parametric survey focused on a prescribed polytropic atmosphere, modified with a uniform vertical magnetic field initially, and varied the field strength, the relative importance of magnetic diffusion, (isotropic) thermal conduction, and viscosity, as well as geometric parameters like the box aspect ratio. Augmented with simple boundary prescriptions fixing the top to bottom temperature contrast and fields, these authors made a systematic parameter study, identifying transitions from essentially two- to three-dimensional behavior, from kinematic to more magnetically influenced cases, from ordered to chaotic regimes. A detailed analysis of the (loss of) symmetry in the convecting endstates could benefit from group theoretical classifications using the linear eigenfunction behaviors. This allowed to obtain bifurcation diagrams, serving to classify the large variety of steady to unsteady magnetoconvection patterns.
Here we will adopt two realisations, one in the steady regime and one in the unsteady chaotic parameter regime.  

In a 3D Cartesian box $[0,\lambda]\times[0,1],\times[0,\lambda]$ with gravity along the negative $y$-direction and periodic horizontal $x$ and $z$ directions, we initialize density and pressure as
\begin{eqnarray}
\rho(y) & = & \left[1+\theta \left(1-y\right)\right] \,,\\
p(y) & = & \left[1+\theta \left(1-y\right)\right]^2 \,,
\end{eqnarray}
such that the dimensionless temperatures at top $T(1)=1$ and bottom $T(0)=1+\theta$ are fixed when $\theta=10$. This dimensionalization uses the top layer temperature $T_0$ and density $\rho_0$, together with the layer depth $d_0$, to set dimensionless profiles and parameter values. Specifically, the atmosphere obeys hydrostatic equilibrium with dimensionless gravity parameter $\bar{g}=g\,d_0/R_*T_0=2 \theta$ for gas constant $R_*$. Similarly, non-ideal parameters enter for viscosity $\bar{\mu}=\mu/\rho_0 d_0 \sqrt{R_*T_0}$, thermal conduction $\bar{\kappa}=\kappa/\rho_0 d_0 R_*\sqrt{R_* T_0}$ and resistivity $\bar{\eta}=\eta/d_0\sqrt{R_*T_0}$. The original study fixed the Prandtl parameter $\sigma=\frac{\bar{\mu}}{\bar{\kappa}}\frac{\gamma}{\gamma-1}=1$ (with the ratio of specific heats $\gamma=c_p/c_v=5/3$ where $R_*=c_p-c_v$), and then varied the initial settings through a Chandrasekhar number $Q$ and a Rayleigh number $R$. The ratio of magnetic to thermal diffusivities is dimensionally fixed by $\zeta=\frac{\eta\rho_0 c_p}{\kappa}=\frac{\bar{\eta}}{\bar{\kappa}}\frac{\gamma}{\gamma-1}$, and we will focus on cases where the related mid-layer value $\zeta_m=\zeta (1+\theta/2)$ is set to 1.2 (the `astrophysically relevant situation', as stated in~\cite{rucklidgeWeiss2000}). The Chandrasekhar number was always computed from $Q=1200/\zeta_m=1000$, and our dimensionalization yields the initial dimensionless magnetic field strength as $B_y(t=0)=\sqrt{Q\bar{\mu}\bar{\eta}}$. All parameters then become fixed by the value of the Rayleigh number $R$, whose value at mid-layer in essence determines thermal conduction through the relation
\begin{equation}
R = 2 \left[1-2\frac{\gamma-1}{\gamma}\right] \left(1+\theta/2\right) \frac{\theta^2 \gamma^2}{\sigma \bar{\kappa}^2 (\gamma-1)^2} \,.
\end{equation}
Note that we here employ an isotropic thermal conduction (in analogy with the original study, although \amrvac allows for anisotropic heat conduction physics as exploited in~\cite{xia2012} and~\cite{fangXia2013}), together with uniform resistivity and full tensorial viscosity.
We used a deterministic incompressible velocity perturbation found from
\begin{eqnarray}
\mathbf{v} & = & \nabla \times \mathbf{\Psi} \,,\\
\Psi_z(x,y) & = & \sum_{j=1}^{N_z} \frac{0.05}{j} \cos(\frac{2\pi j x}{\lambda}+\phi_j^z)\exp(-\left(\frac{y-0.5}{0.2}\right)^2) \,,\\
\Psi_x(y,z) & = & \sum_{j=1}^{N_x} \frac{0.05}{j} \cos(\frac{2\pi j z}{\lambda}+\phi_j^x)\exp(-\left(\frac{y-0.5}{0.2}\right)^2) \,,
\end{eqnarray}
where $N_z=N_x=6$ modes with specific phases $\phi_j^{z,x}$ were used. Boundary conditions are double periodic sideways, while both top and bottom use symmetric conditions for density $\rho$, velocity components $v_x$, $v_z$, with asymmetry for $v_y$ and $B_x$, $B_z$. The pressure is set in the ghost cells to $p=\rho T$ with the fixed top $T(1)=1$ and bottom $T(0)=1+\theta$ values. A second order central differencing formula on the solenoidal constraint is used to extrapolate $B_y$, while the GLM scalar $\psi=0$ in the ghost cells.

We use a multi-stage (ssprk54), finite difference scheme with MP5 reconstruction. Two cases are shown below, one (top part of Fig.~\ref{f-convect}) is at aspect ratio $\lambda=8/3$ and $R=45000$, a case known to allow steady state solutions with irregular hexagons consisting of a fixed number of uprising plumes. While both an 8 and 9 plume solution were reported in~\cite{rucklidgeWeiss2000}, we found a 7 plume pattern which can be safely quantified as a true steady-state solution in a domain decomposition run of overal resolution $160\times 60\times 160$. Figure~\ref{f-convect} shows the magnetic pressure $B^2/2$ pattern in the endstate (with flow field vectors on the sidepanels), and the temporal evolution of the residual, reaching a value of $2\times10^{-8}$ after time $t=150$. The slightly erratic oscillations between this value and $4\times10^{-7}$ thereafter are influenced by IO operations, which have e.g. switched conservative to primitive variables in place at selected save times.  

 Another case is shown in the lower panel of Fig.~\ref{f-convect}, for parameters $\lambda=8$ and $R=100000$ at resolution $240\times 60\times 240$. In this parameter regime with a very wide box, one witnesses flux separation where narrow strong field lanes surround patches that are almost field-free with vigorous convective motions. The field-free regions merge and split in a continuously evolving fashion. We show a snapshot taken at time $t=50$, where the bottom and top planes are colored by magnetic pressure, the two sidepanels quantify the instantaneous temperature difference $T(t)-T(t=0)$, and the velocity field is shown as arrows in the midplane, colored by this latter quantity. It shows the close relation between up versus down flows and the local temperature variations. 

\begin{figure}
\centering
\includegraphics[width=0.48\textwidth]{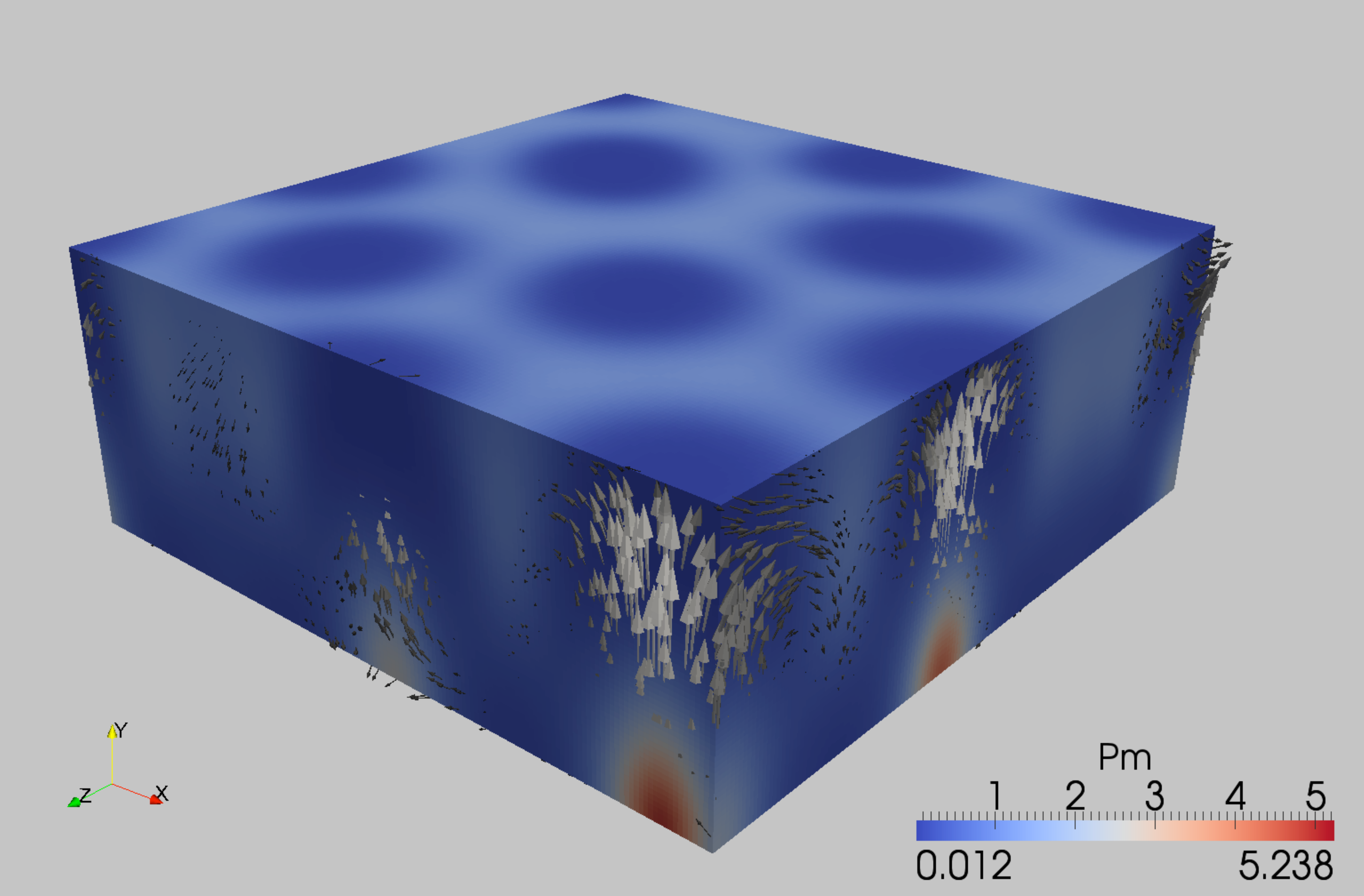}
\includegraphics[width=0.48\textwidth]{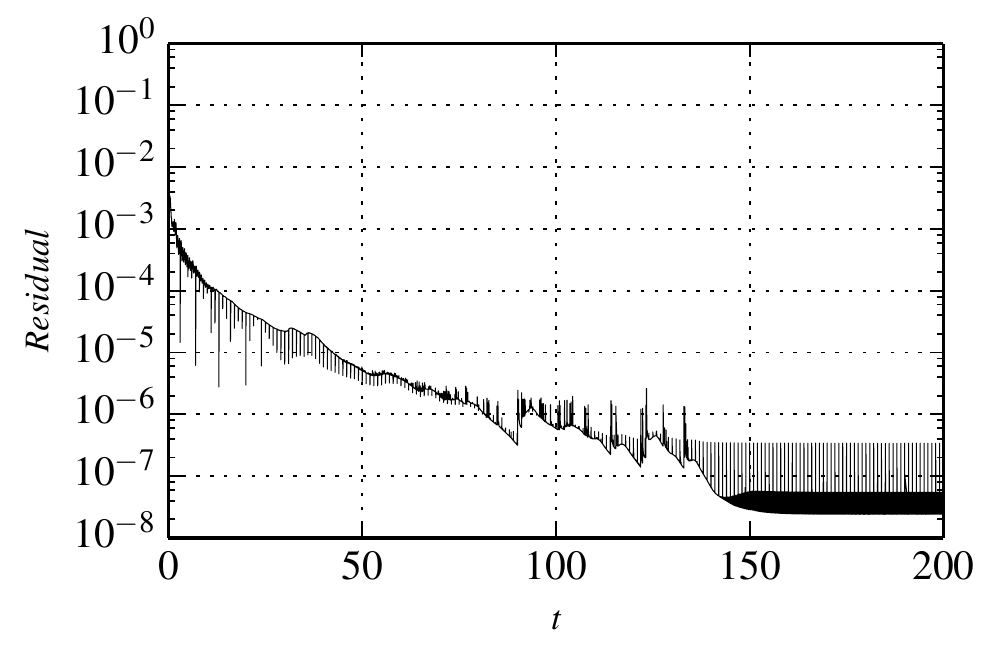}
\includegraphics[width=0.9\textwidth]{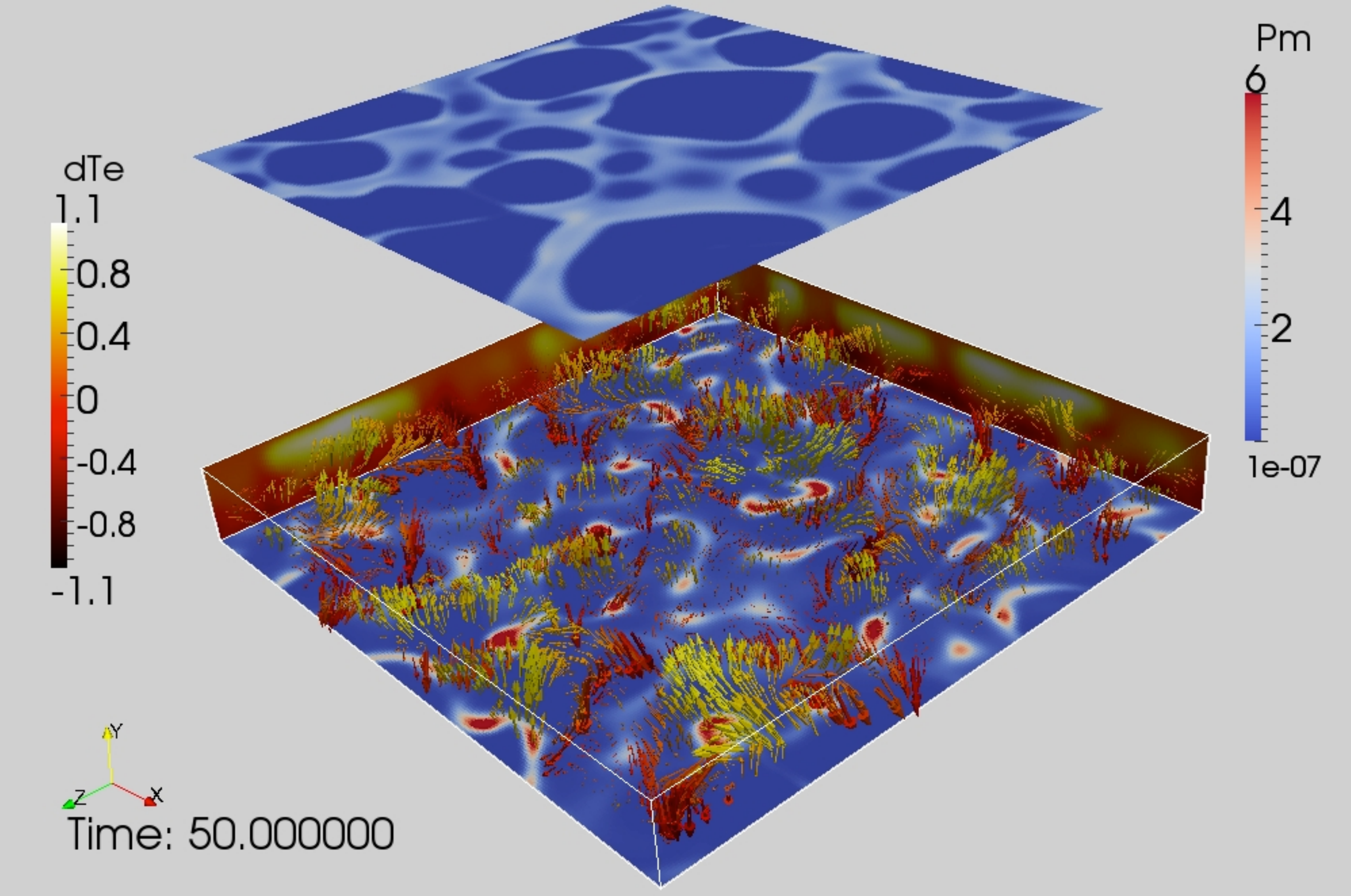}
\caption{Magnetoconvection simulations, in a parameter regime allowing for steady (top) versus unsteady (bottom) behavior. See text for a discussion.}
\label{f-convect}
\end{figure}

\section{Scaling experiments}

Here, we report on results of scaling experiments for \amrvac performed on various supercomputing platforms.

\subsection{Weak scaling}
We start with a weak scaling experiment
of the \amrvac code on the BGQ Fermi computer, as quantified in Figure~\ref{fermi}.
The setup actually realizes a 3D, compressible MHD setup inspired by the discussion in Longcope-Strauss~\citep{longcope}, where the authors argue for near-singular current sheets developing from coalescence instability. Our setup has four `magnetic islands', which have purely planar ($B_z=0$) magnetic fields from $B_x=B_0 \sin(2\pi x)\cos(2\pi y)$ and $B_y=-B_0\cos(2\pi x) \sin(2\pi y)$  initially, in a 3D unit-sized triple periodic box. The pressure varies with $(x,y)$ to realize an equilibrium, and the temperature is uniform initially.  The velocity perturbation takes a small amplitude incompressible planar $v_x\propto \sin (2\pi y)$, $v_y\propto\sin(2\pi x)$, with an extra perturbation for velocity component $v_z \propto \sin(z)$ in the $z$-dimension. We use resistive MHD on a uniform grid in this domain decomposition parallel scaling experiment. We use the HLLC scheme for the spatial discretisation and a third order \u{C}ada limiter~\citep{cada} for the time advance. To perform the weak scaling, we set up the problem with a fixed number of grid blocks per CPU (i.e. 4 blocks, each having $32^3$ cells, excluding ghost cells). When we increase the number of CPUs, the resolution of the simulation is increased as well, keeping the number of blocks per CPU fixed. For the smallest number of processors available on Fermi, which is 1024, the resolution is thus $512^3$. At the highest number of CPUs, 31250 (almost 1 rack at Fermi), the resolution is $1600^3$. For each setup, we calculate 100 iterations. For all numbers of processors this takes about 1074 seconds wall clock time proving excellent weak scaling in the domain decomposition case. The obtained efficiency is plotted in the figure, and the snapshot shows a visualization of the density and current sheet structure. The four flux tubes, with initial predominant poloidal magnetic field, are susceptible to kink instability, while the central current sheet formation happens as before. An ongoing study will further investigate its fully nonlinear evolution.

\begin{figure}[ht]
\centering
\includegraphics[width=0.8\textwidth]{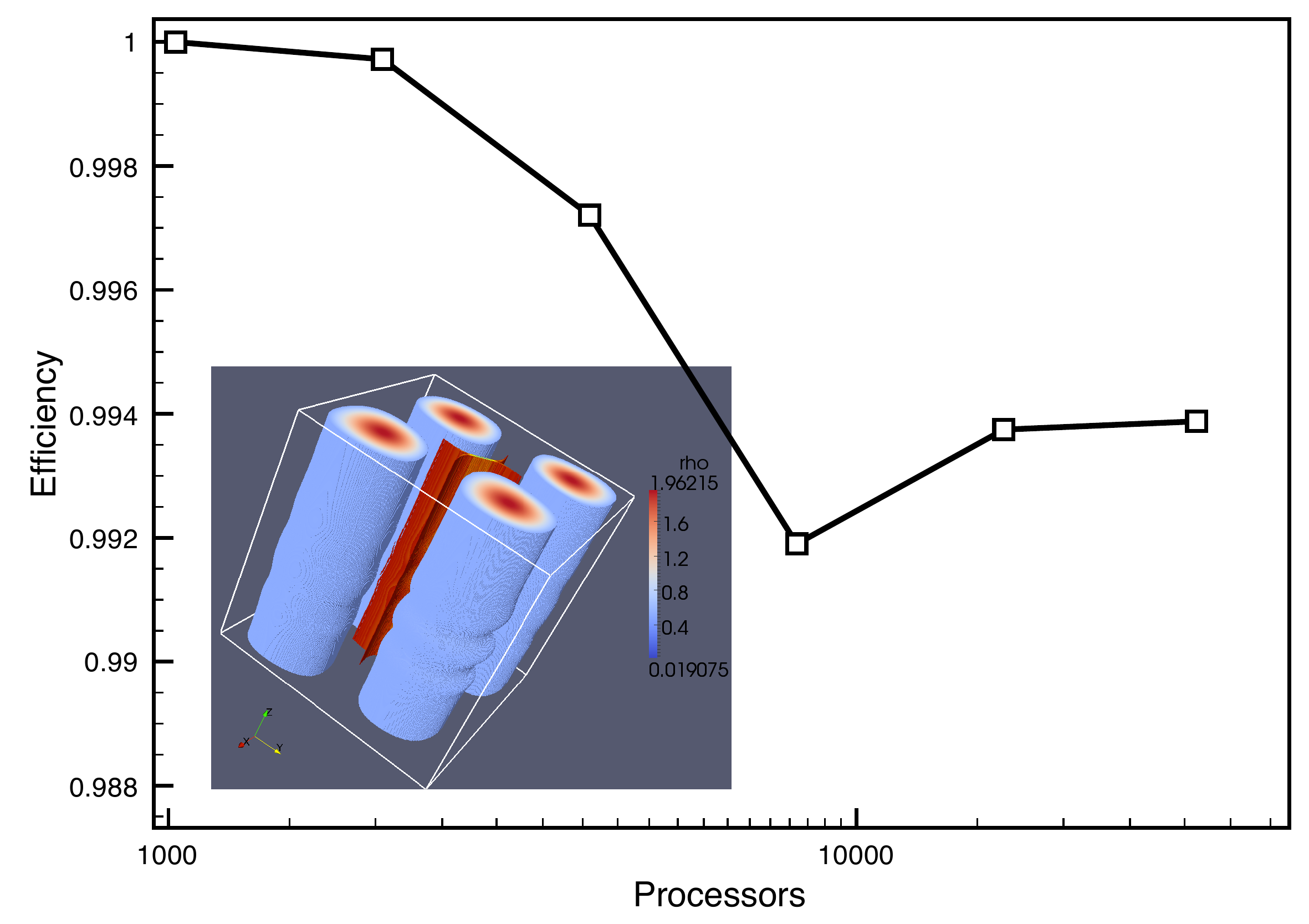}
\caption{Obtained efficiency on Fermi supercomputer -- weak scaling. From 1024 to 31250 CPUs, we maintain better than 99 \% efficient in this weak 3D MHD domain decomposition scaling experiment.}
\label{fermi}
\end{figure}

\subsection{Strong scaling with AMR}

Strong scaling of \amrvac using adaptive mesh refinement has been investigated on the Jade supercomputer\footnote{Centre Informatique National de l' Enseignement Sup\'erieur: http://www.cines.fr} with the relativistic jet-formation scenario discussed in \cite{porth13}.  
For this test, we simulate one physical time-unit starting from a snapshot roughly at midpoint of the total simulation time, giving a reasonable estimate of the average workload of the simulation.  The AMR-blocksize for this test is $12^3$ cells and two ghost cells are used on each side of the blocks.  
Efficiency quantification of a $160\times10^6$ cell, five level production setup (case 160M) as well as small domain case with $40\times 10^6$ cells and four levels (case 40M) is shown in figure \ref{fig:jade}.  We normalise the efficiency to the lowest processor number used, corresponding to 128 processors for case 160M and 64 processors for case 40M.\footnote{Here, ``processor'' is used synonymous with ``core'' and thus denotes the atomic compute unit.}  At the lowest processor number, the simulations perform $29\,437$ (160M) and $30\,964$ (40M) cell-updates per second per core.  
\begin{figure}[ht]
\centering
\includegraphics[width=0.8\textwidth]{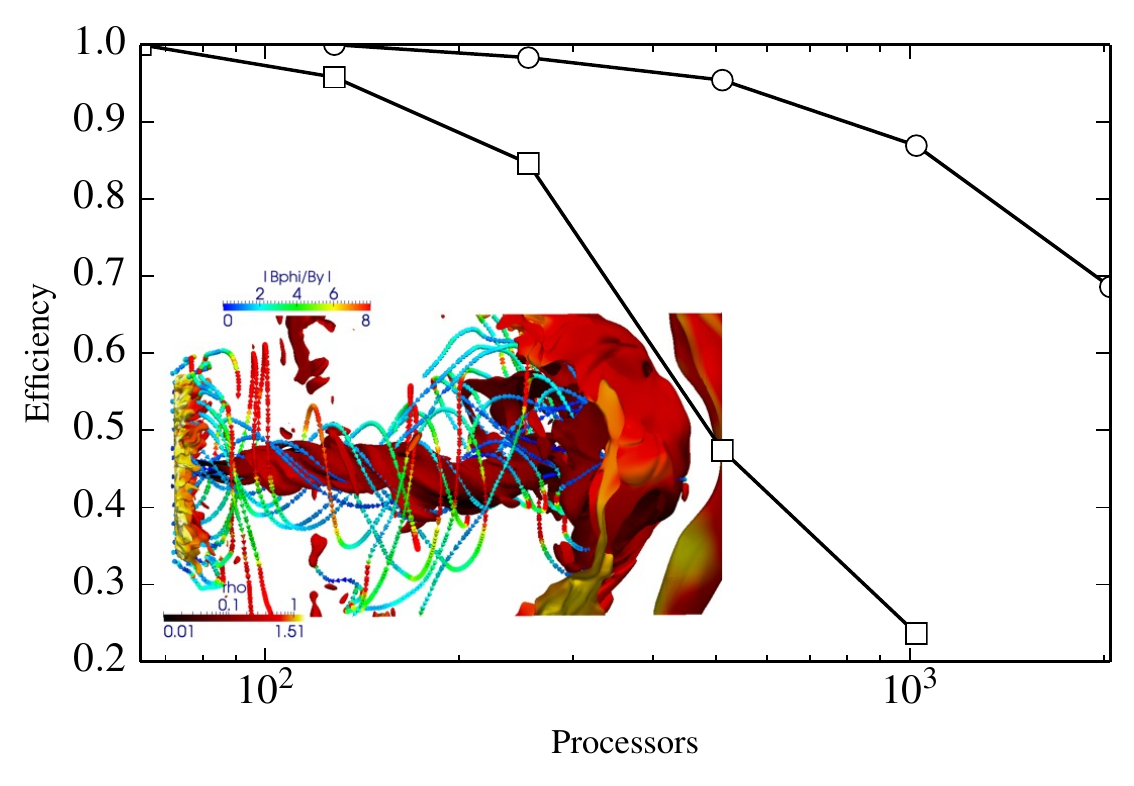}
\caption{Obtained efficiency on Jade supercomputer -- strong scaling with AMR.  We show a realisation with $160 \times 10^6$ grid cells with five grid levels (circles) as well as a case with only $40\times 10^6$ cells realised by four levels (squares).  Both cases show an efficiency of $\sim85\ \%$ as long as at least $150\,000$ cells per processor are used.  At 2048 processors we still obtain $\sim70\ \%$ efficiency in the large domain case.  The inset shows an exemplary rendering of iso-pressure contours and field lines at the end of the simulation.  }
\label{fig:jade}
\end{figure}

For case 40M, we quantified the AMR speedup by restarting from a corresponding snapshot where all cells were refined to the highest level.  
This yields a total of $65\,536$ blocks, to be compared to $21\,216$ blocks on the highest level of case 40M.  We would thus expect a speedup due to a more efficient space-filling by a factor of $3.1$.  
The observed run time comparison at 256 processors agrees roughly with this estimate and yields a speedup by a factor of $2.8$ when AMR is used, resulting from an AMR-overhead of approximately $10\%$.  \\

\subsection{Strong scaling without AMR}

A strong scaling test on a uniform grid version of the cloud shock test with one dust species (discussed in section \ref{cloudshock}) was performed on the SuperMUC cluster\footnote{Leibniz-Rechenzentrum, Garching: http://www.lrz.de}. We now adopt cartesian coordinates in three dimensions and use a uniform gridsize of $480^3$ cells (110M cells) divided in $8^3$ blocks (totaling 216k blocks) with two ghostcell layers on each block-face. The result of the scaling test for one ``island'' of SuperMuc is shown in figure \ref{fig:smuc}.  For this setup, we find the efficiency to actually \textit{increases} with processor number up to 2048 processors.  At 2048 processors we obtain the peak performance of $\sim150$k cell updates per processor per second.  We suspect that this \textit{super-linear scaling} is a result of the network configuration of the SuperMUC cluster: the tree topology allows direct communication between each individual computational node within one island.  This causes the total communication bandwidth to increase when more nodes are used.  
As a consequence, the time spend in the routine which communicates the boundary conditions between the blocks on different nodes (see figure \ref{fig:smuc}) is decreased dramatically. When more than 2048 cores are used, the efficiency drops again, as communication time is becoming predominantly latency-limited.  Note that at 8192 cores, the efficiency is still $96\%$ as compared to 1024 cores.

\begin{figure}[ht]
\centering
\includegraphics[width=0.6\textwidth]{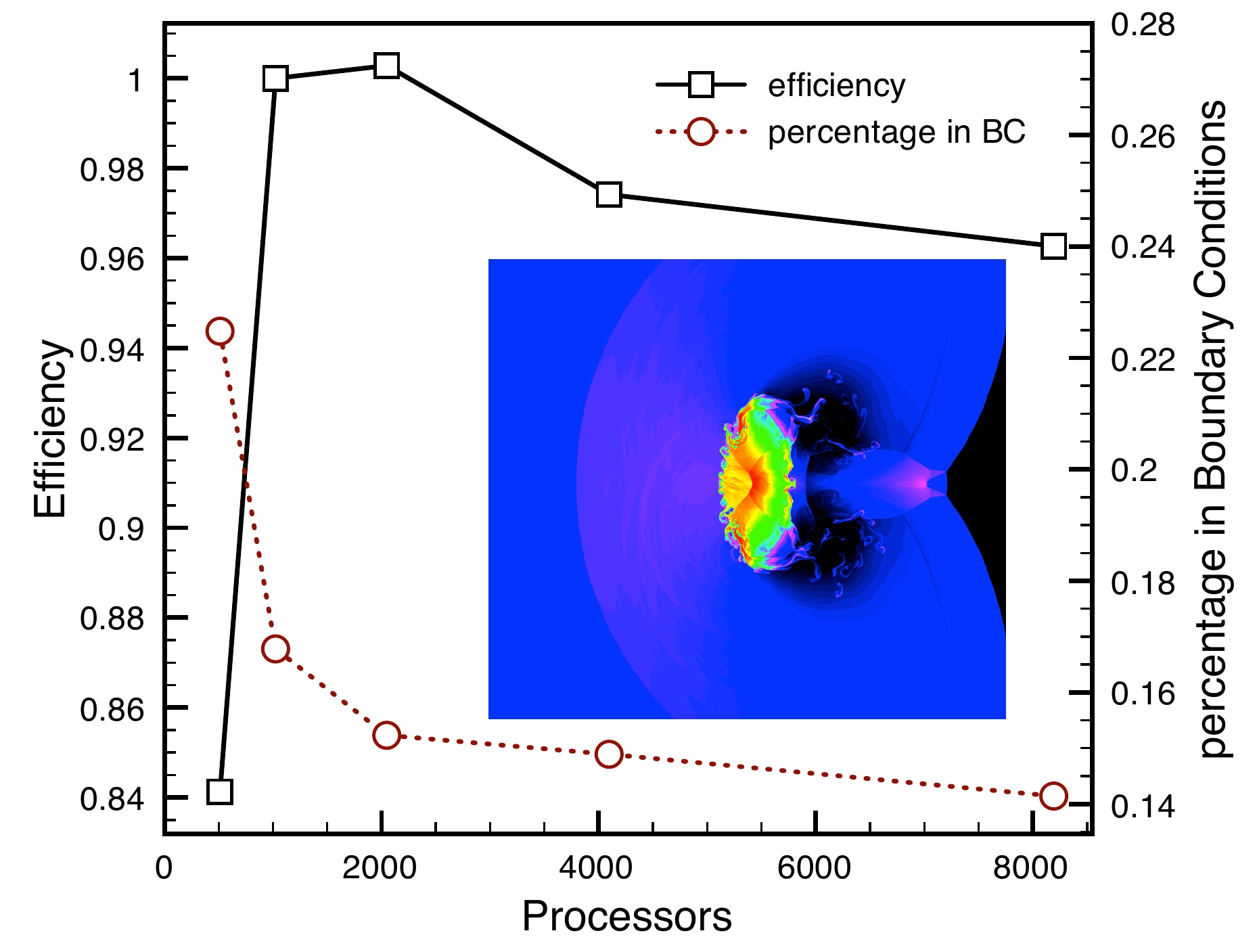}
\caption{Strong scaling without AMR on the cloud shock test with one dust species. A total of 110M cells are used. The efficiency is scaled to the runtime at 1024 processors. The efficiency can be seen to increase up to 2048 due to \textit{superscaling}, which is caused by the relative reduction of time spend in the boundary conditions routine, which communicates the boundaries between different blocks (possibly on different computational nodes). }
\label{fig:smuc}
\end{figure}

\section{Summary and outlook}

We provided an update on the \amrvac development, with a focus on the latest additions to our open source repository (see {\tt http://gitorious.org/amrvac}). The block-AMR, fully parallel software has many possibilities for gas dynamical and plasma physical applications, inspired by concrete astrophysical or solar physical observations. The presentation here emphasized the newest additions to the non-relativistic physics modules, although the high order FD schemes and time steppers are directly available to all physics modules, including the relativistic HD and MHD ones. In the appendices, we present details on how we  produce sliced or collapsed views that fully respect the AMR structure during runtime, as well as the added benefits of being able to distinguish between active and passive grid blocks. These can be of generic interest to all complementary coding efforts on open source, grid adaptive, parallel software for astrophysical applications. 

In the future, we plan to extend recent works~\citep{marle11,marle12} on circumstellar wind interactions to cases where stellar winds interact in binary systems, extending the pure gas dynamical study by~\cite{marlebin} to cases where radiative effects, dust creation and its redistribution over the shocked wind interaction zone are relevant. For solar applications, we intend to continue our work on prominence formation in realistic fluxrope configurations, based on the steps already taken in 2.5D MHD or in 3D isothermal MHD~\citep{xia12,xia14}. Global as well as local solar modeling will use the potential field extrapolation possibilities in ultimately data-driven scenarios, to complement state-of-the-art simulations such as those presented by~\cite{riley,bart14}. These can also aid ongoing efforts on global magnetospheric modeling with \amrvac, such as those for the Jovian case as done by~\cite{chane13}.

\acknowledgments
This research was supported by projects GOA/2015-014 (KU Leuven, 2014-2018), and the Interuniversity Attraction Poles Programme initiated by the Belgian Science Policy Office (IAP P7/08 CHARM). Some of the simulations used the VSC (flemish supercomputer center) funded by the Hercules foundation and the Flemish Government.  
Some of the simulations were carried out on the ARC-1 cluster of the University of Leeds.  
OP acknowledges financial support by the STFC under the standard grant ST/I001816/1.  
CX acknowledges FWO Pegasus funding, SPM is aspirant FWO and acknowledges financial support by the Greek Foundation for Education and European Culture (IPEP). 
Part of the work was performed in the context of grant agreement Swiff (proj. no. 263340) of the EC seventh framework programme (FP7/2007-2013).

\appendix

\section{Slicing a Morton order AMR grid}\label{slice}

Simulations in two and three dimensions frequently lead to large datasets that are not easily visualised.  
For a quick look into the $N$-dimensional data however, it is often sufficient to examine sub-dimensional slices.  \amrvac is capable to slice any of its grids along the coordinate directions and output the $N-1$ dimensional data during runtime.  
The algorithm takes advantage of the tree based grid structure and is described below.  

\begin{enumerate}
\item
Given the direction perpendicular to the resulting slice $d$ and the coordinate-value along $d$, $x^d_{\rm s}$, for each level calculate the grid index indicated by the slice:
\begin{equation}
g^d(l) = int((x^d_{\rm s}-x^d_{\rm min})/\Delta g^d(l)) +1
\end{equation}
where $\Delta g^d(l)$ is the extent of a grid block (in direction $d$ on level $l$) and $x^d_{\rm min}$ is given by the minimal domain boundary.  An exemplary 2D grid is shown in figure \ref{fig:slicegrid}.  
\item
For every base grid at $g^d(l),\, l=1$:
\begin{enumerate}
\item
For every child in direction other than $d$, descend (recursively) into the child:
\begin{equation}
c^d = g^d(l+1) - 2 g^d(l) + 2
\end{equation}
where the child index is defined as $c^d=1$ for the left child and $c^d=2$ for the right child.  
\begin{enumerate}
\item
Upon encounter of a leaf, take the corresponding block and fill a sub-dimensional solution block given the cells closest to $x_s^d$. 
\end{enumerate}
\end{enumerate}
\item
Output sub-dimensional solution block in order of their encounter.  
\end{enumerate}

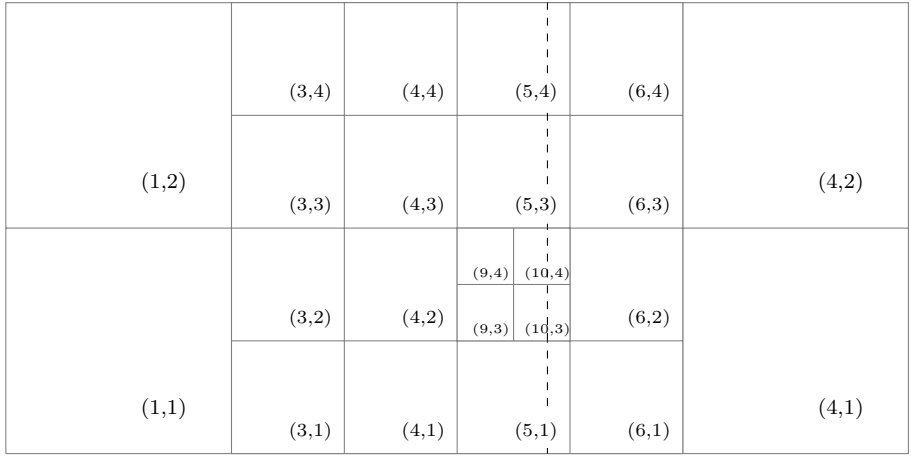
\begin{figure}[htbp]
\centering
\begin{tikzpicture}[scale=3]

\begin{scope}[scale=1]
    \draw [black!50,step=1cm] (0,0) grid (4,2);
    \draw [dashed,black] (2.4,0) -- (2.4,2);
    \foreach \x in {1,2,3,4}
    \foreach \y in {1,2}
    {
      \draw [fill=white,color=black] (\x-0.3,\y-0.8) node[fill=white]{(\x,\y)};
    }
\end{scope}
\begin{scope}[black!0,scale=0.5]
	\fill (2,0) -- (6,0) -- (6,4) -- (2,4);
    \draw [black!50,step=1cm] (2,0) grid (6,4);
    \draw [dashed,black] (4.8,0) -- (4.8,4);
    \foreach \x in {3,4,5,6}
    \foreach \y in {1,2,3,4}
    {
      \draw [font=\scriptsize,black](\x-0.3,\y-0.8) node [fill=white]{(\x,\y)};
    }
\end{scope}

\begin{scope}[black!0,scale=0.25]
	\fill (8,2) -- (10,2) -- (10,4) -- (8,4);
    \draw [black!50,step=1cm] (8,2) grid (10,4);
    \draw [dashed,black] (9.6,2) -- (9.6,4);
    \foreach \x in {9,10}
    \foreach \y in {3,4}
    {
      \draw [black,font=\tiny](\x-0.4,\y-0.8) node[] {(\x,\y)};
    }
\end{scope}

\end{tikzpicture}
\caption{Illustration of the slicing algorithm:  
An exemplary three level 2D grid is shown with the slicing in direction $d=1$ shown by the dashed line.  It intersects grid indices $g^1(2)=5$ and $g^1(3)=10$ for second and third level grids respectively.}
\label{fig:slicegrid}
\end{figure}

The resulting sub-dimensional grid is again in tree-form and reflects the original adaptive mesh which can be valuable for data inspection.  
Due to the recursion, the new sub-dimensional space filling curve is Morton ordered.  
This strong correspondence between the data-structures even allows to restart an AMR run in $N-1$ dimensions without further modifications, implying $\partial_d=0$.  
The presented algorithm can be applied to any dimensionality $N$ and optimally takes advantage of the grid structure since grids relevant for the slice are known a-priori (from step 1.).  The outer loop thus scales as $O(N-1)$, while the complexity added by the recursive inner loop depends on the number of dimensions and levels.  

\section{Collapsing a Morton order AMR grid}\label{collapse}

The dimensionality of the grid can also be reduced by integration along coordinate directions, thus ``collapsing'' the data onto a plane.  This yields surface densities, etc.  
The following algorithm can collapse data onto any level $l$ during runtime while taking advantage of the underlying data-structure.  

\begin{enumerate}
\item
Given the direction of integration $d$ and the target level of the resulting data $l_{\rm t}$, first sum the quantity $q$ for each block
\begin{equation}
Q(\mathbf{i^P},l)=\Delta x^d(l)\sum_{i^d}^{} q(\mathbf{i}) \,,
\end{equation}
where $\mathbf{i}^P=P^d(\mathbf{i})$ is the orthogonal projection of a (index-) vector along $d$.  The projection simply removes the $d$ direction in the arrays, leaving the order of the remaining directions unaffected.  
After this elemental collapse operation, we drop the superscript $P$ and work only in the reduced index space.  
\item
Each processor $j$ allocates the global array for collapsed data $C_{j}(P^d(\mathbf{N}(l_{\rm t})))$, where the number of cells in each direction follows from the number of grid blocks in the target level $\mathbf{N}_g(l_{\rm t})$ and the number of cells within each block $N^i(l_{\rm t})=N^i_g(l_{\rm t})\,N^i_{c}$.  
\item
Add up all collapsed blocks $Q$ in the $C$-array, using block index and level to fill the correct bins:
\begin{equation}
I_c^i(i^i,l) = int((i^i-i_{\rm B}+N_c^i(g^i(l)-1)-1) 2^{l_{\rm t}-l})+1
\end{equation}
where $i^i$ is the local index in the block, $i_{\rm B}$ the number of ghost cells and the term $N_c^i(g^i(l)-1)$ is added to translate from the local block cell-index to the corresponding global cell-index on level $l$.  
Thus for all blocks on the processor $j$, perform 
\begin{equation}
C_j(\mathbf{i}_c) = 
\sum_{\mathbf{i}_c=\mathbf{I}_c(\mathbf{i},l)} Q(\mathbf{i},l)\ \delta S(N,l,l_t)
\ ;
\hspace{1cm}
\delta S(N,l,l_t) = \left\{
\begin{array}{lll}
&2^{(N-1)(l_{\rm t}-l)}&;  l> l_{\rm t}\\
&1 &;  l\le l_{\rm t}
\end{array}
\right.
\end{equation}
where the term $\delta S(N,l,l_t)$ (with $N$ being the original dimensionality) results in an averaging of data on levels higher than the target level.  Each process now holds a version of the $C_j(\mathbf{i}_c)$ array, hence the final step 
\item
Use the MPI reduce operation to obtain
\begin{equation}
C(\mathbf{i}_c) = \sum_j C_j(\mathbf{i}_c).  
\end{equation}
\end{enumerate}
on the head node.  The final array $C(\mathbf{i}_c)$ is then written out either as comma-separated value ascii data-file or in binary {\tt .vti} format.

\section{Dynamic grid activation}

In many applications, parts of the simulation domain can well be described by stationarity or as analytic (e.g. self-similar) solutions, while other parts ask for direct numerical simulation.  
Examples for the ``passive'' regions are injected supersonic stellar winds, jets that have settled to a stationary state (starting near the injection boundary) or simply the static initial configuration that is still unaffected by the dynamical evolution.  
Especially for problems that involve a large separation of scales, as in the case of space weather, the computation can be significantly sped up if these passive regions are taken out of the integration loop.  

We have implemented a scheme that can dynamically (de-) activate grid blocks in an AMR setting, depending on the solution itself or on temporal/spatial properties.  Since the implementation acts only on the grid structure, it can directly be used with all available physics modules.  In the following we describe the strategy and give an example application from recent special relativistic magneto hydrodynamics simulations~\citep{porth14}.  

\emph{
\begin{enumerate}
\item
Loop over all grid blocks and flag grids to be de-activated based on a user-defined criterion.
\item
Loop over the candidate passive blocks and re-activate block if an active neighbour is detected. 
\item
Create final lists of the active and passive blocks.  
\end{enumerate}
}
The second step can be repeated to increase the number of safety blocks.  Note that this procedure seamlessly works across level changes.  
For the ensuing time-integration, only the list of active blocks is advanced.  In principle, a separate loop can then also advance the passive blocks, following for example a self-similar analytic evolution or employ a completely different physics module. 

With active and passive zones present, the parallel load balancing of the code needs some attention.  While normally a balanced load is achieved by cutting of the space filling curve (SFC) such that the number of blocks per processor is balanced for all processors, it is clear that the introduction of passive blocks violates the identity of block and computational load.
Hence we introduce different weights for active and passive blocks.  This allows to better balance the true computational load and at the same time sets a limit for the permitted \emph{memory imbalance}.  In practise the adopted load per block $\mathcal{L}(i_{\rm B})$ is
\begin{align}
\mathcal{L}(i_{\rm B}) = \left\{
\begin{array}{ll}
w_a  & ;\ {\rm Block\ i_B\ is\ active}     \\
w_p  & ;\ {\rm Block\ i_B\ is\ passive}   
\end{array}
\right.
\end{align}
which is balanced by suitably cutting the SFC.  
The imbalance of load $X_{\rm load}$ and memory $X_{\rm mem}$ considering all processors $N_{\rm pe}$ is 
\begin{align}
X_{\rm load} = \frac{\max_{i_{pe}=1\dots N_{pe}}[N_{\rm active}(i_{\rm pe})]}{\min_{i_{pe}=1\dots N_{pe}}[N_{\rm active}(i_{\rm pe}))]}\ ;\hspace{1cm}
X_{\rm mem} = \frac{\max_{i_{pe}=1\dots N_{pe}}[N_{\rm active}(i_{\rm pe})+N_{\rm passive}(i_{\rm pe})]}{\min_{i_{pe}=1\dots N_{pe}}[N_{\rm active}(i_{\rm pe})+N_{\rm passive}(i_{\rm pe})]}\,.
\end{align}
Choosing $w_p=0$ balances the active blocks exactly but could lead to significant memory imbalance.  Given the weights, the maximum permitted memory imbalance defined above is $\hat{X}_{\rm mem}=w_a/w_p$.  We typically adopt $\hat{X}_{\rm mem}=2\dots3$ which gives the best results for the problems and hardware considered so far.  

In figure \ref{activate} we illustrate dynamic grid activation at the example of a relativistic pulsar wind simulation from \cite{PorthKomissarov2013}.  
Cells marked yellow in the left panel of the figure are not advanced in the time loop but hold the stationary solution of the unshocked wind.  Note that in this particular application, the origin was refined up to level 20 to properly resolve the inner regions (the shock is situated on level 9).  For the case that the shock is squeezed back to the pulsar, these grids can be automatically activated.  
In this case, the main speedup is not due to the reduced number of active grids to be advanced, but due to the larger resulting global CFL limited time step.  
\begin{figure}[ht]
\centering
\includegraphics[width=68mm]{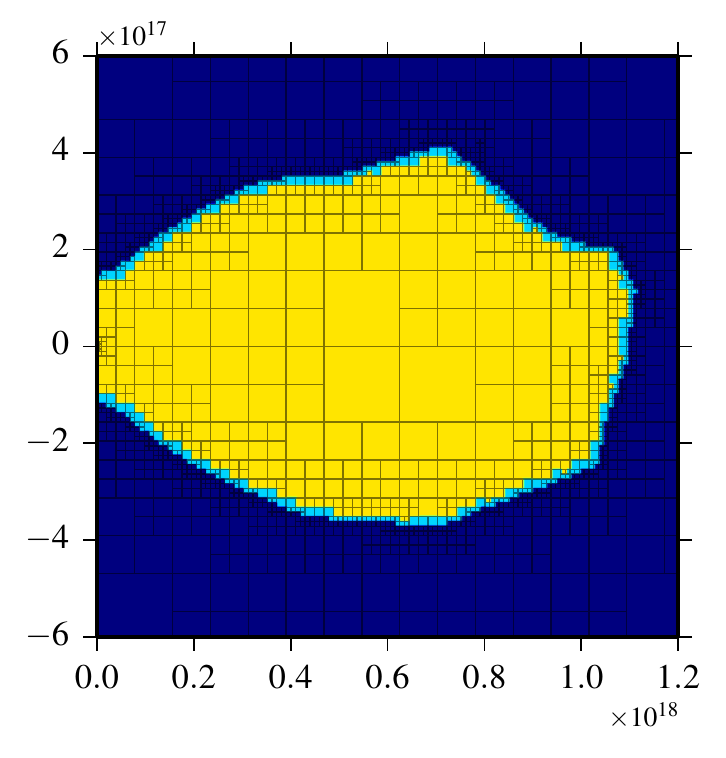}
\includegraphics[width=80mm]{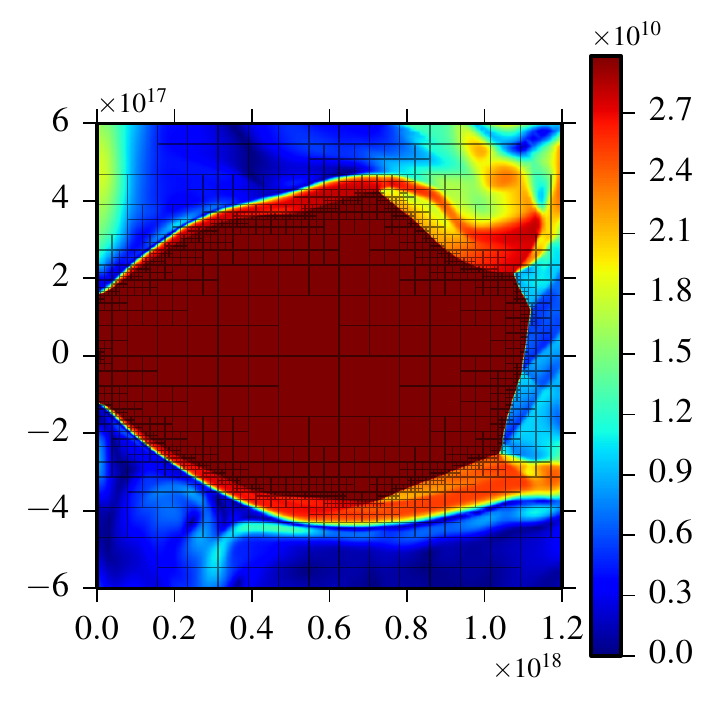}
\caption{Snapshots illustrating dynamic grid activation in the simulation of a relativistic pulsar wind nebula as in \cite{PorthKomissarov2013}.  Shown is a zoom on the termination shock of the analytic wind profile.  Left:  state of the grids, active grids are colored dark blue, while the passive grids of the unshocked wind are colored yellow.  The safety belt is indicated light blue.  Right:  corresponding velocity magnitude of the flow in units of $\rm cm\, s^{-1}$. }
\label{activate}
\end{figure}

\bibliography{astro,bibintro}

\begin{thebibliography}{117}%
\makeatletter
\providecommand \@ifxundefined [1]{%
 \@ifx{#1\undefined}
}%
\providecommand \@ifnum [1]{%
 \ifnum #1\expandafter \@firstoftwo
 \else \expandafter \@secondoftwo
 \fi
}%
\providecommand \@ifx [1]{%
 \ifx #1\expandafter \@firstoftwo
 \else \expandafter \@secondoftwo
 \fi
}%
\providecommand \natexlab [1]{#1}%
\providecommand \enquote  [1]{``#1''}%
\providecommand \bibnamefont  [1]{#1}%
\providecommand \bibfnamefont [1]{#1}%
\providecommand \citenamefont [1]{#1}%
\providecommand \href@noop [0]{\@secondoftwo}%
\providecommand \href [0]{\begingroup \@sanitize@url \@href}%
\providecommand \@href[1]{\@@startlink{#1}\@@href}%
\providecommand \@@href[1]{\endgroup#1\@@endlink}%
\providecommand \@sanitize@url [0]{\catcode `\\12\catcode `\$12\catcode
  `\&12\catcode `\#12\catcode `\^12\catcode `\_12\catcode `\%12\relax}%
\providecommand \@@startlink[1]{}%
\providecommand \@@endlink[0]{}%
\providecommand \url  [0]{\begingroup\@sanitize@url \@url }%
\providecommand \@url [1]{\endgroup\@href {#1}{\urlprefix }}%
\providecommand \urlprefix  [0]{URL }%
\providecommand \Eprint [0]{\href }%
\@ifxundefined \urlstyle {%
  \providecommand \doi  [0]{\begingroup \@sanitize@url \@doi}%
  \providecommand \@doi [1]{\endgroup \@@startlink {\doibase
  #1}doi:\discretionary {}{}{}#1\@@endlink }%
}{%
  \providecommand \doi  [0]{doi:\discretionary{}{}{}\begingroup
  \urlstyle{rm}\Url }%
}%
\providecommand \doibase [0]{http://dx.doi.org/}%
\providecommand \Doi [0]{\begingroup \@sanitize@url \@Doi }%
\providecommand \@Doi  [1]{\endgroup\@@startlink{\doibase#1}\@@Doi}%
\providecommand \@@Doi [1]{#1\@@endlink}%
\providecommand \selectlanguage [0]{\@gobble}%
\providecommand \bibinfo  [0]{\@secondoftwo}%
\providecommand \bibfield  [0]{\@secondoftwo}%
\providecommand \translation [1]{[#1]}%
\providecommand \BibitemOpen [0]{}%
\providecommand \bibitemStop [0]{}%
\providecommand \bibitemNoStop [0]{.\EOS\space}%
\providecommand \EOS [0]{\spacefactor3000\relax}%
\providecommand \BibitemShut  [1]{\csname bibitem#1\endcsname}%
\bibitem [{\citenamefont {{Stone}}\ and\ \citenamefont
  {{Norman}}(1992)}]{zeus}%
  \BibitemOpen
  \bibfield  {author} {\bibinfo {author} {\bibfnamefont {J.~M.}\ \bibnamefont
  {{Stone}}}\ and\ \bibinfo {author} {\bibfnamefont {M.~L.}\ \bibnamefont
  {{Norman}}},\ }\Doi {10.1086/191681} {\bibfield  {journal} {\bibinfo
  {journal} {\apjs},\ }\textbf {\bibinfo {volume} {80}},\ \bibinfo {pages}
  {791} (\bibinfo {year} {1992})}\BibitemShut {NoStop}%
\bibitem [{\citenamefont {{Galsgaard}}\ and\ \citenamefont
  {{Nordlund}}(1996)}]{stagger96}%
  \BibitemOpen
  \bibfield  {author} {\bibinfo {author} {\bibfnamefont {K.}~\bibnamefont
  {{Galsgaard}}}\ and\ \bibinfo {author} {\bibfnamefont {{\AA}.}~\bibnamefont
  {{Nordlund}}},\ }\Doi {10.1029/96JA00428} {\bibfield  {journal} {\bibinfo
  {journal} {\jgr},\ }\textbf {\bibinfo {volume} {101}},\ \bibinfo {pages}
  {13445} (\bibinfo {year} {1996})}\BibitemShut {NoStop}%
\bibitem [{\citenamefont {{T{\'o}th}}(1996)}]{vac}%
  \BibitemOpen
  \bibfield  {author} {\bibinfo {author} {\bibfnamefont {G.}~\bibnamefont
  {{T{\'o}th}}},\ }\href@noop {} {\bibfield  {journal} {\bibinfo  {journal}
  {Astrophysical Letters and Communications},\ }\textbf {\bibinfo {volume}
  {34}},\ \bibinfo {pages} {245} (\bibinfo {year} {1996})}\BibitemShut
  {NoStop}%
\bibitem [{\citenamefont {{T{\'o}th}}\ and\ \citenamefont {{Odstr{\v
  c}il}}(1996)}]{vac96}%
  \BibitemOpen
  \bibfield  {author} {\bibinfo {author} {\bibfnamefont {G.}~\bibnamefont
  {{T{\'o}th}}}\ and\ \bibinfo {author} {\bibfnamefont {D.}~\bibnamefont
  {{Odstr{\v c}il}}},\ }\Doi {10.1006/jcph.1996.0197} {\bibfield  {journal}
  {\bibinfo  {journal} {Journal of Computational Physics},\ }\textbf {\bibinfo
  {volume} {128}},\ \bibinfo {pages} {82} (\bibinfo {year} {1996})}\BibitemShut
  {NoStop}%
\bibitem [{\citenamefont {{T{\'o}th}}(2000)}]{divbgabor}%
  \BibitemOpen
  \bibfield  {author} {\bibinfo {author} {\bibfnamefont {G.}~\bibnamefont
  {{T{\'o}th}}},\ }\Doi {10.1006/jcph.2000.6519} {\bibfield  {journal}
  {\bibinfo  {journal} {Journal of Computational Physics},\ }\textbf {\bibinfo
  {volume} {161}},\ \bibinfo {pages} {605} (\bibinfo {year}
  {2000})}\BibitemShut {NoStop}%
\bibitem [{\citenamefont {{Goedbloed}}\ \emph {et~al.}(2010)\citenamefont
  {{Goedbloed}}, \citenamefont {{Keppens}},\ and\ \citenamefont
  {{Poedts}}}]{book2}%
  \BibitemOpen
  \bibfield  {author} {\bibinfo {author} {\bibfnamefont {J.~P.}\ \bibnamefont
  {{Goedbloed}}}, \bibinfo {author} {\bibfnamefont {R.}~\bibnamefont
  {{Keppens}}}, \ and\ \bibinfo {author} {\bibfnamefont {S.}~\bibnamefont
  {{Poedts}}},\ }\href@noop {} {\emph {\bibinfo {title} {Advanced
  Magnetohydrodynamics, by J.~P.~Goedbloed , Rony Keppens , Stefaan Poedts,
  Cambridge, UK: Cambridge University Press, 2010}}}\ (\bibinfo {year}
  {2010})\BibitemShut {NoStop}%
\bibitem [{\citenamefont {{Balsara}}(2001)}]{riemann}%
  \BibitemOpen
  \bibfield  {author} {\bibinfo {author} {\bibfnamefont {D.~S.}\ \bibnamefont
  {{Balsara}}},\ }\Doi {10.1006/jcph.2001.6917} {\bibfield  {journal} {\bibinfo
   {journal} {Journal of Computational Physics},\ }\textbf {\bibinfo {volume}
  {174}},\ \bibinfo {pages} {614} (\bibinfo {year} {2001})},\ \Eprint
  {http://arxiv.org/abs/astro-ph/0112150} {astro-ph/0112150} \BibitemShut
  {NoStop}%
\bibitem [{\citenamefont {{Powell}}\ \emph {et~al.}(1999)\citenamefont
  {{Powell}}, \citenamefont {{Roe}}, \citenamefont {{Linde}}, \citenamefont
  {{Gombosi}},\ and\ \citenamefont {{De Zeeuw}}}]{batsrus}%
  \BibitemOpen
  \bibfield  {author} {\bibinfo {author} {\bibfnamefont {K.~G.}\ \bibnamefont
  {{Powell}}}, \bibinfo {author} {\bibfnamefont {P.~L.}\ \bibnamefont {{Roe}}},
  \bibinfo {author} {\bibfnamefont {T.~J.}\ \bibnamefont {{Linde}}}, \bibinfo
  {author} {\bibfnamefont {T.~I.}\ \bibnamefont {{Gombosi}}}, \ and\ \bibinfo
  {author} {\bibfnamefont {D.~L.}\ \bibnamefont {{De Zeeuw}}},\ }\Doi
  {10.1006/jcph.1999.6299} {\bibfield  {journal} {\bibinfo  {journal} {Journal
  of Computational Physics},\ }\textbf {\bibinfo {volume} {154}},\ \bibinfo
  {pages} {284} (\bibinfo {year} {1999})}\BibitemShut {NoStop}%
\bibitem [{\citenamefont {{T{\'o}th}}\ \emph {et~al.}(2012)\citenamefont
  {{T{\'o}th}}, \citenamefont {{van der Holst}}, \citenamefont {{Sokolov}},
  \citenamefont {{De Zeeuw}}, \citenamefont {{Gombosi}}, \citenamefont
  {{Fang}}, \citenamefont {{Manchester}}, \citenamefont {{Meng}}, \citenamefont
  {{Najib}}, \citenamefont {{Powell}}, \citenamefont {{Stout}}, \citenamefont
  {{Glocer}}, \citenamefont {{Ma}},\ and\ \citenamefont {{Opher}}}]{batsrus12}%
  \BibitemOpen
  \bibfield  {author} {\bibinfo {author} {\bibfnamefont {G.}~\bibnamefont
  {{T{\'o}th}}}, \bibinfo {author} {\bibfnamefont {B.}~\bibnamefont {{van der
  Holst}}}, \bibinfo {author} {\bibfnamefont {I.~V.}\ \bibnamefont
  {{Sokolov}}}, \bibinfo {author} {\bibfnamefont {D.~L.}\ \bibnamefont {{De
  Zeeuw}}}, \bibinfo {author} {\bibfnamefont {T.~I.}\ \bibnamefont
  {{Gombosi}}}, \bibinfo {author} {\bibfnamefont {F.}~\bibnamefont {{Fang}}},
  \bibinfo {author} {\bibfnamefont {W.~B.}\ \bibnamefont {{Manchester}}},
  \bibinfo {author} {\bibfnamefont {X.}~\bibnamefont {{Meng}}}, \bibinfo
  {author} {\bibfnamefont {D.}~\bibnamefont {{Najib}}}, \bibinfo {author}
  {\bibfnamefont {K.~G.}\ \bibnamefont {{Powell}}}, \bibinfo {author}
  {\bibfnamefont {Q.~F.}\ \bibnamefont {{Stout}}}, \bibinfo {author}
  {\bibfnamefont {A.}~\bibnamefont {{Glocer}}}, \bibinfo {author}
  {\bibfnamefont {Y.-J.}\ \bibnamefont {{Ma}}}, \ and\ \bibinfo {author}
  {\bibfnamefont {M.}~\bibnamefont {{Opher}}},\ }\Doi
  {10.1016/j.jcp.2011.02.006} {\bibfield  {journal} {\bibinfo  {journal}
  {Journal of Computational Physics},\ }\textbf {\bibinfo {volume} {231}},\
  \bibinfo {pages} {870} (\bibinfo {year} {2012})}\BibitemShut {NoStop}%
\bibitem [{\citenamefont {{T{\'o}th}}\ \emph {et~al.}(2005)\citenamefont
  {{T{\'o}th}}, \citenamefont {{Sokolov}}, \citenamefont {{Gombosi}},
  \citenamefont {{Chesney}}, \citenamefont {{Clauer}}, \citenamefont {{de
  Zeeuw}}, \citenamefont {{Hansen}}, \citenamefont {{Kane}}, \citenamefont
  {{Manchester}}, \citenamefont {{Oehmke}}, \citenamefont {{Powell}},
  \citenamefont {{Ridley}}, \citenamefont {{Roussev}}, \citenamefont {{Stout}},
  \citenamefont {{Volberg}}, \citenamefont {{Wolf}}, \citenamefont {{Sazykin}},
  \citenamefont {{Chan}}, \citenamefont {{Yu}},\ and\ \citenamefont
  {{K{\'o}ta}}}]{swmf}%
  \BibitemOpen
  \bibfield  {author} {\bibinfo {author} {\bibfnamefont {G.}~\bibnamefont
  {{T{\'o}th}}}, \bibinfo {author} {\bibfnamefont {I.~V.}\ \bibnamefont
  {{Sokolov}}}, \bibinfo {author} {\bibfnamefont {T.~I.}\ \bibnamefont
  {{Gombosi}}}, \bibinfo {author} {\bibfnamefont {D.~R.}\ \bibnamefont
  {{Chesney}}}, \bibinfo {author} {\bibfnamefont {C.~R.}\ \bibnamefont
  {{Clauer}}}, \bibinfo {author} {\bibfnamefont {D.~L.}\ \bibnamefont {{de
  Zeeuw}}}, \bibinfo {author} {\bibfnamefont {K.~C.}\ \bibnamefont {{Hansen}}},
  \bibinfo {author} {\bibfnamefont {K.~J.}\ \bibnamefont {{Kane}}}, \bibinfo
  {author} {\bibfnamefont {W.~B.}\ \bibnamefont {{Manchester}}}, \bibinfo
  {author} {\bibfnamefont {R.~C.}\ \bibnamefont {{Oehmke}}}, \bibinfo {author}
  {\bibfnamefont {K.~G.}\ \bibnamefont {{Powell}}}, \bibinfo {author}
  {\bibfnamefont {A.~J.}\ \bibnamefont {{Ridley}}}, \bibinfo {author}
  {\bibfnamefont {I.~I.}\ \bibnamefont {{Roussev}}}, \bibinfo {author}
  {\bibfnamefont {Q.~F.}\ \bibnamefont {{Stout}}}, \bibinfo {author}
  {\bibfnamefont {O.}~\bibnamefont {{Volberg}}}, \bibinfo {author}
  {\bibfnamefont {R.~A.}\ \bibnamefont {{Wolf}}}, \bibinfo {author}
  {\bibfnamefont {S.}~\bibnamefont {{Sazykin}}}, \bibinfo {author}
  {\bibfnamefont {A.}~\bibnamefont {{Chan}}}, \bibinfo {author} {\bibfnamefont
  {B.}~\bibnamefont {{Yu}}}, \ and\ \bibinfo {author} {\bibfnamefont
  {J.}~\bibnamefont {{K{\'o}ta}}},\ }\Doi {10.1029/2005JA011126} {\bibfield
  {journal} {\bibinfo  {journal} {Journal of Geophysical Research (Space
  Physics)},\ }\textbf {\bibinfo {volume} {110}},\ \bibinfo {eid} {A12226}
  (\bibinfo {year} {2005})}\BibitemShut {NoStop}%
\bibitem [{\citenamefont {{Ziegler}}(2008)}]{nirvana}%
  \BibitemOpen
  \bibfield  {author} {\bibinfo {author} {\bibfnamefont {U.}~\bibnamefont
  {{Ziegler}}},\ }\Doi {10.1016/j.cpc.2008.02.017} {\bibfield  {journal}
  {\bibinfo  {journal} {Computer Physics Communications},\ }\textbf {\bibinfo
  {volume} {179}},\ \bibinfo {pages} {227} (\bibinfo {year}
  {2008})}\BibitemShut {NoStop}%
\bibitem [{\citenamefont {{Fromang}}\ \emph {et~al.}(2006)\citenamefont
  {{Fromang}}, \citenamefont {{Hennebelle}},\ and\ \citenamefont
  {{Teyssier}}}]{ramses}%
  \BibitemOpen
  \bibfield  {author} {\bibinfo {author} {\bibfnamefont {S.}~\bibnamefont
  {{Fromang}}}, \bibinfo {author} {\bibfnamefont {P.}~\bibnamefont
  {{Hennebelle}}}, \ and\ \bibinfo {author} {\bibfnamefont {R.}~\bibnamefont
  {{Teyssier}}},\ }\Doi {10.1051/0004-6361:20065371} {\bibfield  {journal}
  {\bibinfo  {journal} {\aap},\ }\textbf {\bibinfo {volume} {457}},\ \bibinfo
  {pages} {371} (\bibinfo {year} {2006})},\ \Eprint
  {http://arxiv.org/abs/astro-ph/0607230} {astro-ph/0607230} \BibitemShut
  {NoStop}%
\bibitem [{\citenamefont {{Cunningham}}\ \emph {et~al.}(2009)\citenamefont
  {{Cunningham}}, \citenamefont {{Frank}}, \citenamefont {{Varni{\`e}re}},
  \citenamefont {{Mitran}},\ and\ \citenamefont {{Jones}}}]{astrobear}%
  \BibitemOpen
  \bibfield  {author} {\bibinfo {author} {\bibfnamefont {A.~J.}\ \bibnamefont
  {{Cunningham}}}, \bibinfo {author} {\bibfnamefont {A.}~\bibnamefont
  {{Frank}}}, \bibinfo {author} {\bibfnamefont {P.}~\bibnamefont
  {{Varni{\`e}re}}}, \bibinfo {author} {\bibfnamefont {S.}~\bibnamefont
  {{Mitran}}}, \ and\ \bibinfo {author} {\bibfnamefont {T.~W.}\ \bibnamefont
  {{Jones}}},\ }\Doi {10.1088/0067-0049/182/2/519} {\bibfield  {journal}
  {\bibinfo  {journal} {\apjs},\ }\textbf {\bibinfo {volume} {182}},\ \bibinfo
  {pages} {519} (\bibinfo {year} {2009})},\ \Eprint
  {http://arxiv.org/abs/0710.0424} {arXiv:0710.0424} \BibitemShut {NoStop}%
\bibitem [{\citenamefont {{Mignone}}\ \emph {et~al.}(2007)\citenamefont
  {{Mignone}}, \citenamefont {{Bodo}}, \citenamefont {{Massaglia}},
  \citenamefont {{Matsakos}}, \citenamefont {{Tesileanu}}, \citenamefont
  {{Zanni}},\ and\ \citenamefont {{Ferrari}}}]{pluto}%
  \BibitemOpen
  \bibfield  {author} {\bibinfo {author} {\bibfnamefont {A.}~\bibnamefont
  {{Mignone}}}, \bibinfo {author} {\bibfnamefont {G.}~\bibnamefont {{Bodo}}},
  \bibinfo {author} {\bibfnamefont {S.}~\bibnamefont {{Massaglia}}}, \bibinfo
  {author} {\bibfnamefont {T.}~\bibnamefont {{Matsakos}}}, \bibinfo {author}
  {\bibfnamefont {O.}~\bibnamefont {{Tesileanu}}}, \bibinfo {author}
  {\bibfnamefont {C.}~\bibnamefont {{Zanni}}}, \ and\ \bibinfo {author}
  {\bibfnamefont {A.}~\bibnamefont {{Ferrari}}},\ }\Doi {10.1086/513316}
  {\bibfield  {journal} {\bibinfo  {journal} {\apjs},\ }\textbf {\bibinfo
  {volume} {170}},\ \bibinfo {pages} {228} (\bibinfo {year} {2007})},\ \Eprint
  {http://arxiv.org/abs/astro-ph/0701854} {astro-ph/0701854} \BibitemShut
  {NoStop}%
\bibitem [{\citenamefont {{Mignone}}\ \emph {et~al.}(2012)\citenamefont
  {{Mignone}}, \citenamefont {{Zanni}}, \citenamefont {{Tzeferacos}},
  \citenamefont {{van Straalen}}, \citenamefont {{Colella}},\ and\
  \citenamefont {{Bodo}}}]{plutoamr}%
  \BibitemOpen
  \bibfield  {author} {\bibinfo {author} {\bibfnamefont {A.}~\bibnamefont
  {{Mignone}}}, \bibinfo {author} {\bibfnamefont {C.}~\bibnamefont {{Zanni}}},
  \bibinfo {author} {\bibfnamefont {P.}~\bibnamefont {{Tzeferacos}}}, \bibinfo
  {author} {\bibfnamefont {B.}~\bibnamefont {{van Straalen}}}, \bibinfo
  {author} {\bibfnamefont {P.}~\bibnamefont {{Colella}}}, \ and\ \bibinfo
  {author} {\bibfnamefont {G.}~\bibnamefont {{Bodo}}},\ }\Doi
  {10.1088/0067-0049/198/1/7} {\bibfield  {journal} {\bibinfo  {journal}
  {\apjs},\ }\textbf {\bibinfo {volume} {198}},\ \bibinfo {eid} {7} (\bibinfo
  {year} {2012})},\ \Eprint {http://arxiv.org/abs/1110.0740} {arXiv:1110.0740
  [astro-ph.HE]} \BibitemShut {NoStop}%
\bibitem [{\citenamefont {{Lukin}}\ and\ \citenamefont
  {{Linton}}(2011)}]{hifi11}%
  \BibitemOpen
  \bibfield  {author} {\bibinfo {author} {\bibfnamefont {V.~S.}\ \bibnamefont
  {{Lukin}}}\ and\ \bibinfo {author} {\bibfnamefont {M.~G.}\ \bibnamefont
  {{Linton}}},\ }\Doi {10.5194/npg-18-871-2011} {\bibfield  {journal} {\bibinfo
   {journal} {Nonlinear Processes in Geophysics},\ }\textbf {\bibinfo {volume}
  {18}},\ \bibinfo {pages} {871} (\bibinfo {year} {2011})}\BibitemShut
  {NoStop}%
\bibitem [{\citenamefont {{Bryan}}\ \emph {et~al.}(2014)\citenamefont
  {{Bryan}}, \citenamefont {{Norman}}, \citenamefont {{O'Shea}}, \citenamefont
  {{Abel}}, \citenamefont {{Wise}}, \citenamefont {{Turk}}, \citenamefont
  {{Reynolds}}, \citenamefont {{Collins}}, \citenamefont {{Wang}},
  \citenamefont {{Skillman}}, \citenamefont {{Smith}}, \citenamefont
  {{Harkness}}, \citenamefont {{Bordner}}, \citenamefont {{Kim}}, \citenamefont
  {{Kuhlen}}, \citenamefont {{Xu}}, \citenamefont {{Goldbaum}}, \citenamefont
  {{Hummels}}, \citenamefont {{Kritsuk}}, \citenamefont {{Tasker}},
  \citenamefont {{Skory}}, \citenamefont {{Simpson}}, \citenamefont {{Hahn}},
  \citenamefont {{Oishi}}, \citenamefont {{So}}, \citenamefont {{Zhao}},
  \citenamefont {{Cen}}, \citenamefont {{Li}},\ and\ \citenamefont {{The Enzo
  Collaboration}}}]{enzo14}%
  \BibitemOpen
  \bibfield  {author} {\bibinfo {author} {\bibfnamefont {G.~L.}\ \bibnamefont
  {{Bryan}}}, \bibinfo {author} {\bibfnamefont {M.~L.}\ \bibnamefont
  {{Norman}}}, \bibinfo {author} {\bibfnamefont {B.~W.}\ \bibnamefont
  {{O'Shea}}}, \bibinfo {author} {\bibfnamefont {T.}~\bibnamefont {{Abel}}},
  \bibinfo {author} {\bibfnamefont {J.~H.}\ \bibnamefont {{Wise}}}, \bibinfo
  {author} {\bibfnamefont {M.~J.}\ \bibnamefont {{Turk}}}, \bibinfo {author}
  {\bibfnamefont {D.~R.}\ \bibnamefont {{Reynolds}}}, \bibinfo {author}
  {\bibfnamefont {D.~C.}\ \bibnamefont {{Collins}}}, \bibinfo {author}
  {\bibfnamefont {P.}~\bibnamefont {{Wang}}}, \bibinfo {author} {\bibfnamefont
  {S.~W.}\ \bibnamefont {{Skillman}}}, \bibinfo {author} {\bibfnamefont
  {B.}~\bibnamefont {{Smith}}}, \bibinfo {author} {\bibfnamefont {R.~P.}\
  \bibnamefont {{Harkness}}}, \bibinfo {author} {\bibfnamefont
  {J.}~\bibnamefont {{Bordner}}}, \bibinfo {author} {\bibfnamefont {J.-h.}\
  \bibnamefont {{Kim}}}, \bibinfo {author} {\bibfnamefont {M.}~\bibnamefont
  {{Kuhlen}}}, \bibinfo {author} {\bibfnamefont {H.}~\bibnamefont {{Xu}}},
  \bibinfo {author} {\bibfnamefont {N.}~\bibnamefont {{Goldbaum}}}, \bibinfo
  {author} {\bibfnamefont {C.}~\bibnamefont {{Hummels}}}, \bibinfo {author}
  {\bibfnamefont {A.~G.}\ \bibnamefont {{Kritsuk}}}, \bibinfo {author}
  {\bibfnamefont {E.}~\bibnamefont {{Tasker}}}, \bibinfo {author}
  {\bibfnamefont {S.}~\bibnamefont {{Skory}}}, \bibinfo {author} {\bibfnamefont
  {C.~M.}\ \bibnamefont {{Simpson}}}, \bibinfo {author} {\bibfnamefont
  {O.}~\bibnamefont {{Hahn}}}, \bibinfo {author} {\bibfnamefont {J.~S.}\
  \bibnamefont {{Oishi}}}, \bibinfo {author} {\bibfnamefont {G.~C.}\
  \bibnamefont {{So}}}, \bibinfo {author} {\bibfnamefont {F.}~\bibnamefont
  {{Zhao}}}, \bibinfo {author} {\bibfnamefont {R.}~\bibnamefont {{Cen}}},
  \bibinfo {author} {\bibfnamefont {Y.}~\bibnamefont {{Li}}}, \ and\ \bibinfo
  {author} {\bibnamefont {{The Enzo Collaboration}}},\ }\Doi
  {10.1088/0067-0049/211/2/19} {\bibfield  {journal} {\bibinfo  {journal}
  {\apjs},\ }\textbf {\bibinfo {volume} {211}},\ \bibinfo {eid} {19} (\bibinfo
  {year} {2014})},\ \Eprint {http://arxiv.org/abs/1307.2265} {arXiv:1307.2265
  [astro-ph.IM]} \BibitemShut {NoStop}%
\bibitem [{\citenamefont {{Del Zanna}}\ \emph {et~al.}(2007)\citenamefont {{Del
  Zanna}}, \citenamefont {{Zanotti}}, \citenamefont {{Bucciantini}},\ and\
  \citenamefont {{Londrillo}}}]{echo}%
  \BibitemOpen
  \bibfield  {author} {\bibinfo {author} {\bibfnamefont {L.}~\bibnamefont {{Del
  Zanna}}}, \bibinfo {author} {\bibfnamefont {O.}~\bibnamefont {{Zanotti}}},
  \bibinfo {author} {\bibfnamefont {N.}~\bibnamefont {{Bucciantini}}}, \ and\
  \bibinfo {author} {\bibfnamefont {P.}~\bibnamefont {{Londrillo}}},\ }\Doi
  {10.1051/0004-6361:20077093} {\bibfield  {journal} {\bibinfo  {journal}
  {\aap},\ }\textbf {\bibinfo {volume} {473}},\ \bibinfo {pages} {11} (\bibinfo
  {year} {2007})},\ \Eprint {http://arxiv.org/abs/0704.3206} {arXiv:0704.3206}
  \BibitemShut {NoStop}%
\bibitem [{\citenamefont {{Fryxell}}\ \emph {et~al.}(2000)\citenamefont
  {{Fryxell}}, \citenamefont {{Olson}}, \citenamefont {{Ricker}}, \citenamefont
  {{Timmes}}, \citenamefont {{Zingale}}, \citenamefont {{Lamb}}, \citenamefont
  {{MacNeice}}, \citenamefont {{Rosner}}, \citenamefont {{Truran}},\ and\
  \citenamefont {{Tufo}}}]{flash}%
  \BibitemOpen
  \bibfield  {author} {\bibinfo {author} {\bibfnamefont {B.}~\bibnamefont
  {{Fryxell}}}, \bibinfo {author} {\bibfnamefont {K.}~\bibnamefont {{Olson}}},
  \bibinfo {author} {\bibfnamefont {P.}~\bibnamefont {{Ricker}}}, \bibinfo
  {author} {\bibfnamefont {F.~X.}\ \bibnamefont {{Timmes}}}, \bibinfo {author}
  {\bibfnamefont {M.}~\bibnamefont {{Zingale}}}, \bibinfo {author}
  {\bibfnamefont {D.~Q.}\ \bibnamefont {{Lamb}}}, \bibinfo {author}
  {\bibfnamefont {P.}~\bibnamefont {{MacNeice}}}, \bibinfo {author}
  {\bibfnamefont {R.}~\bibnamefont {{Rosner}}}, \bibinfo {author}
  {\bibfnamefont {J.~W.}\ \bibnamefont {{Truran}}}, \ and\ \bibinfo {author}
  {\bibfnamefont {H.}~\bibnamefont {{Tufo}}},\ }\Doi {10.1086/317361}
  {\bibfield  {journal} {\bibinfo  {journal} {\apjs},\ }\textbf {\bibinfo
  {volume} {131}},\ \bibinfo {pages} {273} (\bibinfo {year}
  {2000})}\BibitemShut {NoStop}%
\bibitem [{\citenamefont {{Dreher}}\ and\ \citenamefont
  {{Grauer}}(2006)}]{racoon}%
  \BibitemOpen
  \bibfield  {author} {\bibinfo {author} {\bibfnamefont {J.}~\bibnamefont
  {{Dreher}}}\ and\ \bibinfo {author} {\bibfnamefont {R.}~\bibnamefont
  {{Grauer}}},\ }\href@noop {} {\bibfield  {journal} {\bibinfo  {journal} {Par.
  Comput.},\ }\textbf {\bibinfo {volume} {31}},\ \bibinfo {pages} {913}
  (\bibinfo {year} {2006})},\ \Eprint {http://arxiv.org/abs/physics/0602004}
  {physics/0602004} \BibitemShut {NoStop}%
\bibitem [{\citenamefont {{Freytag}}\ \emph {et~al.}(2012)\citenamefont
  {{Freytag}}, \citenamefont {{Steffen}}, \citenamefont {{Ludwig}},
  \citenamefont {{Wedemeyer-B{\"o}hm}}, \citenamefont {{Schaffenberger}},\ and\
  \citenamefont {{Steiner}}}]{co5bold}%
  \BibitemOpen
  \bibfield  {author} {\bibinfo {author} {\bibfnamefont {B.}~\bibnamefont
  {{Freytag}}}, \bibinfo {author} {\bibfnamefont {M.}~\bibnamefont
  {{Steffen}}}, \bibinfo {author} {\bibfnamefont {H.-G.}\ \bibnamefont
  {{Ludwig}}}, \bibinfo {author} {\bibfnamefont {S.}~\bibnamefont
  {{Wedemeyer-B{\"o}hm}}}, \bibinfo {author} {\bibfnamefont {W.}~\bibnamefont
  {{Schaffenberger}}}, \ and\ \bibinfo {author} {\bibfnamefont
  {O.}~\bibnamefont {{Steiner}}},\ }\Doi {10.1016/j.jcp.2011.09.026} {\bibfield
   {journal} {\bibinfo  {journal} {Journal of Computational Physics},\ }\textbf
  {\bibinfo {volume} {231}},\ \bibinfo {pages} {919} (\bibinfo {year}
  {2012})},\ \Eprint {http://arxiv.org/abs/1110.6844} {arXiv:1110.6844
  [astro-ph.SR]} \BibitemShut {NoStop}%
\bibitem [{\citenamefont {{Stone}}\ \emph {et~al.}(2008)\citenamefont
  {{Stone}}, \citenamefont {{Gardiner}}, \citenamefont {{Teuben}},
  \citenamefont {{Hawley}},\ and\ \citenamefont {{Simon}}}]{athena}%
  \BibitemOpen
  \bibfield  {author} {\bibinfo {author} {\bibfnamefont {J.~M.}\ \bibnamefont
  {{Stone}}}, \bibinfo {author} {\bibfnamefont {T.~A.}\ \bibnamefont
  {{Gardiner}}}, \bibinfo {author} {\bibfnamefont {P.}~\bibnamefont
  {{Teuben}}}, \bibinfo {author} {\bibfnamefont {J.~F.}\ \bibnamefont
  {{Hawley}}}, \ and\ \bibinfo {author} {\bibfnamefont {J.~B.}\ \bibnamefont
  {{Simon}}},\ }\Doi {10.1086/588755} {\bibfield  {journal} {\bibinfo
  {journal} {\apjs},\ }\textbf {\bibinfo {volume} {178}},\ \bibinfo {pages}
  {137} (\bibinfo {year} {2008})},\ \Eprint {http://arxiv.org/abs/0804.0402}
  {arXiv:0804.0402} \BibitemShut {NoStop}%
\bibitem [{\citenamefont {{Arber}}\ \emph {et~al.}(2001)\citenamefont
  {{Arber}}, \citenamefont {{Longbottom}}, \citenamefont {{Gerrard}},\ and\
  \citenamefont {{Milne}}}]{lare3d}%
  \BibitemOpen
  \bibfield  {author} {\bibinfo {author} {\bibfnamefont {T.~D.}\ \bibnamefont
  {{Arber}}}, \bibinfo {author} {\bibfnamefont {A.~W.}\ \bibnamefont
  {{Longbottom}}}, \bibinfo {author} {\bibfnamefont {C.~L.}\ \bibnamefont
  {{Gerrard}}}, \ and\ \bibinfo {author} {\bibfnamefont {A.~M.}\ \bibnamefont
  {{Milne}}},\ }\Doi {10.1006/jcph.2001.6780} {\bibfield  {journal} {\bibinfo
  {journal} {Journal of Computational Physics},\ }\textbf {\bibinfo {volume}
  {171}},\ \bibinfo {pages} {151} (\bibinfo {year} {2001})}\BibitemShut
  {NoStop}%
\bibitem [{\citenamefont {{Brandenburg}}\ and\ \citenamefont
  {{Dobler}}(2002)}]{pencil}%
  \BibitemOpen
  \bibfield  {author} {\bibinfo {author} {\bibfnamefont {A.}~\bibnamefont
  {{Brandenburg}}}\ and\ \bibinfo {author} {\bibfnamefont {W.}~\bibnamefont
  {{Dobler}}},\ }\Doi {10.1016/S0010-4655(02)00334-X} {\bibfield  {journal}
  {\bibinfo  {journal} {Computer Physics Communications},\ }\textbf {\bibinfo
  {volume} {147}},\ \bibinfo {pages} {471} (\bibinfo {year} {2002})},\ \Eprint
  {http://arxiv.org/abs/astro-ph/0111569} {astro-ph/0111569} \BibitemShut
  {NoStop}%
\bibitem [{\citenamefont {{Walder}}\ and\ \citenamefont
  {{Folini}}(2000)}]{amaze}%
  \BibitemOpen
  \bibfield  {author} {\bibinfo {author} {\bibfnamefont {R.}~\bibnamefont
  {{Walder}}}\ and\ \bibinfo {author} {\bibfnamefont {D.}~\bibnamefont
  {{Folini}}},\ }in\ \href@noop {} {\emph {\bibinfo {booktitle} {Thermal and
  Ionization Aspects of Flows from Hot Stars}}},\ \bibinfo {series}
  {Astronomical Society of the Pacific Conference Series}, Vol.\ \bibinfo
  {volume} {204},\ \bibinfo {editor} {edited by\ \bibinfo {editor}
  {\bibfnamefont {H.}~\bibnamefont {{Lamers}}}\ and\ \bibinfo {editor}
  {\bibfnamefont {A.}~\bibnamefont {{Sapar}}}}\ (\bibinfo {year} {2000})\ p.\
  \bibinfo {pages} {281}\BibitemShut {NoStop}%
\bibitem [{\citenamefont {{Gordeev}}\ \emph {et~al.}(2013)\citenamefont
  {{Gordeev}}, \citenamefont {{Facsk{\'o}}}, \citenamefont {{Sergeev}},
  \citenamefont {{Honkonen}}, \citenamefont {{Palmroth}}, \citenamefont
  {{Janhunen}},\ and\ \citenamefont {{Milan}}}]{gumics}%
  \BibitemOpen
  \bibfield  {author} {\bibinfo {author} {\bibfnamefont {E.}~\bibnamefont
  {{Gordeev}}}, \bibinfo {author} {\bibfnamefont {G.}~\bibnamefont
  {{Facsk{\'o}}}}, \bibinfo {author} {\bibfnamefont {V.}~\bibnamefont
  {{Sergeev}}}, \bibinfo {author} {\bibfnamefont {I.}~\bibnamefont
  {{Honkonen}}}, \bibinfo {author} {\bibfnamefont {M.}~\bibnamefont
  {{Palmroth}}}, \bibinfo {author} {\bibfnamefont {P.}~\bibnamefont
  {{Janhunen}}}, \ and\ \bibinfo {author} {\bibfnamefont {S.}~\bibnamefont
  {{Milan}}},\ }\Doi {10.1002/jgra.50359} {\bibfield  {journal} {\bibinfo
  {journal} {Journal of Geophysical Research (Space Physics)},\ }\textbf
  {\bibinfo {volume} {118}},\ \bibinfo {pages} {3138} (\bibinfo {year}
  {2013})}\BibitemShut {NoStop}%
\bibitem [{\citenamefont {{Shelyag}}\ \emph {et~al.}(2008)\citenamefont
  {{Shelyag}}, \citenamefont {{Fedun}},\ and\ \citenamefont
  {{Erd{\'e}lyi}}}]{sac}%
  \BibitemOpen
  \bibfield  {author} {\bibinfo {author} {\bibfnamefont {S.}~\bibnamefont
  {{Shelyag}}}, \bibinfo {author} {\bibfnamefont {V.}~\bibnamefont {{Fedun}}},
  \ and\ \bibinfo {author} {\bibfnamefont {R.}~\bibnamefont {{Erd{\'e}lyi}}},\
  }\Doi {10.1051/0004-6361:200809800} {\bibfield  {journal} {\bibinfo
  {journal} {\aap},\ }\textbf {\bibinfo {volume} {486}},\ \bibinfo {pages}
  {655} (\bibinfo {year} {2008})}\BibitemShut {NoStop}%
\bibitem [{\citenamefont {{V{\"o}gler}}\ \emph {et~al.}(2005)\citenamefont
  {{V{\"o}gler}}, \citenamefont {{Shelyag}}, \citenamefont {{Sch{\"u}ssler}},
  \citenamefont {{Cattaneo}}, \citenamefont {{Emonet}},\ and\ \citenamefont
  {{Linde}}}]{muram05}%
  \BibitemOpen
  \bibfield  {author} {\bibinfo {author} {\bibfnamefont {A.}~\bibnamefont
  {{V{\"o}gler}}}, \bibinfo {author} {\bibfnamefont {S.}~\bibnamefont
  {{Shelyag}}}, \bibinfo {author} {\bibfnamefont {M.}~\bibnamefont
  {{Sch{\"u}ssler}}}, \bibinfo {author} {\bibfnamefont {F.}~\bibnamefont
  {{Cattaneo}}}, \bibinfo {author} {\bibfnamefont {T.}~\bibnamefont
  {{Emonet}}}, \ and\ \bibinfo {author} {\bibfnamefont {T.}~\bibnamefont
  {{Linde}}},\ }\Doi {10.1051/0004-6361:20041507} {\bibfield  {journal}
  {\bibinfo  {journal} {\aap},\ }\textbf {\bibinfo {volume} {429}},\ \bibinfo
  {pages} {335} (\bibinfo {year} {2005})}\BibitemShut {NoStop}%
\bibitem [{\citenamefont {{Hayek}}\ \emph {et~al.}(2010)\citenamefont
  {{Hayek}}, \citenamefont {{Asplund}}, \citenamefont {{Carlsson}},
  \citenamefont {{Trampedach}}, \citenamefont {{Collet}}, \citenamefont
  {{Gudiksen}}, \citenamefont {{Hansteen}},\ and\ \citenamefont
  {{Leenaarts}}}]{bifrost}%
  \BibitemOpen
  \bibfield  {author} {\bibinfo {author} {\bibfnamefont {W.}~\bibnamefont
  {{Hayek}}}, \bibinfo {author} {\bibfnamefont {M.}~\bibnamefont {{Asplund}}},
  \bibinfo {author} {\bibfnamefont {M.}~\bibnamefont {{Carlsson}}}, \bibinfo
  {author} {\bibfnamefont {R.}~\bibnamefont {{Trampedach}}}, \bibinfo {author}
  {\bibfnamefont {R.}~\bibnamefont {{Collet}}}, \bibinfo {author}
  {\bibfnamefont {B.~V.}\ \bibnamefont {{Gudiksen}}}, \bibinfo {author}
  {\bibfnamefont {V.~H.}\ \bibnamefont {{Hansteen}}}, \ and\ \bibinfo {author}
  {\bibfnamefont {J.}~\bibnamefont {{Leenaarts}}},\ }\Doi
  {10.1051/0004-6361/201014210} {\bibfield  {journal} {\bibinfo  {journal}
  {\aap},\ }\textbf {\bibinfo {volume} {517}},\ \bibinfo {eid} {A49} (\bibinfo
  {year} {2010})}\BibitemShut {NoStop}%
\bibitem [{\citenamefont {{Giacomazzo}}\ and\ \citenamefont
  {{Rezzolla}}(2007)}]{whiskymhd}%
  \BibitemOpen
  \bibfield  {author} {\bibinfo {author} {\bibfnamefont {B.}~\bibnamefont
  {{Giacomazzo}}}\ and\ \bibinfo {author} {\bibfnamefont {L.}~\bibnamefont
  {{Rezzolla}}},\ }\Doi {10.1088/0264-9381/24/12/S16} {\bibfield  {journal}
  {\bibinfo  {journal} {Classical and Quantum Gravity},\ }\textbf {\bibinfo
  {volume} {24}},\ \bibinfo {pages} {235} (\bibinfo {year} {2007})},\ \Eprint
  {http://arxiv.org/abs/gr-qc/0701109} {gr-qc/0701109} \BibitemShut {NoStop}%
\bibitem [{\citenamefont {{Gammie}}\ \emph {et~al.}(2003)\citenamefont
  {{Gammie}}, \citenamefont {{McKinney}},\ and\ \citenamefont
  {{T{\'o}th}}}]{harm}%
  \BibitemOpen
  \bibfield  {author} {\bibinfo {author} {\bibfnamefont {C.~F.}\ \bibnamefont
  {{Gammie}}}, \bibinfo {author} {\bibfnamefont {J.~C.}\ \bibnamefont
  {{McKinney}}}, \ and\ \bibinfo {author} {\bibfnamefont {G.}~\bibnamefont
  {{T{\'o}th}}},\ }\Doi {10.1086/374594} {\bibfield  {journal} {\bibinfo
  {journal} {\apj},\ }\textbf {\bibinfo {volume} {589}},\ \bibinfo {pages}
  {444} (\bibinfo {year} {2003})},\ \Eprint
  {http://arxiv.org/abs/astro-ph/0301509} {astro-ph/0301509} \BibitemShut
  {NoStop}%
\bibitem [{\citenamefont {{Keppens}}\ \emph {et~al.}(2003)\citenamefont
  {{Keppens}}, \citenamefont {{Nool}}, \citenamefont {{T{\'o}th}},\ and\
  \citenamefont {{Goedbloed}}}]{amrvac03}%
  \BibitemOpen
  \bibfield  {author} {\bibinfo {author} {\bibfnamefont {R.}~\bibnamefont
  {{Keppens}}}, \bibinfo {author} {\bibfnamefont {M.}~\bibnamefont {{Nool}}},
  \bibinfo {author} {\bibfnamefont {G.}~\bibnamefont {{T{\'o}th}}}, \ and\
  \bibinfo {author} {\bibfnamefont {J.~P.}\ \bibnamefont {{Goedbloed}}},\ }\Doi
  {10.1016/S0010-4655(03)00139-5} {\bibfield  {journal} {\bibinfo  {journal}
  {Computer Physics Communications},\ }\textbf {\bibinfo {volume} {153}},\
  \bibinfo {pages} {317} (\bibinfo {year} {2003})},\ \Eprint
  {http://arxiv.org/abs/astro-ph/0403124} {astro-ph/0403124} \BibitemShut
  {NoStop}%
\bibitem [{\citenamefont {{van der Holst}}\ and\ \citenamefont
  {{Keppens}}(2007)}]{amrvac07}%
  \BibitemOpen
  \bibfield  {author} {\bibinfo {author} {\bibfnamefont {B.}~\bibnamefont {{van
  der Holst}}}\ and\ \bibinfo {author} {\bibfnamefont {R.}~\bibnamefont
  {{Keppens}}},\ }\Doi {10.1016/j.jcp.2007.05.007} {\bibfield  {journal}
  {\bibinfo  {journal} {Journal of Computational Physics},\ }\textbf {\bibinfo
  {volume} {226}},\ \bibinfo {pages} {925} (\bibinfo {year}
  {2007})}\BibitemShut {NoStop}%
\bibitem [{\citenamefont {{Keppens}}\ and\ \citenamefont
  {{Porth}}(2014)}]{jcam14}%
  \BibitemOpen
  \bibfield  {author} {\bibinfo {author} {\bibfnamefont {R.}~\bibnamefont
  {{Keppens}}}\ and\ \bibinfo {author} {\bibfnamefont {O.}~\bibnamefont
  {{Porth}}},\ }\Doi {10.1016/j.cam.2014.01.017} {\bibfield  {journal}
  {\bibinfo  {journal} {Journal of Computational and Applied Mathematics},\
  }\textbf {\bibinfo {volume} {266}},\ \bibinfo {pages} {87} (\bibinfo {year}
  {2014})}\BibitemShut {NoStop}%
\bibitem [{\citenamefont {{T{\'o}th}}(1997)}]{lasy}%
  \BibitemOpen
  \bibfield  {author} {\bibinfo {author} {\bibfnamefont {G.}~\bibnamefont
  {{T{\'o}th}}},\ }\Doi {10.1006/jcph.1997.5813} {\bibfield  {journal}
  {\bibinfo  {journal} {Journal of Computational Physics},\ }\textbf {\bibinfo
  {volume} {138}},\ \bibinfo {pages} {981} (\bibinfo {year}
  {1997})}\BibitemShut {NoStop}%
\bibitem [{\citenamefont {{Meliani}}\ \emph {et~al.}(2007)\citenamefont
  {{Meliani}}, \citenamefont {{Keppens}}, \citenamefont {{Casse}},\ and\
  \citenamefont {{Giannios}}}]{amrvaczak07}%
  \BibitemOpen
  \bibfield  {author} {\bibinfo {author} {\bibfnamefont {Z.}~\bibnamefont
  {{Meliani}}}, \bibinfo {author} {\bibfnamefont {R.}~\bibnamefont
  {{Keppens}}}, \bibinfo {author} {\bibfnamefont {F.}~\bibnamefont {{Casse}}},
  \ and\ \bibinfo {author} {\bibfnamefont {D.}~\bibnamefont {{Giannios}}},\
  }\Doi {10.1111/j.1365-2966.2007.11500.x} {\bibfield  {journal} {\bibinfo
  {journal} {\mnras},\ }\textbf {\bibinfo {volume} {376}},\ \bibinfo {pages}
  {1189} (\bibinfo {year} {2007})},\ \Eprint
  {http://arxiv.org/abs/astro-ph/0701434} {astro-ph/0701434} \BibitemShut
  {NoStop}%
\bibitem [{\citenamefont {{van der Holst}}\ \emph {et~al.}(2008)\citenamefont
  {{van der Holst}}, \citenamefont {{Keppens}},\ and\ \citenamefont
  {{Meliani}}}]{amrvac08}%
  \BibitemOpen
  \bibfield  {author} {\bibinfo {author} {\bibfnamefont {B.}~\bibnamefont {{van
  der Holst}}}, \bibinfo {author} {\bibfnamefont {R.}~\bibnamefont
  {{Keppens}}}, \ and\ \bibinfo {author} {\bibfnamefont {Z.}~\bibnamefont
  {{Meliani}}},\ }\Doi {10.1016/j.cpc.2008.05.005} {\bibfield  {journal}
  {\bibinfo  {journal} {Computer Physics Communications},\ }\textbf {\bibinfo
  {volume} {179}},\ \bibinfo {pages} {617} (\bibinfo {year} {2008})},\ \Eprint
  {http://arxiv.org/abs/0807.0713} {arXiv:0807.0713} \BibitemShut {NoStop}%
\bibitem [{\citenamefont {{Keppens}}\ \emph {et~al.}(2012)\citenamefont
  {{Keppens}}, \citenamefont {{Meliani}}, \citenamefont {{van Marle}},
  \citenamefont {{Delmont}}, \citenamefont {{Vlasis}},\ and\ \citenamefont
  {{van der Holst}}}]{amrvac12}%
  \BibitemOpen
  \bibfield  {author} {\bibinfo {author} {\bibfnamefont {R.}~\bibnamefont
  {{Keppens}}}, \bibinfo {author} {\bibfnamefont {Z.}~\bibnamefont
  {{Meliani}}}, \bibinfo {author} {\bibfnamefont {A.~J.}\ \bibnamefont {{van
  Marle}}}, \bibinfo {author} {\bibfnamefont {P.}~\bibnamefont {{Delmont}}},
  \bibinfo {author} {\bibfnamefont {A.}~\bibnamefont {{Vlasis}}}, \ and\
  \bibinfo {author} {\bibfnamefont {B.}~\bibnamefont {{van der Holst}}},\ }\Doi
  {10.1016/j.jcp.2011.01.020} {\bibfield  {journal} {\bibinfo  {journal}
  {Journal of Computational Physics},\ }\textbf {\bibinfo {volume} {231}},\
  \bibinfo {pages} {718} (\bibinfo {year} {2012})}\BibitemShut {NoStop}%
\bibitem [{\citenamefont {{Vlasis}}\ \emph {et~al.}(2011)\citenamefont
  {{Vlasis}}, \citenamefont {{van Eerten}}, \citenamefont {{Meliani}},\ and\
  \citenamefont {{Keppens}}}]{vlasis11}%
  \BibitemOpen
  \bibfield  {author} {\bibinfo {author} {\bibfnamefont {A.}~\bibnamefont
  {{Vlasis}}}, \bibinfo {author} {\bibfnamefont {H.~J.}\ \bibnamefont {{van
  Eerten}}}, \bibinfo {author} {\bibfnamefont {Z.}~\bibnamefont {{Meliani}}}, \
  and\ \bibinfo {author} {\bibfnamefont {R.}~\bibnamefont {{Keppens}}},\ }\Doi
  {10.1111/j.1365-2966.2011.18696.x} {\bibfield  {journal} {\bibinfo  {journal}
  {\mnras},\ }\textbf {\bibinfo {volume} {415}},\ \bibinfo {pages} {279}
  (\bibinfo {year} {2011})}\BibitemShut {NoStop}%
\bibitem [{\citenamefont {{Meliani}}\ and\ \citenamefont
  {{Keppens}}(2010)}]{melianigrb}%
  \BibitemOpen
  \bibfield  {author} {\bibinfo {author} {\bibfnamefont {Z.}~\bibnamefont
  {{Meliani}}}\ and\ \bibinfo {author} {\bibfnamefont {R.}~\bibnamefont
  {{Keppens}}},\ }\Doi {10.1051/0004-6361/201015423} {\bibfield  {journal}
  {\bibinfo  {journal} {\aap},\ }\textbf {\bibinfo {volume} {520}},\ \bibinfo
  {eid} {L3} (\bibinfo {year} {2010})},\ \Eprint
  {http://arxiv.org/abs/1009.1224} {arXiv:1009.1224 [astro-ph.HE]} \BibitemShut
  {NoStop}%
\bibitem [{\citenamefont {{Monceau-Baroux}}\ \emph {et~al.}(2014)\citenamefont
  {{Monceau-Baroux}}, \citenamefont {{Porth}}, \citenamefont {{Meliani}},\ and\
  \citenamefont {{Keppens}}}]{remi14}%
  \BibitemOpen
  \bibfield  {author} {\bibinfo {author} {\bibfnamefont {R.}~\bibnamefont
  {{Monceau-Baroux}}}, \bibinfo {author} {\bibfnamefont {O.}~\bibnamefont
  {{Porth}}}, \bibinfo {author} {\bibfnamefont {Z.}~\bibnamefont {{Meliani}}},
  \ and\ \bibinfo {author} {\bibfnamefont {R.}~\bibnamefont {{Keppens}}},\
  }\Doi {10.1051/0004-6361/201322682} {\bibfield  {journal} {\bibinfo
  {journal} {\aap},\ }\textbf {\bibinfo {volume} {561}},\ \bibinfo {eid} {A30}
  (\bibinfo {year} {2014})},\ \Eprint {http://arxiv.org/abs/1311.7593}
  {arXiv:1311.7593 [astro-ph.HE]} \BibitemShut {NoStop}%
\bibitem [{\citenamefont {{Keppens}}\ \emph {et~al.}(2008)\citenamefont
  {{Keppens}}, \citenamefont {{Meliani}}, \citenamefont {{van der Holst}},\
  and\ \citenamefont {{Casse}}}]{jet08}%
  \BibitemOpen
  \bibfield  {author} {\bibinfo {author} {\bibfnamefont {R.}~\bibnamefont
  {{Keppens}}}, \bibinfo {author} {\bibfnamefont {Z.}~\bibnamefont
  {{Meliani}}}, \bibinfo {author} {\bibfnamefont {B.}~\bibnamefont {{van der
  Holst}}}, \ and\ \bibinfo {author} {\bibfnamefont {F.}~\bibnamefont
  {{Casse}}},\ }\Doi {10.1051/0004-6361:20079174} {\bibfield  {journal}
  {\bibinfo  {journal} {\aap},\ }\textbf {\bibinfo {volume} {486}},\ \bibinfo
  {pages} {663} (\bibinfo {year} {2008})},\ \Eprint
  {http://arxiv.org/abs/0802.2034} {arXiv:0802.2034} \BibitemShut {NoStop}%
\bibitem [{\citenamefont {{Walg}}\ \emph {et~al.}(2013)\citenamefont {{Walg}},
  \citenamefont {{Achterberg}}, \citenamefont {{Markoff}}, \citenamefont
  {{Keppens}},\ and\ \citenamefont {{Meliani}}}]{walg13}%
  \BibitemOpen
  \bibfield  {author} {\bibinfo {author} {\bibfnamefont {S.}~\bibnamefont
  {{Walg}}}, \bibinfo {author} {\bibfnamefont {A.}~\bibnamefont
  {{Achterberg}}}, \bibinfo {author} {\bibfnamefont {S.}~\bibnamefont
  {{Markoff}}}, \bibinfo {author} {\bibfnamefont {R.}~\bibnamefont
  {{Keppens}}}, \ and\ \bibinfo {author} {\bibfnamefont {Z.}~\bibnamefont
  {{Meliani}}},\ }\Doi {10.1093/mnras/stt823} {\bibfield  {journal} {\bibinfo
  {journal} {\mnras},\ }\textbf {\bibinfo {volume} {433}},\ \bibinfo {pages}
  {1453} (\bibinfo {year} {2013})},\ \Eprint {http://arxiv.org/abs/1305.2157}
  {arXiv:1305.2157 [astro-ph.HE]} \BibitemShut {NoStop}%
\bibitem [{\citenamefont {{Porth}}(2013)}]{porth13}%
  \BibitemOpen
  \bibfield  {author} {\bibinfo {author} {\bibfnamefont {O.}~\bibnamefont
  {{Porth}}},\ }\Doi {10.1093/mnras/sts519} {\bibfield  {journal} {\bibinfo
  {journal} {\mnras},\ }\textbf {\bibinfo {volume} {429}},\ \bibinfo {pages}
  {2482} (\bibinfo {year} {2013})},\ \Eprint {http://arxiv.org/abs/1212.0676}
  {arXiv:1212.0676 [astro-ph.HE]} \BibitemShut {NoStop}%
\bibitem [{\citenamefont {{Porth}}\ \emph {et~al.}(2014)\citenamefont
  {{Porth}}, \citenamefont {{Komissarov}},\ and\ \citenamefont
  {{Keppens}}}]{porth14}%
  \BibitemOpen
  \bibfield  {author} {\bibinfo {author} {\bibfnamefont {O.}~\bibnamefont
  {{Porth}}}, \bibinfo {author} {\bibfnamefont {S.~S.}\ \bibnamefont
  {{Komissarov}}}, \ and\ \bibinfo {author} {\bibfnamefont {R.}~\bibnamefont
  {{Keppens}}},\ }\Doi {10.1093/mnras/stt2176} {\bibfield  {journal} {\bibinfo
  {journal} {\mnras},\ }\textbf {\bibinfo {volume} {438}},\ \bibinfo {pages}
  {278} (\bibinfo {year} {2014})},\ \Eprint {http://arxiv.org/abs/1310.2531}
  {arXiv:1310.2531 [astro-ph.HE]} \BibitemShut {NoStop}%
\bibitem [{\citenamefont {{van Marle}}\ \emph
  {et~al.}(2011){\natexlab{a}}\citenamefont {{van Marle}}, \citenamefont
  {{Meliani}}, \citenamefont {{Keppens}},\ and\ \citenamefont
  {{Decin}}}]{marle11}%
  \BibitemOpen
  \bibfield  {author} {\bibinfo {author} {\bibfnamefont {A.~J.}\ \bibnamefont
  {{van Marle}}}, \bibinfo {author} {\bibfnamefont {Z.}~\bibnamefont
  {{Meliani}}}, \bibinfo {author} {\bibfnamefont {R.}~\bibnamefont
  {{Keppens}}}, \ and\ \bibinfo {author} {\bibfnamefont {L.}~\bibnamefont
  {{Decin}}},\ }\Doi {10.1088/2041-8205/734/2/L26} {\bibfield  {journal}
  {\bibinfo  {journal} {\apjl},\ }\textbf {\bibinfo {volume} {734}},\ \bibinfo
  {eid} {L26} (\bibinfo {year} {2011}{\natexlab{a}})},\ \Eprint
  {http://arxiv.org/abs/1105.2387} {arXiv:1105.2387 [astro-ph.SR]} \BibitemShut
  {NoStop}%
\bibitem [{\citenamefont {{Meheut}}\ \emph {et~al.}(2012)\citenamefont
  {{Meheut}}, \citenamefont {{Meliani}}, \citenamefont {{Varniere}},\ and\
  \citenamefont {{Benz}}}]{meheut12}%
  \BibitemOpen
  \bibfield  {author} {\bibinfo {author} {\bibfnamefont {H.}~\bibnamefont
  {{Meheut}}}, \bibinfo {author} {\bibfnamefont {Z.}~\bibnamefont {{Meliani}}},
  \bibinfo {author} {\bibfnamefont {P.}~\bibnamefont {{Varniere}}}, \ and\
  \bibinfo {author} {\bibfnamefont {W.}~\bibnamefont {{Benz}}},\ }\Doi
  {10.1051/0004-6361/201219794} {\bibfield  {journal} {\bibinfo  {journal}
  {\aap},\ }\textbf {\bibinfo {volume} {545}},\ \bibinfo {eid} {A134} (\bibinfo
  {year} {2012})},\ \Eprint {http://arxiv.org/abs/1208.4947} {arXiv:1208.4947
  [astro-ph.EP]} \BibitemShut {NoStop}%
\bibitem [{\citenamefont {{Hendrix}}\ and\ \citenamefont
  {{Keppens}}(2014){\natexlab{a}}}]{tom14}%
  \BibitemOpen
  \bibfield  {author} {\bibinfo {author} {\bibfnamefont {T.}~\bibnamefont
  {{Hendrix}}}\ and\ \bibinfo {author} {\bibfnamefont {R.}~\bibnamefont
  {{Keppens}}},\ }\Doi {10.1051/0004-6361/201322322} {\bibfield  {journal}
  {\bibinfo  {journal} {\aap},\ }\textbf {\bibinfo {volume} {562}},\ \bibinfo
  {eid} {A114} (\bibinfo {year} {2014}{\natexlab{a}})},\ \Eprint
  {http://arxiv.org/abs/1401.6774} {arXiv:1401.6774 [astro-ph.GA]} \BibitemShut
  {NoStop}%
\bibitem [{\citenamefont {{Beeck}}\ \emph {et~al.}(2012)\citenamefont
  {{Beeck}}, \citenamefont {{Collet}}, \citenamefont {{Steffen}}, \citenamefont
  {{Asplund}}, \citenamefont {{Cameron}}, \citenamefont {{Freytag}},
  \citenamefont {{Hayek}}, \citenamefont {{Ludwig}},\ and\ \citenamefont
  {{Sch{\"u}ssler}}}]{radhdwith3}%
  \BibitemOpen
  \bibfield  {author} {\bibinfo {author} {\bibfnamefont {B.}~\bibnamefont
  {{Beeck}}}, \bibinfo {author} {\bibfnamefont {R.}~\bibnamefont {{Collet}}},
  \bibinfo {author} {\bibfnamefont {M.}~\bibnamefont {{Steffen}}}, \bibinfo
  {author} {\bibfnamefont {M.}~\bibnamefont {{Asplund}}}, \bibinfo {author}
  {\bibfnamefont {R.~H.}\ \bibnamefont {{Cameron}}}, \bibinfo {author}
  {\bibfnamefont {B.}~\bibnamefont {{Freytag}}}, \bibinfo {author}
  {\bibfnamefont {W.}~\bibnamefont {{Hayek}}}, \bibinfo {author} {\bibfnamefont
  {H.-G.}\ \bibnamefont {{Ludwig}}}, \ and\ \bibinfo {author} {\bibfnamefont
  {M.}~\bibnamefont {{Sch{\"u}ssler}}},\ }\Doi {10.1051/0004-6361/201118252}
  {\bibfield  {journal} {\bibinfo  {journal} {\aap},\ }\textbf {\bibinfo
  {volume} {539}},\ \bibinfo {eid} {A121} (\bibinfo {year} {2012})},\ \Eprint
  {http://arxiv.org/abs/1201.1103} {arXiv:1201.1103 [astro-ph.SR]} \BibitemShut
  {NoStop}%
\bibitem [{\citenamefont {{Keppens}}\ \emph
  {et~al.}(2013){\natexlab{a}}\citenamefont {{Keppens}}, \citenamefont
  {{Porth}}, \citenamefont {{Galsgaard}}, \citenamefont {{Frederiksen}},
  \citenamefont {{Restante}}, \citenamefont {{Lapenta}},\ and\ \citenamefont
  {{Parnell}}}]{pop13}%
  \BibitemOpen
  \bibfield  {author} {\bibinfo {author} {\bibfnamefont {R.}~\bibnamefont
  {{Keppens}}}, \bibinfo {author} {\bibfnamefont {O.}~\bibnamefont {{Porth}}},
  \bibinfo {author} {\bibfnamefont {K.}~\bibnamefont {{Galsgaard}}}, \bibinfo
  {author} {\bibfnamefont {J.~T.}\ \bibnamefont {{Frederiksen}}}, \bibinfo
  {author} {\bibfnamefont {A.~L.}\ \bibnamefont {{Restante}}}, \bibinfo
  {author} {\bibfnamefont {G.}~\bibnamefont {{Lapenta}}}, \ and\ \bibinfo
  {author} {\bibfnamefont {C.}~\bibnamefont {{Parnell}}},\ }\Doi
  {10.1063/1.4820946} {\bibfield  {journal} {\bibinfo  {journal} {Physics of
  Plasmas},\ }\textbf {\bibinfo {volume} {20}},\ \bibinfo {pages} {092109}
  (\bibinfo {year} {2013}{\natexlab{a}})}\BibitemShut {NoStop}%
\bibitem [{\citenamefont {Toro}(1999)}]{Toro1999}%
  \BibitemOpen
  \bibfield  {author} {\bibinfo {author} {\bibfnamefont {E.~F.}\ \bibnamefont
  {Toro}},\ }\href@noop {} {\emph {\bibinfo {title} {Riemann Solvers and
  Numerical Methods for Fluid Dynamics}}}\ (\bibinfo  {publisher} {Springer},\
  \bibinfo {year} {1999})\BibitemShut {NoStop}%
\bibitem [{\citenamefont {Fedkiw}\ \emph {et~al.}(1998)\citenamefont {Fedkiw},
  \citenamefont {Merriman}, \citenamefont {Donat},\ and\ \citenamefont
  {Osher}}]{Fedkiw98thepenultimate}%
  \BibitemOpen
  \bibfield  {author} {\bibinfo {author} {\bibfnamefont {R.~P.}\ \bibnamefont
  {Fedkiw}}, \bibinfo {author} {\bibfnamefont {B.}~\bibnamefont {Merriman}},
  \bibinfo {author} {\bibfnamefont {R.}~\bibnamefont {Donat}}, \ and\ \bibinfo
  {author} {\bibfnamefont {S.}~\bibnamefont {Osher}},\ }\href@noop {} {\emph
  {\bibinfo {title} {The penultimate scheme for systems of conservation laws:
  finite-difference ENO with Marquina's flux splitting}}},\ \bibinfo {type}
  {Tech. Rep.}\ (\bibinfo {year} {1998})\BibitemShut {NoStop}%
\bibitem [{\citenamefont {{Rusanov}}(1961)}]{Rusanov1961}%
  \BibitemOpen
  \bibfield  {author} {\bibinfo {author} {\bibfnamefont {V.~V.}\ \bibnamefont
  {{Rusanov}}},\ }\href@noop {} {\bibfield  {journal} {\bibinfo  {journal} {Zh.
  Vychisl. Mat. Mat. Fiz.},\ }\textbf {\bibinfo {volume} {1}},\ \bibinfo
  {pages} {267} (\bibinfo {year} {1961})}\BibitemShut {NoStop}%
\bibitem [{\citenamefont {{Jiang}}\ and\ \citenamefont
  {{Wu}}(1999)}]{JiangWu1999}%
  \BibitemOpen
  \bibfield  {author} {\bibinfo {author} {\bibfnamefont {G.-S.}\ \bibnamefont
  {{Jiang}}}\ and\ \bibinfo {author} {\bibfnamefont {C.-c.}\ \bibnamefont
  {{Wu}}},\ }\Doi {10.1006/jcph.1999.6207} {\bibfield  {journal} {\bibinfo
  {journal} {Journal of Computational Physics},\ }\textbf {\bibinfo {volume}
  {150}},\ \bibinfo {pages} {561} (\bibinfo {year} {1999})}\BibitemShut
  {NoStop}%
\bibitem [{\citenamefont {{Mignone}}\ \emph {et~al.}(2010)\citenamefont
  {{Mignone}}, \citenamefont {{Tzeferacos}},\ and\ \citenamefont
  {{Bodo}}}]{mignone2010}%
  \BibitemOpen
  \bibfield  {author} {\bibinfo {author} {\bibfnamefont {A.}~\bibnamefont
  {{Mignone}}}, \bibinfo {author} {\bibfnamefont {P.}~\bibnamefont
  {{Tzeferacos}}}, \ and\ \bibinfo {author} {\bibfnamefont {G.}~\bibnamefont
  {{Bodo}}},\ }\Doi {10.1016/j.jcp.2010.04.013} {\bibfield  {journal} {\bibinfo
   {journal} {Journal of Computational Physics},\ }\textbf {\bibinfo {volume}
  {229}},\ \bibinfo {pages} {5896} (\bibinfo {year} {2010})},\ \Eprint
  {http://arxiv.org/abs/1001.2832} {arXiv:1001.2832 [astro-ph.HE]} \BibitemShut
  {NoStop}%
\bibitem [{\citenamefont {{Radice}}\ and\ \citenamefont
  {{Rezzolla}}(2012)}]{radice2012}%
  \BibitemOpen
  \bibfield  {author} {\bibinfo {author} {\bibfnamefont {D.}~\bibnamefont
  {{Radice}}}\ and\ \bibinfo {author} {\bibfnamefont {L.}~\bibnamefont
  {{Rezzolla}}},\ }\Doi {10.1051/0004-6361/201219735} {\bibfield  {journal}
  {\bibinfo  {journal} {\aap},\ }\textbf {\bibinfo {volume} {547}},\ \bibinfo
  {eid} {A26} (\bibinfo {year} {2012})},\ \Eprint
  {http://arxiv.org/abs/1206.6502} {arXiv:1206.6502 [astro-ph.IM]} \BibitemShut
  {NoStop}%
\bibitem [{\citenamefont {{van Leer}}(1982)}]{van-Leer1982}%
  \BibitemOpen
  \bibfield  {author} {\bibinfo {author} {\bibfnamefont {B.}~\bibnamefont {{van
  Leer}}},\ }in\ \Doi {10.1007/3-540-11948-5_66} {\emph {\bibinfo {booktitle}
  {Numerical Methods in Fluid Dynamics}}},\ \bibinfo {series} {Lecture Notes in
  Physics, Berlin Springer Verlag}, Vol.\ \bibinfo {volume} {170},\ \bibinfo
  {editor} {edited by\ \bibinfo {editor} {\bibfnamefont {E.}~\bibnamefont
  {{Krause}}}}\ (\bibinfo {year} {1982})\ pp.\ \bibinfo {pages}
  {507--512}\BibitemShut {NoStop}%
\bibitem [{\citenamefont {{Liou}}\ and\ \citenamefont
  {{Steffen}}(1993)}]{LiouSteffen1993}%
  \BibitemOpen
  \bibfield  {author} {\bibinfo {author} {\bibfnamefont {M.-S.}\ \bibnamefont
  {{Liou}}}\ and\ \bibinfo {author} {\bibfnamefont {C.~J.}\ \bibnamefont
  {{Steffen}}},\ }\Doi {10.1006/jcph.1993.1122} {\bibfield  {journal} {\bibinfo
   {journal} {Journal of Computational Physics},\ }\textbf {\bibinfo {volume}
  {107}},\ \bibinfo {pages} {23} (\bibinfo {year} {1993})}\BibitemShut
  {NoStop}%
\bibitem [{\citenamefont {{{\v C}ada}}\ and\ \citenamefont
  {{Torrilhon}}(2009){\natexlab{a}}}]{cada2009}%
  \BibitemOpen
  \bibfield  {author} {\bibinfo {author} {\bibfnamefont {M.}~\bibnamefont {{{\v
  C}ada}}}\ and\ \bibinfo {author} {\bibfnamefont {M.}~\bibnamefont
  {{Torrilhon}}},\ }\Doi {10.1016/j.jcp.2009.02.020} {\bibfield  {journal}
  {\bibinfo  {journal} {Journal of Computational Physics},\ }\textbf {\bibinfo
  {volume} {228}},\ \bibinfo {pages} {4118} (\bibinfo {year}
  {2009}{\natexlab{a}})}\BibitemShut {NoStop}%
\bibitem [{\citenamefont {{Suresh}}\ and\ \citenamefont
  {{Huynh}}(1997)}]{1997JCoPh.136...83S}%
  \BibitemOpen
  \bibfield  {author} {\bibinfo {author} {\bibfnamefont {A.}~\bibnamefont
  {{Suresh}}}\ and\ \bibinfo {author} {\bibfnamefont {H.~T.}\ \bibnamefont
  {{Huynh}}},\ }\Doi {10.1006/jcph.1997.5745} {\bibfield  {journal} {\bibinfo
  {journal} {Journal of Computational Physics},\ }\textbf {\bibinfo {volume}
  {136}},\ \bibinfo {pages} {83} (\bibinfo {year} {1997})}\BibitemShut
  {NoStop}%
\bibitem [{\citenamefont {{Gottlieb}}\ and\ \citenamefont
  {{Shu}}(1998)}]{gottlieb1998}%
  \BibitemOpen
  \bibfield  {author} {\bibinfo {author} {\bibfnamefont {S.}~\bibnamefont
  {{Gottlieb}}}\ and\ \bibinfo {author} {\bibfnamefont {C.~W.}\ \bibnamefont
  {{Shu}}},\ }\href@noop {} {\bibfield  {journal} {\bibinfo  {journal}
  {Mathematics of Computation},\ }\textbf {\bibinfo {volume} {67}},\ \bibinfo
  {pages} {73} (\bibinfo {year} {1998})}\BibitemShut {NoStop}%
\bibitem [{\citenamefont {Spiteri}\ and\ \citenamefont
  {Ruuth}(2002)}]{SpiteriRuuth2002}%
  \BibitemOpen
  \bibfield  {author} {\bibinfo {author} {\bibfnamefont {R.~J.}\ \bibnamefont
  {Spiteri}}\ and\ \bibinfo {author} {\bibfnamefont {S.~J.}\ \bibnamefont
  {Ruuth}},\ }\Doi {10.1137/S0036142901389025} {\bibfield  {journal} {\bibinfo
  {journal} {SIAM J. Numer. Anal.},\ }\textbf {\bibinfo {volume} {40}},\
  \bibinfo {pages} {469} (\bibinfo {year} {2002})},\ ISSN \bibinfo {issn}
  {0036-1429}\BibitemShut {NoStop}%
\bibitem [{\citenamefont {{van Marle}}\ and\ \citenamefont
  {{Keppens}}(2011)}]{vanmarleRL}%
  \BibitemOpen
  \bibfield  {author} {\bibinfo {author} {\bibfnamefont {A.~J.}\ \bibnamefont
  {{van Marle}}}\ and\ \bibinfo {author} {\bibfnamefont {R.}~\bibnamefont
  {{Keppens}}},\ }\href@noop {} {\bibfield  {journal} {\bibinfo  {journal}
  {Computers \& Fluids},\ }\textbf {\bibinfo {volume} {42}},\ \bibinfo {pages}
  {44} (\bibinfo {year} {2011})}\BibitemShut {NoStop}%
\bibitem [{\citenamefont {{Decin}}(2006)}]{2006A&A...456..549D}%
  \BibitemOpen
  \bibfield  {author} {\bibinfo {author} {\bibfnamefont {L.~e.~a.}\
  \bibnamefont {{Decin}}},\ }\Doi {10.1051/0004-6361:20065230} {\bibfield
  {journal} {\bibinfo  {journal} {\aap},\ }\textbf {\bibinfo {volume} {456}},\
  \bibinfo {pages} {549} (\bibinfo {year} {2006})},\ \Eprint
  {http://arxiv.org/abs/astro-ph/0606299} {astro-ph/0606299} \BibitemShut
  {NoStop}%
\bibitem [{\citenamefont {{Laibe}}\ and\ \citenamefont
  {{Price}}(2012)}]{2012MNRAS.420.2345L}%
  \BibitemOpen
  \bibfield  {author} {\bibinfo {author} {\bibfnamefont {G.}~\bibnamefont
  {{Laibe}}}\ and\ \bibinfo {author} {\bibfnamefont {D.~J.}\ \bibnamefont
  {{Price}}},\ }\Doi {10.1111/j.1365-2966.2011.20202.x} {\bibfield  {journal}
  {\bibinfo  {journal} {\mnras},\ }\textbf {\bibinfo {volume} {420}},\ \bibinfo
  {pages} {2345} (\bibinfo {year} {2012})},\ \Eprint
  {http://arxiv.org/abs/1111.3090} {arXiv:1111.3090 [astro-ph.IM]} \BibitemShut
  {NoStop}%
\bibitem [{\citenamefont {{Sod}}(1978)}]{1978JCoPh..27....1S}%
  \BibitemOpen
  \bibfield  {author} {\bibinfo {author} {\bibfnamefont {G.~A.}\ \bibnamefont
  {{Sod}}},\ }\Doi {10.1016/0021-9991(78)90023-2} {\bibfield  {journal}
  {\bibinfo  {journal} {Journal of Computational Physics},\ }\textbf {\bibinfo
  {volume} {27}},\ \bibinfo {pages} {1} (\bibinfo {year} {1978})}\BibitemShut
  {NoStop}%
\bibitem [{\citenamefont {{Hendrix}}\ and\ \citenamefont
  {{Keppens}}(2014){\natexlab{b}}}]{ASTROproc}%
  \BibitemOpen
  \bibfield  {author} {\bibinfo {author} {\bibfnamefont {T.}~\bibnamefont
  {{Hendrix}}}\ and\ \bibinfo {author} {\bibfnamefont {R.}~\bibnamefont
  {{Keppens}}},\ }in\ \href@noop {} {\emph {\bibinfo {booktitle} {Proc.\ 8th
  International conference of numerical modeling of space plasma flows
  (ASTRONUM, Biarritz, France, July 2013)}}},\ \bibinfo {series and number}
  {ASP Conference Series}\ (\bibinfo  {publisher} {Astronomical Society of the
  Pacific},\ \bibinfo {address} {San Francisco, US},\ \bibinfo {year}
  {2014})\BibitemShut {NoStop}%
\bibitem [{\citenamefont {{Draine}}\ and\ \citenamefont
  {{Lee}}(1984)}]{1984ApJ...285...89D}%
  \BibitemOpen
  \bibfield  {author} {\bibinfo {author} {\bibfnamefont {B.~T.}\ \bibnamefont
  {{Draine}}}\ and\ \bibinfo {author} {\bibfnamefont {H.~M.}\ \bibnamefont
  {{Lee}}},\ }\Doi {10.1086/162480} {\bibfield  {journal} {\bibinfo  {journal}
  {\apj},\ }\textbf {\bibinfo {volume} {285}},\ \bibinfo {pages} {89} (\bibinfo
  {year} {1984})}\BibitemShut {NoStop}%
\bibitem [{\citenamefont {{Laibe}}\ and\ \citenamefont
  {{Price}}(2011)}]{2011MNRAS.418.1491L}%
  \BibitemOpen
  \bibfield  {author} {\bibinfo {author} {\bibfnamefont {G.}~\bibnamefont
  {{Laibe}}}\ and\ \bibinfo {author} {\bibfnamefont {D.~J.}\ \bibnamefont
  {{Price}}},\ }\Doi {10.1111/j.1365-2966.2011.19291.x} {\bibfield  {journal}
  {\bibinfo  {journal} {\mnras},\ }\textbf {\bibinfo {volume} {418}},\ \bibinfo
  {pages} {1491} (\bibinfo {year} {2011})},\ \Eprint
  {http://arxiv.org/abs/1106.1736} {arXiv:1106.1736 [astro-ph.EP]} \BibitemShut
  {NoStop}%
\bibitem [{\citenamefont {{Colella}}\ and\ \citenamefont
  {{Woodward}}(1984)}]{1984JCoPh..54..174C}%
  \BibitemOpen
  \bibfield  {author} {\bibinfo {author} {\bibfnamefont {P.}~\bibnamefont
  {{Colella}}}\ and\ \bibinfo {author} {\bibfnamefont {P.~R.}\ \bibnamefont
  {{Woodward}}},\ }\Doi {10.1016/0021-9991(84)90143-8} {\bibfield  {journal}
  {\bibinfo  {journal} {Journal of Computational Physics},\ }\textbf {\bibinfo
  {volume} {54}},\ \bibinfo {pages} {174} (\bibinfo {year} {1984})}\BibitemShut
  {NoStop}%
\bibitem [{\citenamefont {{Tasker}}\ \emph {et~al.}(2008)\citenamefont
  {{Tasker}}, \citenamefont {{Brunino}}, \citenamefont {{Mitchell}},
  \citenamefont {{Michielsen}}, \citenamefont {{Hopton}}, \citenamefont
  {{Pearce}}, \citenamefont {{Bryan}},\ and\ \citenamefont
  {{Theuns}}}]{2008MNRAS.390.1267T}%
  \BibitemOpen
  \bibfield  {author} {\bibinfo {author} {\bibfnamefont {E.~J.}\ \bibnamefont
  {{Tasker}}}, \bibinfo {author} {\bibfnamefont {R.}~\bibnamefont {{Brunino}}},
  \bibinfo {author} {\bibfnamefont {N.~L.}\ \bibnamefont {{Mitchell}}},
  \bibinfo {author} {\bibfnamefont {D.}~\bibnamefont {{Michielsen}}}, \bibinfo
  {author} {\bibfnamefont {S.}~\bibnamefont {{Hopton}}}, \bibinfo {author}
  {\bibfnamefont {F.~R.}\ \bibnamefont {{Pearce}}}, \bibinfo {author}
  {\bibfnamefont {G.~L.}\ \bibnamefont {{Bryan}}}, \ and\ \bibinfo {author}
  {\bibfnamefont {T.}~\bibnamefont {{Theuns}}},\ }\Doi
  {10.1111/j.1365-2966.2008.13836.x} {\bibfield  {journal} {\bibinfo  {journal}
  {\mnras},\ }\textbf {\bibinfo {volume} {390}},\ \bibinfo {pages} {1267}
  (\bibinfo {year} {2008})},\ \Eprint {http://arxiv.org/abs/0808.1844}
  {arXiv:0808.1844} \BibitemShut {NoStop}%
\bibitem [{\citenamefont {{Sedov}}(1959)}]{Sedov}%
  \BibitemOpen
  \bibfield  {author} {\bibinfo {author} {\bibfnamefont {L.~I.}\ \bibnamefont
  {{Sedov}}},\ }\href@noop {} {\emph {\bibinfo {title} {{Similarity and
  Dimensional Methods in Mechanics}}}}\ (\bibinfo  {publisher} {New York:
  Academic Press},\ \bibinfo {year} {1959})\BibitemShut {NoStop}%
\bibitem [{\citenamefont {{Landau}}\ and\ \citenamefont
  {{Lifshitz}}(1959)}]{Landau}%
  \BibitemOpen
  \bibfield  {author} {\bibinfo {author} {\bibfnamefont {L.~D.}\ \bibnamefont
  {{Landau}}}\ and\ \bibinfo {author} {\bibfnamefont {E.~M.}\ \bibnamefont
  {{Lifshitz}}},\ }\href@noop {} {\emph {\bibinfo {title} {{Course of
  Theoretical Physics}}}},\ Vol.\ \bibinfo {volume} {Vol. 6 : Fluid Mechanics}\
  (\bibinfo  {publisher} {Pergamon Press},\ \bibinfo {year} {1959})\BibitemShut
  {NoStop}%
\bibitem [{\citenamefont {{Patnaude}}\ and\ \citenamefont
  {{Fesen}}(2005)}]{2005ApJ...633..240P}%
  \BibitemOpen
  \bibfield  {author} {\bibinfo {author} {\bibfnamefont {D.~J.}\ \bibnamefont
  {{Patnaude}}}\ and\ \bibinfo {author} {\bibfnamefont {R.~A.}\ \bibnamefont
  {{Fesen}}},\ }\Doi {10.1086/452627} {\bibfield  {journal} {\bibinfo
  {journal} {\apj},\ }\textbf {\bibinfo {volume} {633}},\ \bibinfo {pages}
  {240} (\bibinfo {year} {2005})},\ \Eprint
  {http://arxiv.org/abs/astro-ph/0507330} {astro-ph/0507330} \BibitemShut
  {NoStop}%
\bibitem [{\citenamefont {{Nakamura}}\ \emph {et~al.}(2006)\citenamefont
  {{Nakamura}}, \citenamefont {{McKee}}, \citenamefont {{Klein}},\ and\
  \citenamefont {{Fisher}}}]{2006ApJS..164..477N}%
  \BibitemOpen
  \bibfield  {author} {\bibinfo {author} {\bibfnamefont {F.}~\bibnamefont
  {{Nakamura}}}, \bibinfo {author} {\bibfnamefont {C.~F.}\ \bibnamefont
  {{McKee}}}, \bibinfo {author} {\bibfnamefont {R.~I.}\ \bibnamefont
  {{Klein}}}, \ and\ \bibinfo {author} {\bibfnamefont {R.~T.}\ \bibnamefont
  {{Fisher}}},\ }\Doi {10.1086/501530} {\bibfield  {journal} {\bibinfo
  {journal} {\apjs},\ }\textbf {\bibinfo {volume} {164}},\ \bibinfo {pages}
  {477} (\bibinfo {year} {2006})},\ \Eprint
  {http://arxiv.org/abs/astro-ph/0511016} {astro-ph/0511016} \BibitemShut
  {NoStop}%
\bibitem [{\citenamefont {{Agertz}}\ \emph {et~al.}(2007)\citenamefont
  {{Agertz}}, \citenamefont {{Moore}}, \citenamefont {{Stadel}}, \citenamefont
  {{Potter}}, \citenamefont {{Miniati}}, \citenamefont {{Read}}, \citenamefont
  {{Mayer}}, \citenamefont {{Gawryszczak}}, \citenamefont {{Kravtsov}},
  \citenamefont {{Nordlund}}, \citenamefont {{Pearce}}, \citenamefont
  {{Quilis}}, \citenamefont {{Rudd}}, \citenamefont {{Springel}}, \citenamefont
  {{Stone}}, \citenamefont {{Tasker}}, \citenamefont {{Teyssier}},
  \citenamefont {{Wadsley}},\ and\ \citenamefont
  {{Walder}}}]{2007MNRAS.380..963A}%
  \BibitemOpen
  \bibfield  {author} {\bibinfo {author} {\bibfnamefont {O.}~\bibnamefont
  {{Agertz}}}, \bibinfo {author} {\bibfnamefont {B.}~\bibnamefont {{Moore}}},
  \bibinfo {author} {\bibfnamefont {J.}~\bibnamefont {{Stadel}}}, \bibinfo
  {author} {\bibfnamefont {D.}~\bibnamefont {{Potter}}}, \bibinfo {author}
  {\bibfnamefont {F.}~\bibnamefont {{Miniati}}}, \bibinfo {author}
  {\bibfnamefont {J.}~\bibnamefont {{Read}}}, \bibinfo {author} {\bibfnamefont
  {L.}~\bibnamefont {{Mayer}}}, \bibinfo {author} {\bibfnamefont
  {A.}~\bibnamefont {{Gawryszczak}}}, \bibinfo {author} {\bibfnamefont
  {A.}~\bibnamefont {{Kravtsov}}}, \bibinfo {author} {\bibfnamefont
  {{\AA}.}~\bibnamefont {{Nordlund}}}, \bibinfo {author} {\bibfnamefont
  {F.}~\bibnamefont {{Pearce}}}, \bibinfo {author} {\bibfnamefont
  {V.}~\bibnamefont {{Quilis}}}, \bibinfo {author} {\bibfnamefont
  {D.}~\bibnamefont {{Rudd}}}, \bibinfo {author} {\bibfnamefont
  {V.}~\bibnamefont {{Springel}}}, \bibinfo {author} {\bibfnamefont
  {J.}~\bibnamefont {{Stone}}}, \bibinfo {author} {\bibfnamefont
  {E.}~\bibnamefont {{Tasker}}}, \bibinfo {author} {\bibfnamefont
  {R.}~\bibnamefont {{Teyssier}}}, \bibinfo {author} {\bibfnamefont
  {J.}~\bibnamefont {{Wadsley}}}, \ and\ \bibinfo {author} {\bibfnamefont
  {R.}~\bibnamefont {{Walder}}},\ }\Doi {10.1111/j.1365-2966.2007.12183.x}
  {\bibfield  {journal} {\bibinfo  {journal} {\mnras},\ }\textbf {\bibinfo
  {volume} {380}},\ \bibinfo {pages} {963} (\bibinfo {year} {2007})},\ \Eprint
  {http://arxiv.org/abs/astro-ph/0610051} {astro-ph/0610051} \BibitemShut
  {NoStop}%
\bibitem [{\citenamefont {{Koren}}(1993)}]{Koren}%
  \BibitemOpen
  \bibfield  {author} {\bibinfo {author} {\bibfnamefont {B.}~\bibnamefont
  {{Koren}}},\ }\href@noop {} {\emph {\bibinfo {title} {{A robust upwind
  discretization for advection, diffusion and source terms}}}},\ Vol.\ \bibinfo
  {volume} {Notes on Numerical Fluid Mechanics 45}\ (\bibinfo {year}
  {1993})\BibitemShut {NoStop}%
\bibitem [{\citenamefont {{Dedner}}\ \emph {et~al.}(2002)\citenamefont
  {{Dedner}}, \citenamefont {{Kemm}}, \citenamefont {{Kr{\"o}ner}},
  \citenamefont {{Munz}}, \citenamefont {{Schnitzer}},\ and\ \citenamefont
  {{Wesenberg}}}]{dedner02}%
  \BibitemOpen
  \bibfield  {author} {\bibinfo {author} {\bibfnamefont {A.}~\bibnamefont
  {{Dedner}}}, \bibinfo {author} {\bibfnamefont {F.}~\bibnamefont {{Kemm}}},
  \bibinfo {author} {\bibfnamefont {D.}~\bibnamefont {{Kr{\"o}ner}}}, \bibinfo
  {author} {\bibfnamefont {C.-D.}\ \bibnamefont {{Munz}}}, \bibinfo {author}
  {\bibfnamefont {T.}~\bibnamefont {{Schnitzer}}}, \ and\ \bibinfo {author}
  {\bibfnamefont {M.}~\bibnamefont {{Wesenberg}}},\ }\Doi
  {10.1006/jcph.2001.6961} {\bibfield  {journal} {\bibinfo  {journal} {Journal
  of Computational Physics},\ }\textbf {\bibinfo {volume} {175}},\ \bibinfo
  {pages} {645} (\bibinfo {year} {2002})}\BibitemShut {NoStop}%
\bibitem [{\citenamefont {{Janhunen}}(2000)}]{janhunen}%
  \BibitemOpen
  \bibfield  {author} {\bibinfo {author} {\bibfnamefont {P.}~\bibnamefont
  {{Janhunen}}},\ }\Doi {10.1006/jcph.2000.6479} {\bibfield  {journal}
  {\bibinfo  {journal} {Journal of Computational Physics},\ }\textbf {\bibinfo
  {volume} {160}},\ \bibinfo {pages} {649} (\bibinfo {year}
  {2000})}\BibitemShut {NoStop}%
\bibitem [{\citenamefont {{Xia}}\ \emph {et~al.}(2014)\citenamefont {{Xia}},
  \citenamefont {{Keppens}},\ and\ \citenamefont {{Guo}}}]{xia14}%
  \BibitemOpen
  \bibfield  {author} {\bibinfo {author} {\bibfnamefont {C.}~\bibnamefont
  {{Xia}}}, \bibinfo {author} {\bibfnamefont {R.}~\bibnamefont {{Keppens}}}, \
  and\ \bibinfo {author} {\bibfnamefont {Y.}~\bibnamefont {{Guo}}},\ }\Doi
  {10.1088/0004-637X/780/2/130} {\bibfield  {journal} {\bibinfo  {journal}
  {\apj},\ }\textbf {\bibinfo {volume} {780}},\ \bibinfo {eid} {130} (\bibinfo
  {year} {2014})},\ \Eprint {http://arxiv.org/abs/1311.5478} {arXiv:1311.5478
  [astro-ph.SR]} \BibitemShut {NoStop}%
\bibitem [{\citenamefont {{Yee}}(1989)}]{Yee1989}%
  \BibitemOpen
  \bibfield  {author} {\bibinfo {author} {\bibfnamefont {H.~C.}\ \bibnamefont
  {{Yee}}},\ }\href@noop {} {\emph {\bibinfo {title} {A Class of
  High-Resolution Explicit and Implicit Shock-Capturing Methods}}},\ \bibinfo
  {organization} {NASA} (\bibinfo {year} {1989})\BibitemShut {NoStop}%
\bibitem [{\citenamefont {{Lesur}}\ \emph {et~al.}(2014)\citenamefont
  {{Lesur}}, \citenamefont {{Kunz}},\ and\ \citenamefont
  {{Fromang}}}]{LesurKunz2014}%
  \BibitemOpen
  \bibfield  {author} {\bibinfo {author} {\bibfnamefont {G.}~\bibnamefont
  {{Lesur}}}, \bibinfo {author} {\bibfnamefont {M.~W.}\ \bibnamefont {{Kunz}}},
  \ and\ \bibinfo {author} {\bibfnamefont {S.}~\bibnamefont {{Fromang}}},\
  }\href@noop {} {\bibfield  {journal} {\bibinfo  {journal} {ArXiv e-prints}}
  (\bibinfo {year} {2014})},\ \Eprint {http://arxiv.org/abs/1402.4133}
  {arXiv:1402.4133 [astro-ph.SR]} \BibitemShut {NoStop}%
\bibitem [{\citenamefont {{Bai}}(2012)}]{Bai2012}%
  \BibitemOpen
  \bibfield  {author} {\bibinfo {author} {\bibfnamefont {X.-N.}\ \bibnamefont
  {{Bai}}},\ }\emph {\bibinfo {title} {{Non-ideal magnetohydrodynamic effects
  in protoplanetary disks}}},\ \href@noop {} {Ph.D. thesis},\ \bibinfo
  {school} {Princeton University} (\bibinfo {year} {2012})\BibitemShut
  {NoStop}%
\bibitem [{\citenamefont {{T{\'o}th}}\ \emph {et~al.}(2008)\citenamefont
  {{T{\'o}th}}, \citenamefont {{Ma}},\ and\ \citenamefont
  {{Gombosi}}}]{TothMa2008}%
  \BibitemOpen
  \bibfield  {author} {\bibinfo {author} {\bibfnamefont {G.}~\bibnamefont
  {{T{\'o}th}}}, \bibinfo {author} {\bibfnamefont {Y.}~\bibnamefont {{Ma}}}, \
  and\ \bibinfo {author} {\bibfnamefont {T.~I.}\ \bibnamefont {{Gombosi}}},\
  }\Doi {10.1016/j.jcp.2008.04.010} {\bibfield  {journal} {\bibinfo  {journal}
  {Journal of Computational Physics},\ }\textbf {\bibinfo {volume} {227}},\
  \bibinfo {pages} {6967} (\bibinfo {year} {2008})}\BibitemShut {NoStop}%
\bibitem [{\citenamefont {Davis}(1988)}]{Davis:1988:SSO}%
  \BibitemOpen
  \bibfield  {author} {\bibinfo {author} {\bibfnamefont {S.~F.}\ \bibnamefont
  {Davis}},\ }\Doi {http://dx.doi.org/10.1137/0909030} {\textbf {\bibinfo
  {volume} {9}},\ \bibinfo {pages} {445} (\bibinfo {year} {1988})},\ ISSN
  \bibinfo {issn} {0196-5204}\BibitemShut {NoStop}%
\bibitem [{\citenamefont {{Brio}}\ and\ \citenamefont
  {{Wu}}(1988)}]{BrioWu1988}%
  \BibitemOpen
  \bibfield  {author} {\bibinfo {author} {\bibfnamefont {M.}~\bibnamefont
  {{Brio}}}\ and\ \bibinfo {author} {\bibfnamefont {C.~C.}\ \bibnamefont
  {{Wu}}},\ }\Doi {10.1016/0021-9991(88)90120-9} {\bibfield  {journal}
  {\bibinfo  {journal} {Journal of Computational Physics},\ }\textbf {\bibinfo
  {volume} {75}},\ \bibinfo {pages} {400} (\bibinfo {year} {1988})}\BibitemShut
  {NoStop}%
\bibitem [{\citenamefont {{Ryu}}\ and\ \citenamefont
  {{Jones}}(1995)}]{ryu1995}%
  \BibitemOpen
  \bibfield  {author} {\bibinfo {author} {\bibfnamefont {D.}~\bibnamefont
  {{Ryu}}}\ and\ \bibinfo {author} {\bibfnamefont {T.~W.}\ \bibnamefont
  {{Jones}}},\ }\Doi {10.1086/175437} {\bibfield  {journal} {\bibinfo
  {journal} {\apj},\ }\textbf {\bibinfo {volume} {442}},\ \bibinfo {pages}
  {228} (\bibinfo {year} {1995})},\ \Eprint
  {http://arxiv.org/abs/arXiv:astro-ph/9404074} {arXiv:astro-ph/9404074}
  \BibitemShut {NoStop}%
\bibitem [{\citenamefont {{Hameiri}}\ \emph {et~al.}(2005)\citenamefont
  {{Hameiri}}, \citenamefont {{Ishizawa}},\ and\ \citenamefont
  {{Ishida}}}]{HameiriIshizawa2005}%
  \BibitemOpen
  \bibfield  {author} {\bibinfo {author} {\bibfnamefont {E.}~\bibnamefont
  {{Hameiri}}}, \bibinfo {author} {\bibfnamefont {A.}~\bibnamefont
  {{Ishizawa}}}, \ and\ \bibinfo {author} {\bibfnamefont {A.}~\bibnamefont
  {{Ishida}}},\ }\Doi {10.1063/1.1952887} {\bibfield  {journal} {\bibinfo
  {journal} {Physics of Plasmas},\ }\textbf {\bibinfo {volume} {12}},\ \bibinfo
  {pages} {072109} (\bibinfo {year} {2005})}\BibitemShut {NoStop}%
\bibitem [{\citenamefont {Keppens}\ \emph {et~al.}(2012)\citenamefont
  {Keppens}, \citenamefont {Meliani}, \citenamefont {van Marle}, \citenamefont
  {Delmont}, \citenamefont {Vlasis},\ and\ \citenamefont {van~der
  Holst}}]{Keppens2012718}%
  \BibitemOpen
  \bibfield  {author} {\bibinfo {author} {\bibfnamefont {R.}~\bibnamefont
  {Keppens}}, \bibinfo {author} {\bibfnamefont {Z.}~\bibnamefont {Meliani}},
  \bibinfo {author} {\bibfnamefont {A.}~\bibnamefont {van Marle}}, \bibinfo
  {author} {\bibfnamefont {P.}~\bibnamefont {Delmont}}, \bibinfo {author}
  {\bibfnamefont {A.}~\bibnamefont {Vlasis}}, \ and\ \bibinfo {author}
  {\bibfnamefont {B.}~\bibnamefont {van~der Holst}},\ }\Doi
  {10.1016/j.jcp.2011.01.020} {\bibfield  {journal} {\bibinfo  {journal}
  {Journal of Computational Physics},\ }\textbf {\bibinfo {volume} {231}},\
  \bibinfo {pages} {718 } (\bibinfo {year} {2012})},\ ISSN \bibinfo {issn}
  {0021-9991}\BibitemShut {NoStop}%
\bibitem [{\citenamefont {{Parker}}(1957)}]{1957JGR....62..509P}%
  \BibitemOpen
  \bibfield  {author} {\bibinfo {author} {\bibfnamefont {E.~N.}\ \bibnamefont
  {{Parker}}},\ }\Doi {10.1029/JZ062i004p00509} {\bibfield  {journal} {\bibinfo
   {journal} {\jgr},\ }\textbf {\bibinfo {volume} {62}},\ \bibinfo {pages}
  {509} (\bibinfo {year} {1957})}\BibitemShut {NoStop}%
\bibitem [{\citenamefont {{Keppens}}\ \emph
  {et~al.}(2013){\natexlab{b}}\citenamefont {{Keppens}}, \citenamefont
  {{Porth}}, \citenamefont {{Galsgaard}}, \citenamefont {{Frederiksen}},
  \citenamefont {{Restante}}, \citenamefont {{Lapenta}},\ and\ \citenamefont
  {{Parnell}}}]{2013PhPl...20i2109K}%
  \BibitemOpen
  \bibfield  {author} {\bibinfo {author} {\bibfnamefont {R.}~\bibnamefont
  {{Keppens}}}, \bibinfo {author} {\bibfnamefont {O.}~\bibnamefont {{Porth}}},
  \bibinfo {author} {\bibfnamefont {K.}~\bibnamefont {{Galsgaard}}}, \bibinfo
  {author} {\bibfnamefont {J.~T.}\ \bibnamefont {{Frederiksen}}}, \bibinfo
  {author} {\bibfnamefont {A.~L.}\ \bibnamefont {{Restante}}}, \bibinfo
  {author} {\bibfnamefont {G.}~\bibnamefont {{Lapenta}}}, \ and\ \bibinfo
  {author} {\bibfnamefont {C.}~\bibnamefont {{Parnell}}},\ }\Doi
  {10.1063/1.4820946} {\bibfield  {journal} {\bibinfo  {journal} {Physics of
  Plasmas},\ }\textbf {\bibinfo {volume} {20}},\ \bibinfo {pages} {092109}
  (\bibinfo {year} {2013}{\natexlab{b}})}\BibitemShut {NoStop}%
\bibitem [{\citenamefont {{Ma}}\ and\ \citenamefont
  {{Bhattacharjee}}(2001)}]{ma2001}%
  \BibitemOpen
  \bibfield  {author} {\bibinfo {author} {\bibfnamefont {Z.~W.}\ \bibnamefont
  {{Ma}}}\ and\ \bibinfo {author} {\bibfnamefont {A.}~\bibnamefont
  {{Bhattacharjee}}},\ }\Doi {10.1029/1999JA001004} {\bibfield  {journal}
  {\bibinfo  {journal} {\jgr},\ }\textbf {\bibinfo {volume} {106}},\ \bibinfo
  {pages} {3773} (\bibinfo {year} {2001})}\BibitemShut {NoStop}%
\bibitem [{\citenamefont {{Lohner}}(1987)}]{lohner1987}%
  \BibitemOpen
  \bibfield  {author} {\bibinfo {author} {\bibfnamefont {R.}~\bibnamefont
  {{Lohner}}},\ }\Doi {10.1016/0045-7825(87)90098-3} {\bibfield  {journal}
  {\bibinfo  {journal} {Computer Methods in Applied Mechanics and
  Engineering},\ }\textbf {\bibinfo {volume} {61}},\ \bibinfo {pages} {323}
  (\bibinfo {year} {1987})}\BibitemShut {NoStop}%
\bibitem [{\citenamefont {{Shay}}\ \emph {et~al.}(2001)\citenamefont {{Shay}},
  \citenamefont {{Drake}}, \citenamefont {{Rogers}},\ and\ \citenamefont
  {{Denton}}}]{ShayDrake2001}%
  \BibitemOpen
  \bibfield  {author} {\bibinfo {author} {\bibfnamefont {M.~A.}\ \bibnamefont
  {{Shay}}}, \bibinfo {author} {\bibfnamefont {J.~F.}\ \bibnamefont {{Drake}}},
  \bibinfo {author} {\bibfnamefont {B.~N.}\ \bibnamefont {{Rogers}}}, \ and\
  \bibinfo {author} {\bibfnamefont {R.~E.}\ \bibnamefont {{Denton}}},\ }\Doi
  {10.1029/1999JA001007} {\bibfield  {journal} {\bibinfo  {journal} {\jgr},\
  }\textbf {\bibinfo {volume} {106}},\ \bibinfo {pages} {3759} (\bibinfo {year}
  {2001})}\BibitemShut {NoStop}%
\bibitem [{\citenamefont {{Birn}}\ and\ \citenamefont
  {{Hesse}}(2001)}]{birn2001}%
  \BibitemOpen
  \bibfield  {author} {\bibinfo {author} {\bibfnamefont {J.}~\bibnamefont
  {{Birn}}}\ and\ \bibinfo {author} {\bibfnamefont {M.}~\bibnamefont
  {{Hesse}}},\ }\Doi {10.1029/1999JA001001} {\bibfield  {journal} {\bibinfo
  {journal} {\jgr},\ }\textbf {\bibinfo {volume} {106}},\ \bibinfo {pages}
  {3737} (\bibinfo {year} {2001})}\BibitemShut {NoStop}%
\bibitem [{\citenamefont {{Fitzpatrick}}(2004)}]{2004PhPl...11.3961F}%
  \BibitemOpen
  \bibfield  {author} {\bibinfo {author} {\bibfnamefont {R.}~\bibnamefont
  {{Fitzpatrick}}},\ }\Doi {10.1063/1.1768956} {\bibfield  {journal} {\bibinfo
  {journal} {Physics of Plasmas},\ }\textbf {\bibinfo {volume} {11}},\ \bibinfo
  {pages} {3961} (\bibinfo {year} {2004})}\BibitemShut {NoStop}%
\bibitem [{\citenamefont {{Altschuler}}\ and\ \citenamefont
  {{Newkirk}}(1969)}]{1969altschuler}%
  \BibitemOpen
  \bibfield  {author} {\bibinfo {author} {\bibfnamefont {M.~D.}\ \bibnamefont
  {{Altschuler}}}\ and\ \bibinfo {author} {\bibfnamefont {G.}~\bibnamefont
  {{Newkirk}}},\ }\Doi {10.1007/BF00145734} {\bibfield  {journal} {\bibinfo
  {journal} {\solphys},\ }\textbf {\bibinfo {volume} {9}},\ \bibinfo {pages}
  {131} (\bibinfo {year} {1969})}\BibitemShut {NoStop}%
\bibitem [{\citenamefont {{Schatten}}\ \emph {et~al.}(1969)\citenamefont
  {{Schatten}}, \citenamefont {{Wilcox}},\ and\ \citenamefont
  {{Ness}}}]{1969schatten}%
  \BibitemOpen
  \bibfield  {author} {\bibinfo {author} {\bibfnamefont {K.~H.}\ \bibnamefont
  {{Schatten}}}, \bibinfo {author} {\bibfnamefont {J.~M.}\ \bibnamefont
  {{Wilcox}}}, \ and\ \bibinfo {author} {\bibfnamefont {N.~F.}\ \bibnamefont
  {{Ness}}},\ }\Doi {10.1007/BF00146478} {\bibfield  {journal} {\bibinfo
  {journal} {\solphys},\ }\textbf {\bibinfo {volume} {6}},\ \bibinfo {pages}
  {442} (\bibinfo {year} {1969})}\BibitemShut {NoStop}%
\bibitem [{\citenamefont {{Hoeksema}}(1984)}]{1984hoeksema}%
  \BibitemOpen
  \bibfield  {author} {\bibinfo {author} {\bibfnamefont {J.~T.}\ \bibnamefont
  {{Hoeksema}}},\ }\emph {\bibinfo {title} {{Structure and evolution of the
  large scale solar and heliospheric magnetic fields}}},\ \href@noop {} {Ph.D.
  thesis},\ \bibinfo  {school} {Stanford Univ., CA.} (\bibinfo {year}
  {1984})\BibitemShut {NoStop}%
\bibitem [{\citenamefont {{Wang}}\ and\ \citenamefont
  {{Sheeley}}(1992)}]{1992wang}%
  \BibitemOpen
  \bibfield  {author} {\bibinfo {author} {\bibfnamefont {Y.-M.}\ \bibnamefont
  {{Wang}}}\ and\ \bibinfo {author} {\bibfnamefont {N.~R.}\ \bibnamefont
  {{Sheeley}}, \bibfnamefont {Jr.}},\ }\Doi {10.1086/171430} {\bibfield
  {journal} {\bibinfo  {journal} {\apj},\ }\textbf {\bibinfo {volume} {392}},\
  \bibinfo {pages} {310} (\bibinfo {year} {1992})}\BibitemShut {NoStop}%
\bibitem [{\citenamefont {{Schrijver}}\ and\ \citenamefont {{De
  Rosa}}(2003)}]{2003derosa}%
  \BibitemOpen
  \bibfield  {author} {\bibinfo {author} {\bibfnamefont {C.~J.}\ \bibnamefont
  {{Schrijver}}}\ and\ \bibinfo {author} {\bibfnamefont {M.~L.}\ \bibnamefont
  {{De Rosa}}},\ }\Doi {10.1023/A:1022908504100} {\bibfield  {journal}
  {\bibinfo  {journal} {\solphys},\ }\textbf {\bibinfo {volume} {212}},\
  \bibinfo {pages} {165} (\bibinfo {year} {2003})}\BibitemShut {NoStop}%
\bibitem [{\citenamefont {Carpenter}\ and\ \citenamefont
  {Gottlieb}(1995)}]{Carpenter95spectralmethods}%
  \BibitemOpen
  \bibfield  {author} {\bibinfo {author} {\bibfnamefont {M.~H.}\ \bibnamefont
  {Carpenter}}\ and\ \bibinfo {author} {\bibfnamefont {D.}~\bibnamefont
  {Gottlieb}},\ }\href@noop {} {\bibfield  {journal} {\bibinfo  {journal} {J.
  Comput. Phys},\ }\textbf {\bibinfo {volume} {129}},\ \bibinfo {pages} {74}
  (\bibinfo {year} {1995})}\BibitemShut {NoStop}%
\bibitem [{\citenamefont {{Suda}}\ and\ \citenamefont
  {{Takami}}(2002)}]{sudafast2002}%
  \BibitemOpen
  \bibfield  {author} {\bibinfo {author} {\bibfnamefont {R.}~\bibnamefont
  {{Suda}}}\ and\ \bibinfo {author} {\bibfnamefont {M.}~\bibnamefont
  {{Takami}}},\ }\Doi {10.1090/S0025-5718-01-01386-2} {\bibfield  {journal}
  {\bibinfo  {journal} {Mathematics of Computation},\ }\textbf {\bibinfo
  {volume} {71}},\ \bibinfo {pages} {703} (\bibinfo {year} {2002})},\ ISSN
  \bibinfo {issn} {0025-5718}\BibitemShut {NoStop}%
\bibitem [{\citenamefont {{T{\'o}th}}\ \emph {et~al.}(2011)\citenamefont
  {{T{\'o}th}}, \citenamefont {{van der Holst}},\ and\ \citenamefont
  {{Huang}}}]{2011toth}%
  \BibitemOpen
  \bibfield  {author} {\bibinfo {author} {\bibfnamefont {G.}~\bibnamefont
  {{T{\'o}th}}}, \bibinfo {author} {\bibfnamefont {B.}~\bibnamefont {{van der
  Holst}}}, \ and\ \bibinfo {author} {\bibfnamefont {Z.}~\bibnamefont
  {{Huang}}},\ }\Doi {10.1088/0004-637X/732/2/102} {\bibfield  {journal}
  {\bibinfo  {journal} {\apj},\ }\textbf {\bibinfo {volume} {732}},\ \bibinfo
  {eid} {102} (\bibinfo {year} {2011})},\ \Eprint
  {http://arxiv.org/abs/1104.5672} {arXiv:1104.5672 [astro-ph.SR]} \BibitemShut
  {NoStop}%
\bibitem [{\citenamefont {{Chiu}}\ and\ \citenamefont
  {{Hilton}}(1977)}]{1977chiu}%
  \BibitemOpen
  \bibfield  {author} {\bibinfo {author} {\bibfnamefont {Y.~T.}\ \bibnamefont
  {{Chiu}}}\ and\ \bibinfo {author} {\bibfnamefont {H.~H.}\ \bibnamefont
  {{Hilton}}},\ }\Doi {10.1086/155111} {\bibfield  {journal} {\bibinfo
  {journal} {\apj},\ }\textbf {\bibinfo {volume} {212}},\ \bibinfo {pages}
  {873} (\bibinfo {year} {1977})}\BibitemShut {NoStop}%
\bibitem [{\citenamefont {{Rucklidge}}\ \emph {et~al.}(2000)\citenamefont
  {{Rucklidge}}, \citenamefont {{Weiss}}, \citenamefont {{Brownjohn}},
  \citenamefont {{Matthews}},\ and\ \citenamefont
  {{Proctor}}}]{rucklidgeWeiss2000}%
  \BibitemOpen
  \bibfield  {author} {\bibinfo {author} {\bibfnamefont {A.~M.}\ \bibnamefont
  {{Rucklidge}}}, \bibinfo {author} {\bibfnamefont {N.~O.}\ \bibnamefont
  {{Weiss}}}, \bibinfo {author} {\bibfnamefont {D.~P.}\ \bibnamefont
  {{Brownjohn}}}, \bibinfo {author} {\bibfnamefont {P.~C.}\ \bibnamefont
  {{Matthews}}}, \ and\ \bibinfo {author} {\bibfnamefont {M.~R.~E.}\
  \bibnamefont {{Proctor}}},\ }\href@noop {} {\bibfield  {journal} {\bibinfo
  {journal} {Journal of Fluid Mechanics},\ }\textbf {\bibinfo {volume} {419}},\
  \bibinfo {pages} {283} (\bibinfo {year} {2000})}\BibitemShut {NoStop}%
\bibitem [{\citenamefont {{Xia}}\ \emph
  {et~al.}(2012){\natexlab{a}}\citenamefont {{Xia}}, \citenamefont {{Chen}},\
  and\ \citenamefont {{Keppens}}}]{xia2012}%
  \BibitemOpen
  \bibfield  {author} {\bibinfo {author} {\bibfnamefont {C.}~\bibnamefont
  {{Xia}}}, \bibinfo {author} {\bibfnamefont {P.~F.}\ \bibnamefont {{Chen}}}, \
  and\ \bibinfo {author} {\bibfnamefont {R.}~\bibnamefont {{Keppens}}},\
  }\href@noop {} {\bibfield  {journal} {\bibinfo  {journal} {ArXiv e-prints}}
  (\bibinfo {year} {2012}{\natexlab{a}})},\ \Eprint
  {http://arxiv.org/abs/1202.6185} {arXiv:1202.6185 [astro-ph.SR]} \BibitemShut
  {NoStop}%
\bibitem [{\citenamefont {{Fang}}\ \emph {et~al.}(2013)\citenamefont {{Fang}},
  \citenamefont {{Xia}},\ and\ \citenamefont {{Keppens}}}]{fangXia2013}%
  \BibitemOpen
  \bibfield  {author} {\bibinfo {author} {\bibfnamefont {X.}~\bibnamefont
  {{Fang}}}, \bibinfo {author} {\bibfnamefont {C.}~\bibnamefont {{Xia}}}, \
  and\ \bibinfo {author} {\bibfnamefont {R.}~\bibnamefont {{Keppens}}},\ }\Doi
  {10.1088/2041-8205/771/2/L29} {\bibfield  {journal} {\bibinfo  {journal}
  {\apjl},\ }\textbf {\bibinfo {volume} {771}},\ \bibinfo {eid} {L29} (\bibinfo
  {year} {2013})},\ \Eprint {http://arxiv.org/abs/1306.4759} {arXiv:1306.4759
  [astro-ph.SR]} \BibitemShut {NoStop}%
\bibitem [{\citenamefont {{Longcope}}\ and\ \citenamefont
  {{Strauss}}(1994)}]{longcope}%
  \BibitemOpen
  \bibfield  {author} {\bibinfo {author} {\bibfnamefont {D.~W.}\ \bibnamefont
  {{Longcope}}}\ and\ \bibinfo {author} {\bibfnamefont {H.~R.}\ \bibnamefont
  {{Strauss}}},\ }\Doi {10.1086/175045} {\bibfield  {journal} {\bibinfo
  {journal} {\apj},\ }\textbf {\bibinfo {volume} {437}},\ \bibinfo {pages}
  {851} (\bibinfo {year} {1994})}\BibitemShut {NoStop}%
\bibitem [{\citenamefont {{{\v C}ada}}\ and\ \citenamefont
  {{Torrilhon}}(2009){\natexlab{b}}}]{cada}%
  \BibitemOpen
  \bibfield  {author} {\bibinfo {author} {\bibfnamefont {M.}~\bibnamefont {{{\v
  C}ada}}}\ and\ \bibinfo {author} {\bibfnamefont {M.}~\bibnamefont
  {{Torrilhon}}},\ }\Doi {10.1016/j.jcp.2009.02.020} {\bibfield  {journal}
  {\bibinfo  {journal} {Journal of Computational Physics},\ }\textbf {\bibinfo
  {volume} {228}},\ \bibinfo {pages} {4118} (\bibinfo {year}
  {2009}{\natexlab{b}})}\BibitemShut {NoStop}%
\bibitem [{\citenamefont {{van Marle}}\ and\ \citenamefont
  {{Keppens}}(2012)}]{marle12}%
  \BibitemOpen
  \bibfield  {author} {\bibinfo {author} {\bibfnamefont {A.~J.}\ \bibnamefont
  {{van Marle}}}\ and\ \bibinfo {author} {\bibfnamefont {R.}~\bibnamefont
  {{Keppens}}},\ }\Doi {10.1051/0004-6361/201218957} {\bibfield  {journal}
  {\bibinfo  {journal} {\aap},\ }\textbf {\bibinfo {volume} {547}},\ \bibinfo
  {eid} {A3} (\bibinfo {year} {2012})},\ \Eprint
  {http://arxiv.org/abs/1209.4496} {arXiv:1209.4496 [astro-ph.SR]} \BibitemShut
  {NoStop}%
\bibitem [{\citenamefont {{van Marle}}\ \emph
  {et~al.}(2011){\natexlab{b}}\citenamefont {{van Marle}}, \citenamefont
  {{Keppens}},\ and\ \citenamefont {{Meliani}}}]{marlebin}%
  \BibitemOpen
  \bibfield  {author} {\bibinfo {author} {\bibfnamefont {A.~J.}\ \bibnamefont
  {{van Marle}}}, \bibinfo {author} {\bibfnamefont {R.}~\bibnamefont
  {{Keppens}}}, \ and\ \bibinfo {author} {\bibfnamefont {Z.}~\bibnamefont
  {{Meliani}}},\ }\Doi {10.1051/0004-6361/201015517} {\bibfield  {journal}
  {\bibinfo  {journal} {\aap},\ }\textbf {\bibinfo {volume} {527}},\ \bibinfo
  {eid} {A3} (\bibinfo {year} {2011}{\natexlab{b}})},\ \Eprint
  {http://arxiv.org/abs/1011.1734} {arXiv:1011.1734 [astro-ph.GA]} \BibitemShut
  {NoStop}%
\bibitem [{\citenamefont {{Xia}}\ \emph
  {et~al.}(2012){\natexlab{b}}\citenamefont {{Xia}}, \citenamefont {{Chen}},\
  and\ \citenamefont {{Keppens}}}]{xia12}%
  \BibitemOpen
  \bibfield  {author} {\bibinfo {author} {\bibfnamefont {C.}~\bibnamefont
  {{Xia}}}, \bibinfo {author} {\bibfnamefont {P.~F.}\ \bibnamefont {{Chen}}}, \
  and\ \bibinfo {author} {\bibfnamefont {R.}~\bibnamefont {{Keppens}}},\ }\Doi
  {10.1088/2041-8205/748/2/L26} {\bibfield  {journal} {\bibinfo  {journal}
  {\apjl},\ }\textbf {\bibinfo {volume} {748}},\ \bibinfo {eid} {L26} (\bibinfo
  {year} {2012}{\natexlab{b}})},\ \Eprint {http://arxiv.org/abs/1202.6185}
  {arXiv:1202.6185 [astro-ph.SR]} \BibitemShut {NoStop}%
\bibitem [{\citenamefont {{Riley}}\ \emph {et~al.}(2011)\citenamefont
  {{Riley}}, \citenamefont {{Lionello}}, \citenamefont {{Linker}},
  \citenamefont {{Mikic}}, \citenamefont {{Luhmann}},\ and\ \citenamefont
  {{Wijaya}}}]{riley}%
  \BibitemOpen
  \bibfield  {author} {\bibinfo {author} {\bibfnamefont {P.}~\bibnamefont
  {{Riley}}}, \bibinfo {author} {\bibfnamefont {R.}~\bibnamefont {{Lionello}}},
  \bibinfo {author} {\bibfnamefont {J.~A.}\ \bibnamefont {{Linker}}}, \bibinfo
  {author} {\bibfnamefont {Z.}~\bibnamefont {{Mikic}}}, \bibinfo {author}
  {\bibfnamefont {J.}~\bibnamefont {{Luhmann}}}, \ and\ \bibinfo {author}
  {\bibfnamefont {J.}~\bibnamefont {{Wijaya}}},\ }\Doi
  {10.1007/s11207-010-9698-x} {\bibfield  {journal} {\bibinfo  {journal}
  {\solphys},\ }\textbf {\bibinfo {volume} {274}},\ \bibinfo {pages} {361}
  (\bibinfo {year} {2011})}\BibitemShut {NoStop}%
\bibitem [{\citenamefont {{van der Holst}}\ \emph {et~al.}(2014)\citenamefont
  {{van der Holst}}, \citenamefont {{Sokolov}}, \citenamefont {{Meng}},
  \citenamefont {{Jin}}, \citenamefont {{Manchester}}, \citenamefont
  {{T{\'o}th}},\ and\ \citenamefont {{Gombosi}}}]{bart14}%
  \BibitemOpen
  \bibfield  {author} {\bibinfo {author} {\bibfnamefont {B.}~\bibnamefont {{van
  der Holst}}}, \bibinfo {author} {\bibfnamefont {I.~V.}\ \bibnamefont
  {{Sokolov}}}, \bibinfo {author} {\bibfnamefont {X.}~\bibnamefont {{Meng}}},
  \bibinfo {author} {\bibfnamefont {M.}~\bibnamefont {{Jin}}}, \bibinfo
  {author} {\bibfnamefont {W.~B.}\ \bibnamefont {{Manchester}}, \bibfnamefont
  {IV}}, \bibinfo {author} {\bibfnamefont {G.}~\bibnamefont {{T{\'o}th}}}, \
  and\ \bibinfo {author} {\bibfnamefont {T.~I.}\ \bibnamefont {{Gombosi}}},\
  }\Doi {10.1088/0004-637X/782/2/81} {\bibfield  {journal} {\bibinfo  {journal}
  {\apj},\ }\textbf {\bibinfo {volume} {782}},\ \bibinfo {eid} {81} (\bibinfo
  {year} {2014})},\ \Eprint {http://arxiv.org/abs/1311.4093} {arXiv:1311.4093
  [astro-ph.SR]} \BibitemShut {NoStop}%
\bibitem [{\citenamefont {{Chan{\'e}}}\ \emph {et~al.}(2013)\citenamefont
  {{Chan{\'e}}}, \citenamefont {{Saur}},\ and\ \citenamefont
  {{Poedts}}}]{chane13}%
  \BibitemOpen
  \bibfield  {author} {\bibinfo {author} {\bibfnamefont {E.}~\bibnamefont
  {{Chan{\'e}}}}, \bibinfo {author} {\bibfnamefont {J.}~\bibnamefont {{Saur}}},
  \ and\ \bibinfo {author} {\bibfnamefont {S.}~\bibnamefont {{Poedts}}},\ }\Doi
  {10.1002/jgra.50258} {\bibfield  {journal} {\bibinfo  {journal} {Journal of
  Geophysical Research (Space Physics)},\ }\textbf {\bibinfo {volume} {118}},\
  \bibinfo {pages} {2157} (\bibinfo {year} {2013})}\BibitemShut {NoStop}%
\bibitem [{\citenamefont {{Porth}}\ \emph {et~al.}(2013)\citenamefont
  {{Porth}}, \citenamefont {{Komissarov}},\ and\ \citenamefont
  {{Keppens}}}]{PorthKomissarov2013}%
  \BibitemOpen
  \bibfield  {author} {\bibinfo {author} {\bibfnamefont {O.}~\bibnamefont
  {{Porth}}}, \bibinfo {author} {\bibfnamefont {S.~S.}\ \bibnamefont
  {{Komissarov}}}, \ and\ \bibinfo {author} {\bibfnamefont {R.}~\bibnamefont
  {{Keppens}}},\ }\Doi {10.1093/mnrasl/slt006} {\bibfield  {journal} {\bibinfo
  {journal} {\mnras},\ }\textbf {\bibinfo {volume} {431}},\ \bibinfo {pages}
  {L48} (\bibinfo {year} {2013})},\ \Eprint {http://arxiv.org/abs/1212.1382}
  {arXiv:1212.1382 [astro-ph.HE]} \BibitemShut {NoStop}%
\end{thebibliography}%
\clearpage

\end{document}